\documentclass[twocolumn,superscriptaddress,amsmath,amssymb,floatfix,aps,showpacs]{revtex4}
\usepackage{amsmath}    

\usepackage{graphicx}   
\usepackage{verbatim}   
\usepackage{color}      
\usepackage{subfigure}  
\usepackage{hyperref}   

\usepackage{amsmath}    
\usepackage{graphicx}   
\usepackage{verbatim}   
\usepackage{color}      
\usepackage{subfigure}  
\usepackage{hyphenat}
\usepackage[normalem]{ulem}
\usepackage{array}

\usepackage{color}      
\usepackage{soul}

\newcommand{\fref}[1]{Fig.~\ref{fig:#1}}

\newcommand{\frefstwo}[2]{Figs.~\ref{fig:#1}~and~\ref{fig:#2}}

\newcommand{\flabel}[1]{\label{fig:#1}}
\newcommand{\eref}[1]{Eq.~\ref{eqn:#1}}

\newcommand{\erefstwo}[2]{Eqs.~\ref{eqn:#1}~and~\ref{eqn:#2}}

\newcommand{\erefsrange}[2]{Eqs.~\ref{eqn:#1}-\ref{eqn:#2}}
\newcommand{\elabel}[1]{\label{eqn:#1}}

\newcommand{\avg}[1]{\langle #1\rangle}

\renewcommand{\b}[1]{{#1}}

\begin{document}
\bibliographystyle{prsty} 

\title{Robustness of clocks to input noise}
\author{Michele Monti} \affiliation{FOM Institute AMOLF,
  Science Park 104, 1098 XE Amsterdam, The Netherlands}
\author{David K. Lubensky}\affiliation{Department of Physics,
    University of Michigan, Ann Arbor, MI 48109-1040}
 \author{Pieter
  Rein ten Wolde} \affiliation{FOM Institute AMOLF, Science Park 104,
  1098 XE Amsterdam, The Netherlands}

\begin{abstract}
  To estimate the time, many organisms, ranging from cyanobacteria to
  animals, employ a circadian clock which is based on a limit-cycle
  oscillator that can tick autonomously with a nearly 24h period. Yet,
  a limit-cycle oscillator is not essential for knowing the time, as
  exemplified by bacteria that possess an ``hourglass'': a system that
  when forced by an oscillatory light input exhibits robust
  oscillations from which the organism can infer the time, but that in
  the absence of driving relaxes to a stable fixed point. Here,
  using models of the Kai system of cyanobacteria, we compare a
  limit-cycle oscillator with two hourglass models, one that without
  driving relaxes exponentially and one that does so in an oscillatory
  fashion.  In the limit of low input noise, all three systems are
  equally informative on time, yet in the regime of high input-noise
  the limit-cycle oscillator is far superior. The same behavior is
  found in the Stuart-Landau model, indicating that our result is
  universal.
\end{abstract}

\maketitle
\section{Introduction}
Many organisms, ranging from animals, plants, insects, to even
bacteria, \b{possess a circadian clock, which is a biochemical
  oscillator that can tick autonomously with a nearly 24h
  period. Competition experiments on cyanobacteria have demonstrated
  that these clocks can confer a fitness benefit to organisms that
  live in a rhythmic environment with a 24h period
  \cite{Ouyang:1998wp,Woelfle:2004cq}. Clocks enable organisms to
  estimate the time of day, allowing them to anticipate, rather than
  respond to, the daily changes in the environment.}  While it is
clear that circadian clocks which are entrained to their environment
make it possible to estimate the time, it is far less obvious that
they are the only or best means to do so
\cite{Roenneberg:2002tt,Ma:2016ca}. The oscillatory environmental
input could, for example, also be used to drive a system which in the
absence of any driving would relax to a stable fixed point rather than
exhibit a limit cycle. The driving would then generate oscillations
from which the organism could infer the time.  It thus remains an open
question what the benefits of circadian clocks are in estimating the
time of day.

This question is highlighted by the timekeeping mechanisms of
prokaryotes. While circadian clocks are ubiquitous in eukaryotes, the
only known prokaryotes to possess circadian clocks are cyanobacteria,
which exhibit photosynthesis. The best characterized clock
is that of the cyanobacterium {\em Synechococcus elongatus}, which
consists of three proteins, KaiA, KaiB, and KaiC
\cite{Ishiura:1998vc}.  The central clock component is KaiC, which
forms a hexamer that is phosphorylated and dephosphorylated in a
cyclical fashion under the influence of KaiA and KaiB. This
phosphorylation cycle can be reconstitued in the test tube, forming
 a bonafide circadian clock that ticks autonomously in
the absence of any oscillatory driving with a period of nearly 24
hours \cite{Nakajima2005}.  However, {\it S. elongatus} is not the
only cyanobacterial species. {\it Prochlorococcus marinus} possesses
{\it kaiB} and {\it kaiC}, but lacks (functional) KaiA. Interestingly,
this species exhibits daily rhythms in gene expression under
light-dark (LD) cycles but not in constant conditions
\cite{Holtzendorff:2008dj,Zinser:2009js}.  Recently, Johnson and
coworkers made similar observations for the purple bacterium {\it
  Rhodopseudomonas palustris}, which harbors homologs of KaiB and
KaiC. Its growth rate depends on the KaiC homolog in LD but not
constant conditions \cite{Ma:2016ca}, suggesting that the
bacterium uses its Kai system to keep time. Moreover, this species too
does not exhibit sustained rhythms in constant conditions, but does
show daily rhythms in e.g. nitrogen fixation in cyclic
conditions. {\it P. marinus} and {\it R. palustris} thus appear to
keep time via an ``hourglass'' mechanism that relies on oscillatory
driving \cite{Holtzendorff:2008dj,Zinser:2009js,Ma:2016ca}.  These
observations raise the question why some bacterial species like {\it
  S. elongatus} have evolved a bonafide clock that can run freely,
while others have evolved an hourglass timekeeping system.

Troein et al. studied the evolution of timekeeping systems in silico
\cite{Troein:2009bm}. They found that only in the presence of seasonal
variations {\em and} stochastic fluctuations in the input signal did
systems evolve that can also oscillate autonomously. However,
organisms near the equator have evolved self-sustained oscillations
\cite{Ma:2016ca}, showing that seasonal variations cannot be
essential. \b{Pfeuty {\it et al.}  suggest that limit-cycle
  oscillators have evolved because they enable timekeepers that ignore
  the uninformative light-intensity fluctuations during the day
  (corresponding to a deadzone in the phase-response curve), yet
  selectively respond to the more informative intensity changes around
  dawn and dusk \cite{Pfeuty:2011em}}.

\begin{figure*}[t]
\centering
\includegraphics[width=2 \columnwidth]{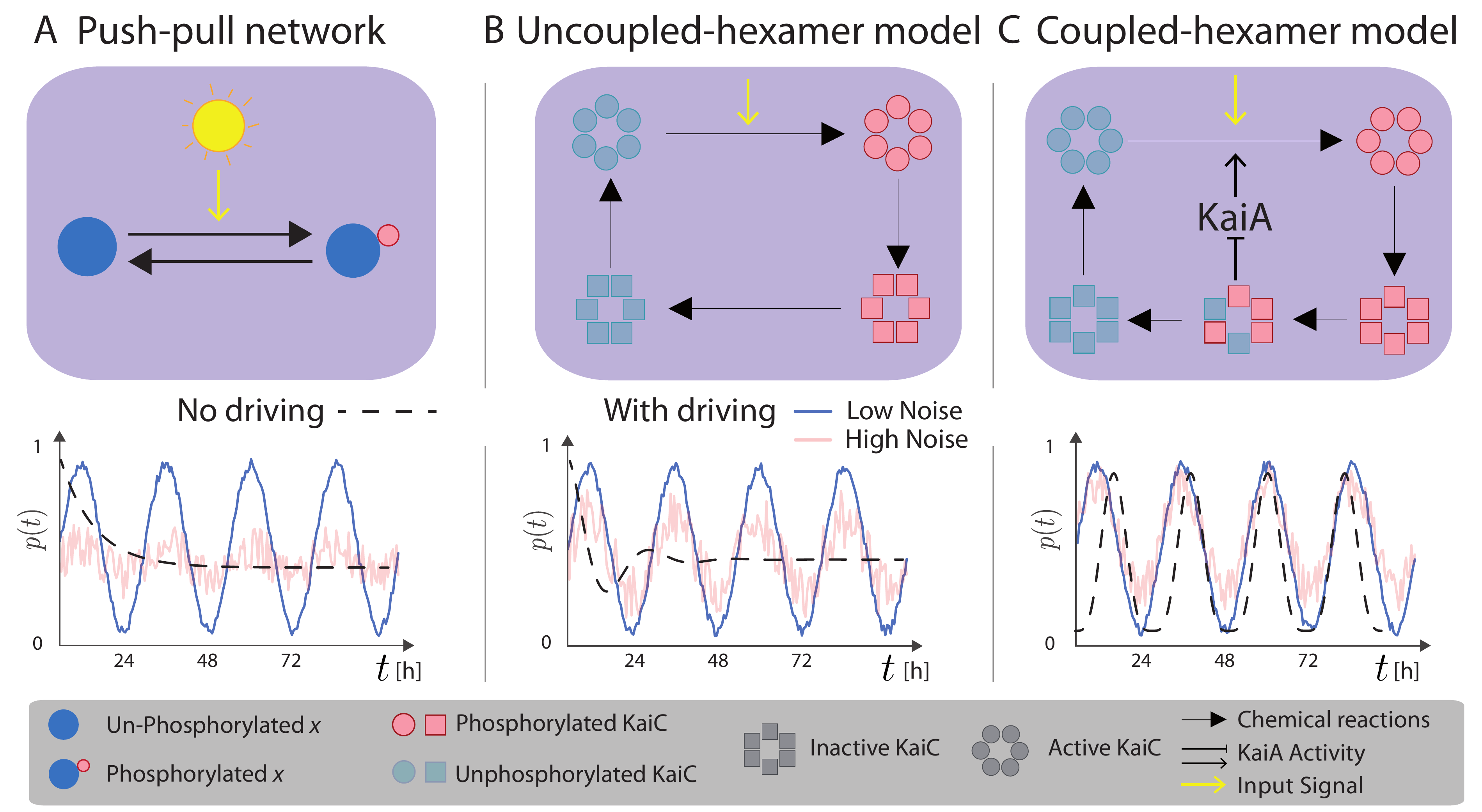}
\caption{Overview different timekeeping systems. (A) A push-pull
  network (PPN). Each protein can switch between a phosphorylated and
  an unphosphorylated state, and the input signal enhances the
  phosphorylation rate.  In the absence of driving, the PPN relaxes
  exponentially to a steady state (middle panel). Yet, in the presence
  of an oscillatory input, e.g. sunlight, the system exhibits
  oscillations from which the
  time can be inferred (lower panel). (B) The uncoupled-hexamer model
  (UHM), inspired by the Kai system of {\it P. marinus}. It consists
  of KaiC hexamers which can switch between an active state in which
  the phosphorylation level tends to rise and an inactive one in which
  it tends to fall. The phosphorylation rate is, via changes in the
  ATP/ADP ratio, enhanced by the light input
  \cite{Rust2011,Pattanayak:2015jm}. The system is akin to a harmonic
  oscillator, with an intrinsic frequency $\omega_0$, resulting from
  the hexamer phosphorylation cycle. However, the hexamers are not
  coupled via KaiA as in the CHM shown in panel C, so it cannot
  sustain autonomous oscillations; in the absence of driving, it relaxes
  in an oscillatory fashion to a stable fixed point (middle
  panel). (C) The coupled-hexamer model (CHM), inspired by the Kai
  system of {\it S. elongatus}. Like the UHM, it consists of KaiC
  hexamers, which tend to be phosphorylated cyclically. However, in
  contrast to the UHM, the hexamers are synchronized via KaiA, such
  that the system can exhibit limit-cycle oscillations in the absence
  of driving (middle panel). In all models, time is estimated from the
  protein phosphorylation fraction $p(t)$.  \flabel{Models}}
\end{figure*}

Here, we hypothesize that the optimal design of the readout system
that maximizes the reliability by which cells can estimate the time
depends on the noise in the input signal. To test this idea, we study
three different network designs from which the cell can infer time
(\fref{Models}): 1) a simple push-pull network (PPN), in which a
readout protein switches between a phosphorylated and an
unphosphorylated state (\fref{Models}A). Because the phosphorylation
rate increases with the light intensity, the phosphorylation level
oscillates in the presence of oscillatory driving, enabling the cell
to estimate the time. This network lacks an intrinsic oscillation
frequency, and in the absence of driving it relaxes to a stable fixed
point in an exponential fashion; 2) an uncoupled hexamer model (UHM),
which is inspired by the Kai system of {\it P. marinus}
(\fref{Models}B). This model consists of KaiC hexamers which each have
an inherent propensity to proceed through a phosphorylation
cycle. However, the phosphorylation cycles of the hexamers are not
coupled among each other, and without a common forcing the cycles will
therefore desynchronize, leading to the loss of macroscopic
oscillations.  In contrast to the proteins of the PPN, each hexamer is
a tiny oscillator with an intrinsic frequency $\omega_0$, which means
that an ensemble of hexamers that has been synchronized initially,
will, in the absence of driving, relax to its fixed point in an
oscillatory manner. 3) a coupled hexamer model (CHM), which is
inspired by the Kai system of {\it S. elongatus} (\fref{Models}C). As
in the previous UHM, each KaiC hexamer has an intrinsic capacity to
proceed through a phosphorylation cycle, but, in contrast to that
system, the cycles of the hexamers are coupled and synchronized via
KaiA, as described further below. Consequently, this system exhibits a
limit cycle, yielding macroscopic oscillations with intrinsic
frequency $\omega_0$ even in the absence of any driving.

\b{Here we are interested in the question how the precision of
  time estimation is limited by the noise in the input signal, and how
  this limit depends on the architecture of the readout system. We
  thus focus on the regime in which the input noise dominates over the
  internal noise \cite{Monti:2018hs} and model the different systems
  using mean-field (deterministic) chemical rate equations. In
  \cite{SI}, we also consider internal noise, and show that, at least
  for {\it S. elongatus}, the input-noise dominated regime is the
 relevant limit.}

The chemical rate equation of the PPN is: $\dot{x}_p = k_{\rm f}
s(t) (x_{\rm T}- x_p(t)) - k_{\rm b} x_p(t)$, where $x_p(t)$ is the
concentration of phosphorylated protein, $x_{\rm T}$ is the total
concentration, $k_{\rm f} s(t)$ is the phosphorylation rate $k_{\rm
  f}$ times the input signal $s(t)$, and $k_{\rm b}$ is the
dephosphorylation rate. The uncoupled (UHM) and coupled (CHM) hexamer
model are based on the Kai system
\cite{VanZon2007,Rust2007,Clodong2007,Mori2007a,Zwicker2010,Lin2014,Paijmans:2017gx,Paijmans:2017gp}. In
both models, KaiC switches between an active conformation in which the
phosphorylation level tends to rise and an inactive one in which it
tends to fall \cite{VanZon2007,Lin2014}. Experiments indicate that the
main Zeitgeber is the ATP/ADP ratio
\cite{Rust2011,Pattanayak:2015jm}, meaning the clock
predominantly couples to the input $s(t)$ during the phosphorylation
phase of the oscillations \cite{Rust2011,Paijmans:2017gp}. In both the
UHM and the CHM, $s(t)$ therefore modulates the phosphorylation rate
of active KaiC.
The principal difference between the UHM and CHM is KaiA: (functional)
KaiA is absent in {\it P. marinus} and hence in the UHM
\cite{Holtzendorff:2008dj,Zinser:2009js}. In contrast, in {\it
  S. elongatus} and hence the CHM, KaiA phosphorylates active KaiC,
yet inactive KaiC can bind and sequester KaiA. This gives
rise to the synchronisation mechanism of differential affinity
\cite{VanZon2007,Rust2007,SI}.  In all three models, the input is
modeled as a sinusoidal signal with mean $\bar{s}$ and driving
frequency $\omega=2\pi/T$ plus additive noise $\eta_s(t)$: $s(t) =
\sin(\omega t) + \bar{s} + \eta_s(t)$. The noise is uncorrelated with
the mean signal, and has strength $\sigma^2_s$ and correlation time
$\tau_c$, $\avg{\eta_s(t) \eta_s(t^\prime)} = \sigma_s^2
e^{-|t-t^\prime|/\tau_c}$.  A detailed description of the
models is given in \cite{SI}.

As a performance measure for the accuracy of estimating time, we use
the mutual information $I(p;t)$ between the time $t$ and the
phosphorylation level $p(t)$ \cite{Monti:2016bp,Monti:2018hs}:
\begin{align}
I(p;t) = \int_0^T dt \int_0^1 dp P(p,t) \log_2 \frac{P(p,t)}{P(p)P(t)}.
\end{align}
Here $P(p,t)$ is the joint probability distribution while $P(p)$ and
$P(t)=1/T$ are the marginal distributions of $p$ and
$t$. \b{The quantity $2^{I(p;t)}$ corresponds to the number of
  time points that can be inferred uniquely from $p(t)$;
  $I(p;t)=1{\rm bit}$ means that from $p(t)$ the cell can reliably
  distinguish between day and
  night \cite{Walczak:1324157}.} The distributions are obtained from
running long simulations of the chemical rate equations of the
different models \cite{SI}.

\b{For each system, to maximize the mutual information we first
  optimized over all parameters except the coupling strength. For the
  CHM, the coupling strength $\rho$ was taken to be comparable to that
  of {\it S. elongatus} \cite{SI}, and for the PPN and the UHM $\rho$
  was set to an arbitrary low value, because in the relevant
  weak-coupling regime the mutual information is independent of
  $\rho$, as elucidated below and in \cite{SI}}.  For the PPN, there
exists an optimal response time $\tau_r \sim 1/ k_{\rm b}$ that
maximizes $I(p;t)$, arising from a trade-off between maximizing the
amplitude of $p(t)$, which increases with decreasing $\tau_r$, and
minimizing the noise in $p(t)$, which decreases with increasing
$\tau_r$ because of time averaging \cite{Becker:2015iu,SI}. Similarly,
for the UHM, there exists an optimal intrinsic frequency $\omega_0$ of
the individual hexamers.  The UHM is linear and similar to a harmonic
oscillator. Analyzing this system shows that while the amplitude $A$
of the output $x(t)$ is maximized at resonance, $\omega_0\to\omega$,
the standard deviation $\sigma_x$ of $x$ is maximized when
$\omega_0\to 0$, such that the signal-to-noise ratio $A/\sigma_x$
peaks for $\omega_0 > \omega$ \cite{SI}. Interestingly, also the CHM
exhibits a maximum in $A/\sigma_x$ for intrinsic frequencies that are
slightly off-resonance \cite{SI}.

\begin{figure}[t]
\includegraphics[width=\columnwidth]{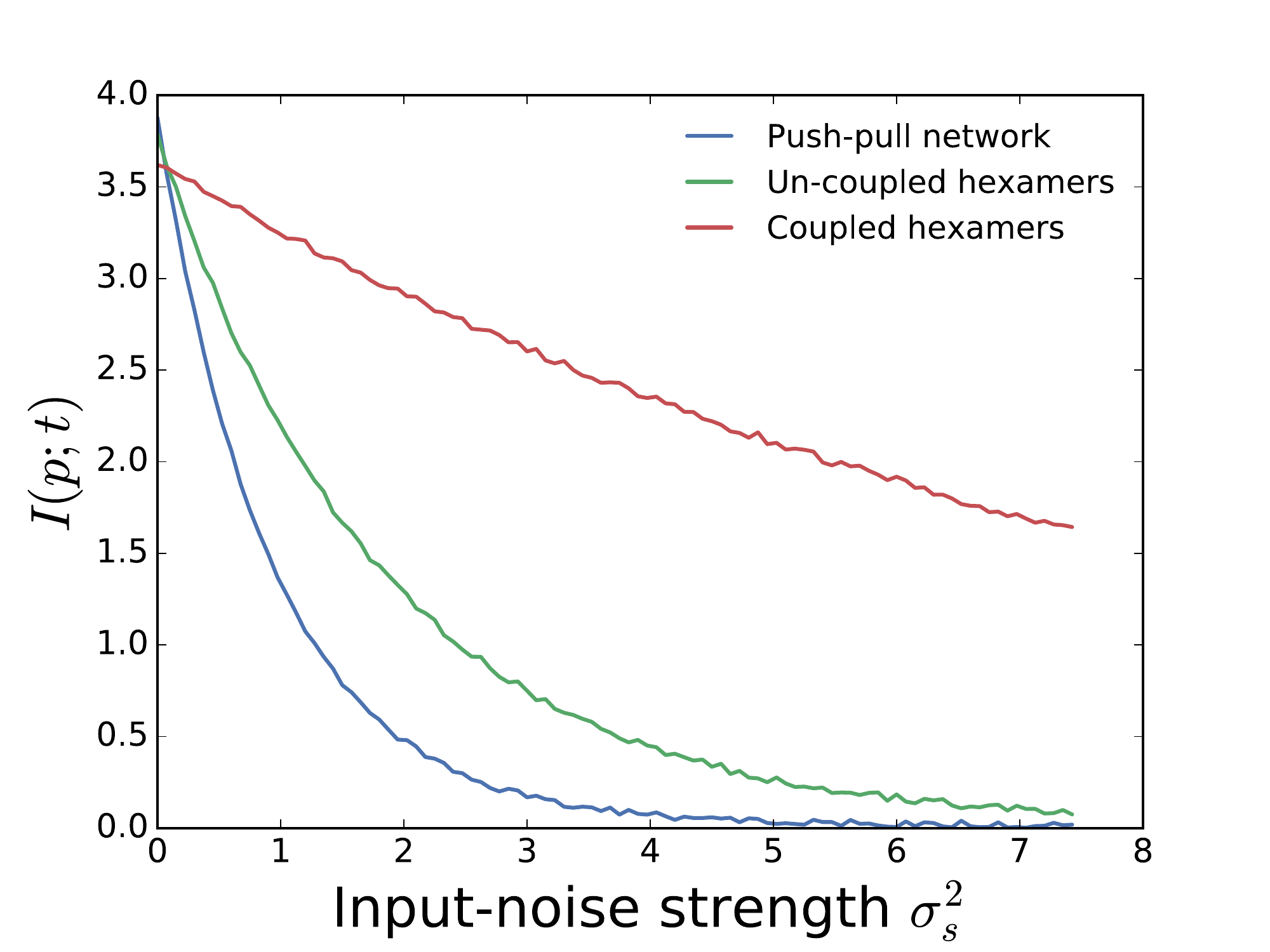}
\caption{The mutual information $I(p;t)$ as a function of the
  input-noise strength $\sigma^2_s$, for the push-pull network (PPN),
  the uncoupled-hexamer model (UHM) and the coupled-hexamer model
  (CHM), see \fref{Models}. In the limit of low input noise, all
  systems are equally informative on time, but in the high-noise
  regime the CHM is most accurate. The parameters have been optimized
  to maximize $I(p;t)$; since these are (nearly) independent of
  $\sigma^2_s$ (Figs.S1-S3), they are fixed (Table S1 \cite{SI}).
  \flabel{I_CPM}}
\end{figure}

\fref{I_CPM} shows the mutual information $I(p;t)$ as a function of
the input-noise strength $\sigma^2_s$ for the three systems.  In the
regime that $\sigma^2_s$ is small, $I(p;t)$ is essentially the same for
all systems. However, the figure also shows that as $\sigma^2_s$ rises,
$I(p;t)$ of the UHM and especially the PPN decrease very rapidly,
while that of the CHM falls much more slowly. For $\sigma^2_s
\approx 3$, $I(p;t)$ of the CHM is still above 2 bits, while $I(p;t)$
of the PPN and UHM have already dropped below 1 bit, meaning
the cell would no longer be able to distinguish
between day and night. Indeed, this figure shows that in the regime of
high input noise, a bonafide clock that can tick autonomously is a
much better time-keeper than a system which relies on oscillatory
driving to show oscillations. This is the principal result of our
paper. It is observed for other values of $\tau_c$ and other types of
input, such as a truncated sinusoid corresponding to no
driving at night (Fig. S6 \cite{SI}).

\begin{figure}[t]
\includegraphics[width=\columnwidth]{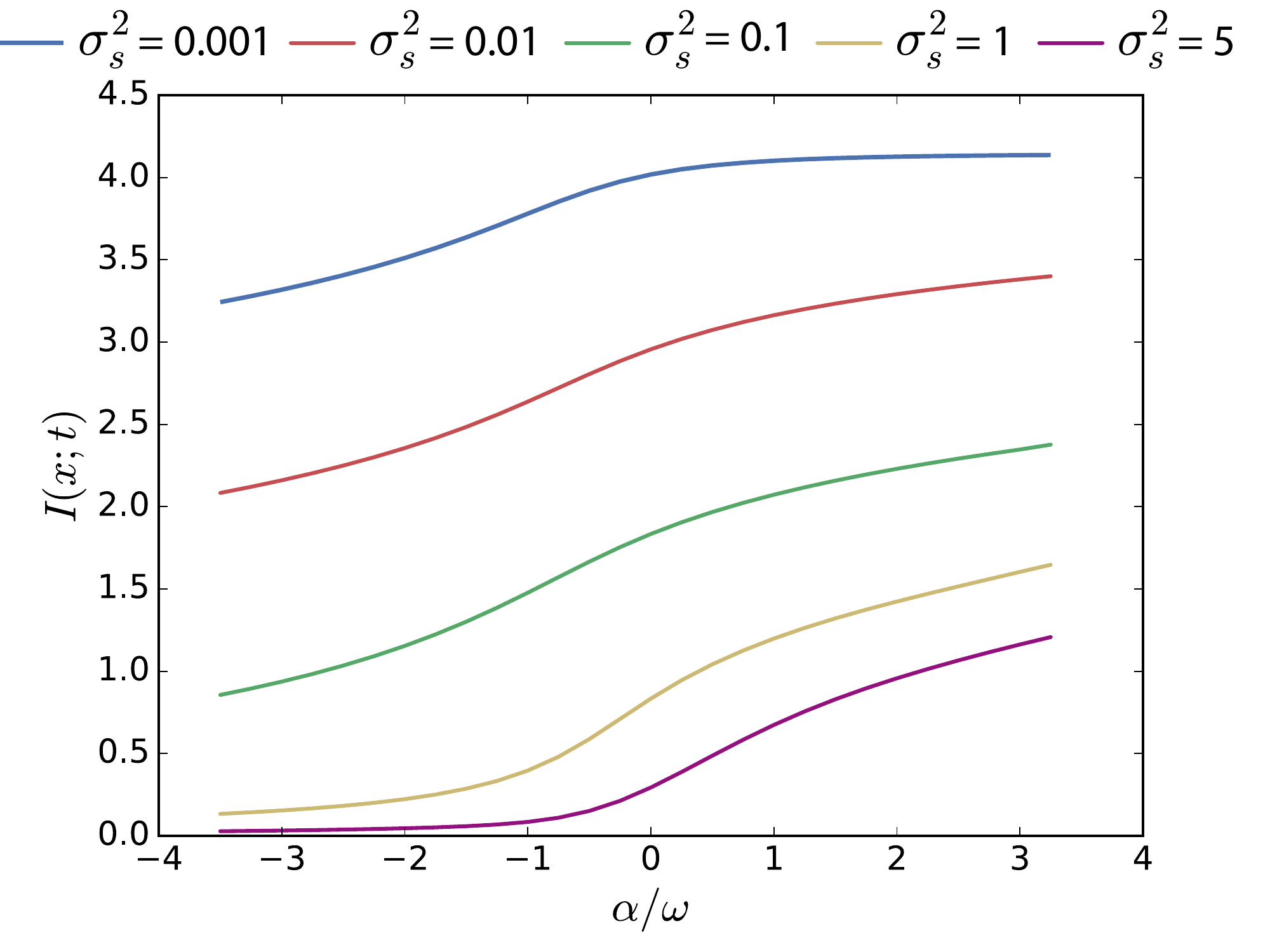}
\caption{The mutual information $I(p;t)$ as a function of $\alpha$ of
  the Stuart-Landau model (\eref{dotASL}), for different strengths
  of the input noise $\sigma^2_s$. Clearly, $I(p;t)$ rises as the
  system is changed from a damped oscillator like the UHM
  ($\alpha<0)$ to a limit-cycle oscillator like the CHM
  $(\alpha>0)$. Moreover, the increase is most pronounced when
  $\sigma^2_s$ is large, as also observed for the UHM and CHM, see
  \fref{I_CPM}. Parameters: $\nu=0$; $\beta = \omega$;
  $\epsilon=0.5\omega$; $\sigma^2_s$ in units of $\omega$.
  \flabel{I_SL}}
\end{figure}

The robustness of our observation that bonafide clocks are more
reliable timekeepers, suggests it is a universal phenomenon,
independent of the details of the system. We therefore analyzed a
generic minimal model, the Stuart-Landau model. It allows us to study
how the capacity to infer time changes as a system is altered from a
damped (nearly) linear oscillator, which has a characteristic
frequency but cannot sustain oscillations in the absence of driving,
to a non-linear oscillator that can sustain autonomous oscillations
\cite{SI}.  Near a Hopf bifurcation where a limit cycle appears the
effect of the non-linearity is weak, so that the solution $x(t)$ is
close to that of a harmonic oscillator,
 $x(t) = 1/2(A(t) e^{i\omega t} + c.c.)$, where
$A(t)$ is a complex amplitude that can be time-dependent
\cite{Pikovsky2003}. The dynamics of $A(t)$ is then given by
\begin{align}
\dot{A} &= -i \nu A + \alpha A - \beta
|A|^2 A - \epsilon E, \elabel{dotASL}
\end{align}
where $\nu \equiv (\omega^2-\omega_0^2) / (2\omega)$ with $\omega_0$
the intrinsic frequency, $\alpha$ and $\beta$
govern the linear and non-linear growth and decay of oscillations, $E$
is the first harmonic of $s(t)$ and $\epsilon\equiv \rho / (2 \omega)$
is the coupling strength. \eref{dotASL}
gives a universal description of a driven weakly non-linear oscillator
near a supercritical Hopf bifurcation \cite{Pikovsky2003}.
 
The non-driven system exhibits a Hopf bifurcation at $\alpha = 0$. By
varying $\alpha$ we can thus change the system from a {\em damped
  oscillator} ($\alpha<0)$ which in the absence of driving
exhibits oscillations that decay, to a {\em limit-cycle oscillator}
($\alpha>0$) that shows free-running oscillations. The driven damped
oscillator ($\alpha<0$) always has one stable fixed point with $|A|>0$
corresponding to sinusoidal oscillations that are synchronized with
the driving. The driven limit-cycle oscillator ($\alpha>0$), however,
can exhibit several distinct dynamical regimes \cite{Pikovsky2003}. Here, we
limit ourselves to the case of perfect synchronization, where $x(t)$
has a constant amplitude $A$ and phase shift with respect to $s(t)$.

To compute $I(x,t)$, we use an approach inspired by the
linear-noise approximation
\cite{Monti:2018hs}.  It assumes $P(x|t)$ is a
Gaussian distribution with variance $\sigma^2_x(t)$ centered at the
deterministic solution $x(t)= 1/2 (A e^{i \omega t}+c.c.)$, where $A$
is obtained by solving \eref{dotASL} in steady state. To find
$\sigma^2_x$, we first compute $\sigma^2_A$ from \eref{dotASL} by
adding Gaussian white-noise of strength $\sigma^2_s$ to $E$ and
expanding $A$ to linear order around its fixed point; $\sigma^2_x(t)$
is then obtained from $\sigma^2_A$ via a coordinate transformation
\cite{SI}.

\fref{I_SL} shows the mutual information $I(x;t)$ as a function
$\alpha$, for different values of $\sigma^2_s$. The figure shows that
$I(x;t)$ rises as the system is changed from a damped oscillator
($\alpha<0$) to a self-sustained oscillator ($\alpha>0$). Moreover, the
increase is most pronounced when the input noise $\sigma^2_s$ is
large. \b{The Stuart-Landau model can thus reproduce the qualitative
  behavior of our computational models, indicating that our principal
  result is generic. Interestingly, the CHM is even more
  robust to input noise than the Stuart-Landau
  model, likely because the latter is only weakly non-linear.}

\b{To understand why limit-cycle oscillators are
  more robust to input noise, we study in section SIIE \cite{SI}
  analytical models valid in the
    limit of weak coupling.  For a damped oscillator
  with a fixed-point attractor (PPN and UHM), we find that the amplitude $A$ of the
  harmonic oscillations (the signal) increases with the coupling strength
  $\rho$, $A\sim \rho$.
  The noise in the output signal $\sigma_x$ scales with $\rho$,
  $\sigma_x \sim \rho$, because the coupling amplifies not only the
  input signal, but also the input noise. Hence, the signal-to-noise
  ratio $A/\sigma_x$ is independent of $\rho$: an oscillator based on
  a fixed-point attractor faces a fundamental trade-off between gain
  and input noise (section SIIE \cite{SI}). A limit-cycle oscillator
  (CHM) can lift this trade-off: The amplitude is a robust, intrinsic
  property of the system, and essentially independent of $\rho$. The
  output noise $\sigma_x \sim \sqrt{\rho}$, because the coupling not
  only amplifies the input noise proportional to $\rho$, but also
  generates a restoring force that constrains fluctuations, scaling as
  $\sim \sqrt{\rho}$ (SIIE \cite{SI}). Hence, $A/\sigma_x \sim 1 /
  \sqrt{\rho}$. These scaling arguments show that: 1) concerning
  robustness to input noise, the optimal regime is the weak-coupling
  regime; 2) in this regime, a limit-cycle oscillator is generically
  more robust to input noise than a damped oscillator.}

\b{Yet, the coupling cannot be reduced to zero for limit-cycle
  oscillators. When the intrinsic clock period deviates from 24h, as
  it typically will, coupling is essential to phase-lock the clock to
  the driving signal \cite{Monti:2018hs}. Moreover, biochemical
  networks inevitably have some level of internal noise (section SIIF
  \cite{SI}). For the damped oscillator, the output noise $\sigma_x$
  resulting from internal noise is independent of $\rho$, but since
  $A$ increases with $\rho$, $A/\sigma_x \sim \rho$ in the presence of
  internal noise only: coupling helps to lift the signal above the
  internal noise. For the limit-cycle oscillator, the restoring force
  $\sim \sqrt{\rho}$ tames phase diffusion, such that in the presence
  of only internal noise, the output noise $\sigma_x \sim 1 /
  \sqrt{\rho}$ and $A / \sigma_x \sim \sqrt{\rho}$. Hence, also with
  regards to internal noise, a limit-cycle oscillator is superior to a
  damped oscillator in the weak-coupling regime. This analysis also
  shows, however, that this regime is not necessarily optimal, since
  with only internal noise present $A/\sigma_x$ increases with
  $\rho$. In fact, it predicts that in the strong-coupling regime the
  damped oscillator outperforms the limit-cycle oscillator. We
  emphasize, however, that in this regime our weak-coupling analysis breaks
  down and other effects come into play; for example, non-linearities
  arising from the bounded character of $p(t)$ distort the signal,
  reducing information transmission.}

\b{In the presence of both noise sources, we expect an optimal
  coupling that maximizes information transmission (SIIF
  \cite{SI}). For the limit-cycle oscillator the optimum arises from
  the trade-off between minimizing input-noise propagation and
  maximizing internal-noise suppression.  For the damped
  oscillator, $A/\sigma_x$ first rises with $\rho$ because coupling
  helps to lift the signal above the internal noise, but then plateaus
  when the input noise (which increases with $\rho$) dominates over
  the internal noise; for even higher $\rho$, it decreases again
  because of signal distortion. In section SIE \cite{SI} we verify
  these predictions for our computational models using stochastic
  simulations. }

\b{Experiments have shown that the clock of {\it S. elongatus} has a
  strong temporal stability with a correlation time of several months
  \cite{Mihalcescu:2004ch}, suggesting that the internal noise is
  small. Indeed, typical
  input-noise strengths based on weather data \cite{Gu:2001vh} and
  internal-noise strengths based on protein copy numbers in {\it
    S. elongatus} \cite{Kitayama:2003un} indicate that in the
  biologically relevant regime, at least for cyanobacteria, input
  noise dominates over internal noise (Fig. S5 \cite{SI}). In
  this regime, the focus of our paper, the optimal coupling is weak
  and limit-cycle oscillators are generically more robust to input
  noise than damped oscillators.}

This work is part of the research programme of the Netherlands
Organisation for Scientific Research (NWO) and was performed at
AMOLF. DKL acknowledges NSF grant DMR 1056456 and grant PHY 1607611 to
the Aspen Center for Physics, where part of this work was
completed. We thank Jeroen van Zon and Nils Becker for a critical
reading of the manuscript.



\pagebreak
\begin{center}
\textbf{\large Supplemental Material: \\ 
    Robustness of circadian clocks to input noise}
\end{center}

\setcounter{equation}{0}
\setcounter{figure}{0}
\setcounter{table}{0}
\setcounter{page}{1}
\setcounter{section}{0}
\makeatletter
\renewcommand{\theequation}{S\arabic{equation}}
\renewcommand{\thesection}{S\Roman{section}}
\renewcommand{\thefigure}{S\arabic{figure}}
\renewcommand{\thetable}{S\arabic{table}}

This supporting information provides background information on the
computational models and analytical models that we have studied. The
computational models are described in the next section, while the
analytical models are discussed in section \ref{sec:ANA}.

\section{Computational Models}

In this section, we describe the three computational models that we
have considered in this study: the push-pull network; the
uncoupled-hexamer model; and the coupled-hexamer model. We also
describe how we have modeled the input signal and how the systems are
coupled to the input. \b{As described in the main text, we are
  interested in the question how the robustness to input noise depends
  on the architecture of the readout system; we therefore model these
  systems with deterministic mean-field chemical rate
  equations. However, here in the {\it Supporting Information} we also
  test how robust our findings are, not only to the shape of the input
  signal, but also to the presence of internal noise.}

In the next section, we first describe how we have modeled the input
signal. In the subsequent sections, we then describe the
\b{deterministic} computational models, how they are coupled to the
input, and how we have set their parameters. Table \ref{tab:Models}
lists the values of all the parameters of all the models. In section
\b{\ref{sec:CompIntNoise} we show that the principal findings of
Fig.2 are robust to the presence of internal noise and in section
\ref{sec:Robustness} we show that they are
robust to the type of input signal and the noise correlation time}.

\subsection{Input signal}
\label{sec:Input}
The input signal is modeled as a sinusoidal oscillation with additive noise:
\begin{align}
s(t) = \sin(\omega t) + \bar{s} + \eta_s(t),
\elabel{s_t}
\end{align}
where $\bar{s}$ is the mean input signal and $\eta_s(t)$ describes the
input noise. The noise in the input is assumed to be uncorrelated with
the mean input signal $s(t)$. Moreover, we assume that the input noise
has strength $\sigma^2_s$ and is colored, relaxing exponentially with
correlation time $\tau_c$: $\avg{\eta_s(t)
  \eta_s(t^\prime)}=\sigma^2_s e^{-|t-t^\prime|/\tau_c}$.

The input signal $s(t)$ is coupled to the system by modulating the
phosphorylation rate $k_{\alpha}$ of the core clock protein, as we
describe in detail for the respective computational models in the next
sections. Here, $k_\alpha =k_{\rm f}, k_{\rm ps}, k_i$, depending
on the computational model. As we will see, the net phosphorylation
rate is given by
\begin{align}
k_\alpha s(t) &= k_\alpha s(t)\\
&=k_\alpha \bar{s} + k_\alpha \left (\sin(\omega t)
  + \eta_s\right).\elabel{kfs}
\end{align}
This expression shows that in the presence of oscillatory driving, the
mean phosphorylation rate averaged over a period is set by $k_\alpha
\bar{s}$, while the amplitude of the oscillation in the
phosphorylation rate, which sets the strength of the forcing, is given
by $k_\alpha$. We also note that $k_\alpha$ amplifies not only the
``true'' signal $\sin(\omega t)$, but also the noise $\eta_s$, the
consequences of which will be discussed below. Lastly, the absence of
any oscillatory driving is modeled by taking $s(t) = \bar{s}$, such
that the net phosphorylation rate is then $k_\alpha \bar{s}$. The
phosphorylation rate in the presence of stochastic driving is thus
characterized by the following parameters: the mean phosphorylation
rate $k_\alpha \bar{s}$, the amplitude of the phosphorylation-rate
oscillations $k_\alpha$, and the noise $\eta_s(t)$, characterized by
the noise strength $\sigma^2_s$ and correlation time $\tau_c$. We will
vary $\sigma^2_s$ and $\tau_c$ systematically, while $\bar{s}$ and
$k_\alpha$, together with the other system parameters, will be
optimized to maximize the mutual information, as described below.

\b{While we will vary $\sigma^2_s$, weather data gives us ball-park
  estimates for the typical input-noise strengths. The weather data of
  \cite{Gu:2001vh} indicates that the average relative noise
    intensity at noon is around $\avg{\delta I^2}/\avg{I}^2 \approx
    0.2 - 0.3$, which corresponds to $\sigma^2_s / \bar{s}^2$ in our
    model, yielding $\sigma^2_s \approx 1 - 2$ for the baseline
    parameter value of the mean signal $\bar{s}=2$ (see Table
    \ref{tab:Models}).  Because there will be variations in the
    fluctuations in the light intensity from day-to-day, we will also
    study higher values of the input noise.}

In the simulations, realisations of $\eta_s(t)$ are generated via the
Ornstein-Uhlenbeck process
\begin{align}
\dot{\eta}_s &= - \eta_s  / \tau_c + \xi (t),
\end{align}
where $\xi(t)$ is Gaussian white noise $\avg{\xi(t)\xi(t^\prime)} =
\avg{\xi^2}\delta (t-t^\prime)$. This generates colored noise of
$\eta_s(t)$, $\avg{\eta_s(t) \eta_s (t^\prime)}= \sigma^2_s
e^{-|t-t^\prime|/\tau_c}$, where $\sigma^2_s = \avg{\xi^2} \tau_c /
2$.  

The results of Fig. 2 of the main text correspond to $\tau_c = 0.5 /
{\rm h}$, consistent with the weather data of
\cite{Gu:2001vh}. However, we have tested the robustness of the
results by varying the noise correlation time $\tau_c$. In addition,
to test the robustness of our observations to changes in the shape of
the input signal, we have also varied that. These tests are described
in section \ref{sec:Robustness} and the results are shown in
\fref{Robustness}. Clearly, the principal result of Fig. 2 of the main
text is robust to changes in both the noise correlation time $\tau_c$
and the shape of the mean-input signal.

\begin{table*}
\begin{tabular}{ l l l }

\textbf{Parameter}& \textbf{Description} &
\textbf{Value} \\ 
\hline\hline
\underline{\textbf{Push-pull network, \eref{PPN_CPM}}}&&\\
$k_{\rm f}$& Phosphorylation rate & $0.01 / {\rm h}$\\
$k_{\rm b}$ & Dephosphorylation rate (\eref{muopt}) & $0.3 / {\rm h}$\\
\underline{\textbf{Uncoupled-hexamer model, \erefsrange{UHM_F}{UHM_L}}}&&\\
$k_{\rm f}$ & Phosphorylation rate & $0.26 / {\rm h}$\\
$k_{\rm b}$ & Dephosphorylation rate & $0.52 / {\rm h}$\\
$k_{\rm s}$& Conformational switching rate & $100 / {\rm h}$\\
\underline{\textbf{Coupled-hexamer model, \erefsrange{CHM_F}{Afree}}}\\
 $k_{\rm ps}$ & Autophosphorylation rate  &$0.0125 / {\rm h}$ \\
  $k_{\rm b}$ & Dephosphorylation rate  & $0.1875 / {\rm h}$ \\
  $k_{\rm s}$ & Conformational switching rate &$100 / {\rm h}$ \\
 $K_0$ & KaiA dissociation constant ${\rm C}_0$  & $0.0001 $\\
  $K_1$ & KaiA dissociation constant ${\rm C}_1$ & $0.0003$\\
  $K_2$ & KaiA dissociation constant ${\rm C}_2$ & $0.001$\\
  $K_3$ & KaiA aissociation constant ${\rm C}_3$ & $0.003$\\
$K_4$ & KaiA dissociation constant ${\rm C}_4$ & $0.01$\\
$K_5$ & KaiA dissociation constant ${\rm C}_5$ & $0.03$\\
  $k_0$ & KaiA-stimulated phosphorylation rate ${\rm C}_0$   & $0.5 /
  {\rm h}$\\
  $k_1$ & KaiA-stimulated phosphorylation rate ${\rm C}_1$& $0.5 /
  {\rm h}$\\
  $k_2$ & KaiA-stimulated phosphorylation rate ${\rm C}_2$ &
  $0.5 / {\rm h}$\\
  $k_3$ & KaiA-stimulated phosphorylation rate ${\rm C}_3$&
  $0.5 / {\rm h}$\\
  $k_4$ & KaiA-stimulated phosphorylation rate ${\rm C}_4$&
  $0.5 / {\rm h}$\\
  $k_5$ & KaiA-stimulated phosphorylation rate ${\rm C}_5$&
  $0.5 / {\rm h}$ \\
  $\tilde{b}_{2-4}$& Number KaiA dimers  sequestered by $\tilde{\rm C}_{1-4}$ & $2$ \\
$\tilde{b}_{0,5,6}$ & Number KaiA dimers sequestered by $\tilde{\rm C}_{0,5,6}$ & 0\\
$\tilde{K}_{1-4}$& KaiA dissociation constant $\tilde{\rm C}_{1-4}$&
0.000001\\
$\tilde{K}_{0,5,6}$&KaiA dissociation constant $\tilde{\rm C}_{0,5,6}$&$\infty$\\
  $c_{\rm T}$ & Total concentration of KaiC  & $1$ \\
  $A_{\rm T}$ & Total concentration of KaiA  & $1$ \\
    \hline
\end{tabular}
\caption{Parameter values of all the three computational models
  studied in the main text. The parameter values listed are those that
  maximize the mutual information $I(p;t)$ between the phosphorylation level
  $p$ and time $t$; these values are nearly
  independent of the input-noise strength $\sigma^2_s$, and thus kept
  constant as $\sigma^2_s$ is varied in the simulations corresponding
  to Fig. 2 of the main text. For these optimal parameters values, the intrinsic
  period of the uncoupled-hexamer model is $T_0^{\rm opt} \approx
  23.1 {\rm h}$ while that of the coupled-hexamer model is $T_0^{\rm
    opt} \approx 25.1 {\rm h}$. All three models are coupled to the input
  by multiplying the phosphorylation rates with $s(t) = \sin(\omega) +
  \bar{s} + \eta_s(t)$, where $\bar{s}=2$ and $\eta_s(t)$ describes
  colored noise with strength $\sigma^2_s$ and correlation time
  $\tau_c$, $\avg{\eta_s(t)\eta_s(t^\prime)}=\sigma^2_s
  e^{-|t-t^\prime|/\tau_c}$. For Fig. 2 of the main text, $\tau_c =
  0.5 {\rm h}$.  Dissociation constants and protein concentrations are in
  units of the total KaiC concentration. Note that in the absence of
  oscillatory driving $s(t) = \bar{s}=2$, meaning that in simulations
  of the non-driven systems the
  phosphorylation rates $k_{\rm f}$, $k_i$, $k_{\rm ps}$, still have to be
  multiplied by $\bar{s}=2$. \label{tab:Models}}
\end{table*}

\begin{figure*}[t]
\includegraphics[width=2\columnwidth]{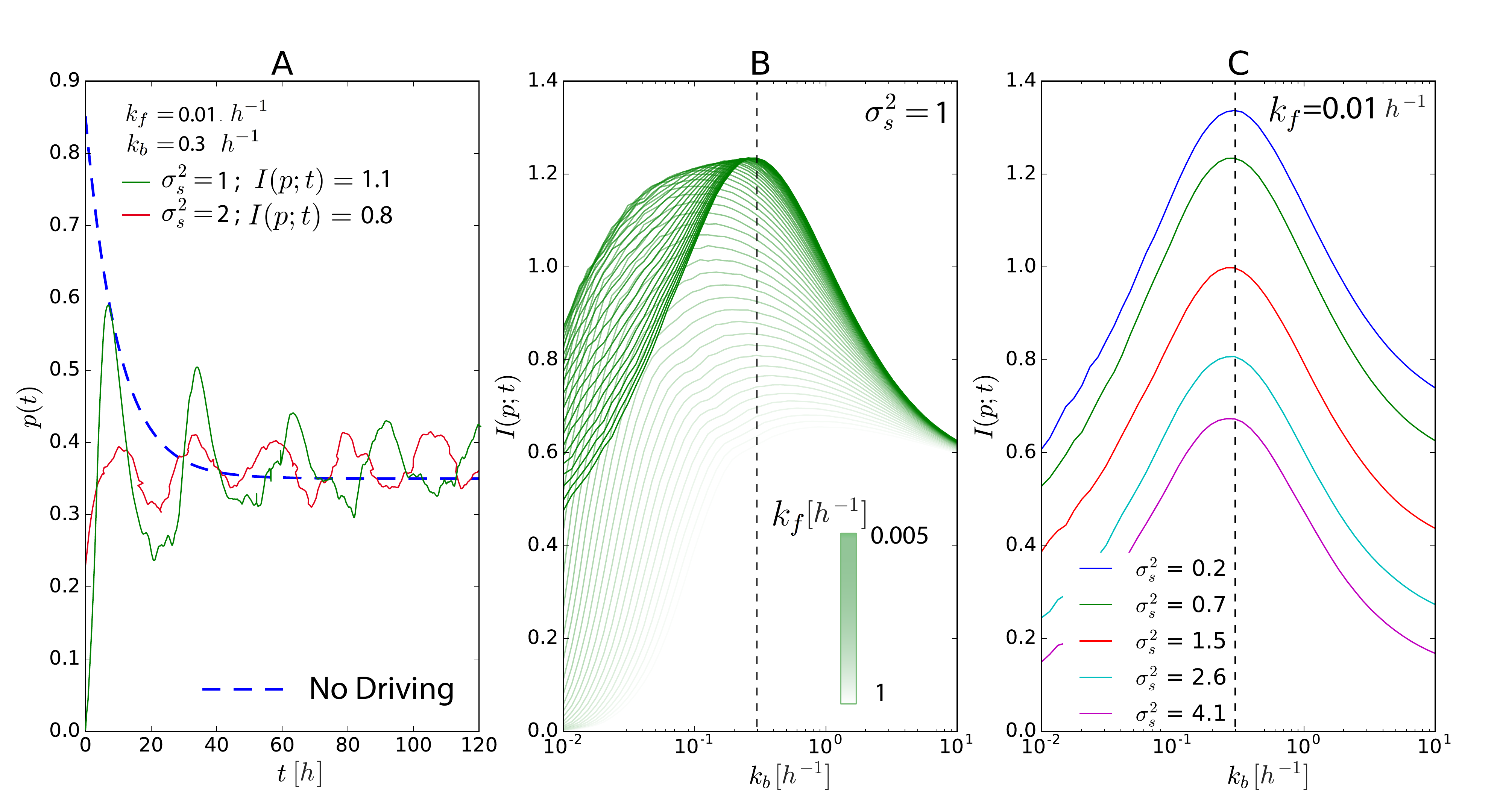}
\caption{The \b{deterministic} push-pull network. (A) Time traces of
  $p(t)$ in the absence of driving (dashed line) and in the presence
  of driving (solid lines), for two different values of the
  input-noise strength $\sigma^2_s$; the corresponding values of the mutual
  information $I(p;t)$ are also shown. Note that in the absence of
  driving, the system relaxes in an exponential fashion to a stable
  fixed point. (B) The mutual information $I(p;t)$ as a function of
  $k_{\rm b}$ for different values of $k_{\rm f}$ (see
  \eref{PPN_CPM}), for $\sigma^2_s = 1$. It is seen that for each
  phosphorylation rate $k_{\rm f}$ there is an optimal
  dephosphorylation rate $k_{\rm b}$ that maximizes the mutual
  information $I(p;t)$. Moreover, $I(p;t)$ increases as $k_{\rm f}$
  decreases, but then saturates and hence becomes independent of
  $k_{\rm f}$ as the system enters the regime in which it responds
  linearly to the input $s$.  The dashed line shows the optimal value
  of $k_{\rm b}^{\rm opt}\approx 0.3/{\rm h}$, as predicted by
  \eref{muopt}. (C) The mutual information $I(p;t)$ as a function of
  the dephosphorylation rate $k_{\rm b}$, for different values of the
  input-noise strength $\sigma^2_s$, keeping the phosphorylation rate
  fixed at $k_{\rm f}=0.01/{\rm h}$. The optimal dephosphorylation
  rate $k_{\rm b}^{\rm opt}\approx 0.3 / {\rm h}$ (dashed line) is
  independent of $\sigma^2_s$, as predicted by \eref{muopt}. The
  input-noise correlation time $\tau_c = 0.5 {\rm h}$.  \flabel{PPN}}
\end{figure*}

\subsection{Push-pull network}
\label{sec:CPM_PPN}
The \b{deterministic} push-pull network is described by the following reaction
\begin{align}
\dot{x}_p = k_{\rm f} s(t) (x_{\rm T} - x_p(t)) - k_{\rm b} x_p(t),
\elabel{PPN_CPM}
\end{align}
where $x_{\rm T}=x+x_p$ is the total protein concentration, $x_p$ is
the concentration of phosphorylated protein, $k_{\rm f} s(t)$ is the
phosphorylation rate $k_{\rm f}$ times the input signal $s(t)$ (see
\eref{s_t}) and $k_{\rm b}$ is the dephosphorylation rate.
\fref{PPN}A shows a time trace of both a driven and a non-driven
push-pull network.

\noindent {\bf Setting the parameters}\\
The steady-state mean phosphorylation level is set by $\bar{p} =
\bar{x}_p /x_T = 
k_{\rm f}\bar{s} / (k_{\rm f} \bar{s} + k_{\rm b})$. We anticipated,
based on the analytical calculations described in section
\ref{sec:ANA_PPN}, that a key timescale is $k_{\rm
  b}$ and that the system should operate in the regime in which it
responds linearly to changes in the mean input $\bar{s}$. This means
that for a given $k_{\rm b}$, $k_{\rm f}$ and $\bar{s}$ cannot be too
large. We have chosen $\bar{s}=2$, and then varied $k_{\rm f}$ and
$k_{\rm b}$ to optimize the mutual information. We then verified a
posteriori that the value of $\bar{s}=2$ indeed puts the system in the
optimal linear regime.
 
{\bf Optimal dephosphorylation rate} Specifically, the parameters
$k_{\rm f}$ and $k_{\rm b}$ are set as follows: for a given input
noise strength $\sigma^2_s=1.0$, we first fix the phosphorylation rate
$k_{\rm f}$ and compute the mutual information $I(p;t)$ between the
phosphorylated fraction $p(t) = x_p(t) / x_T$ and time $t$ as
a function of the dephosphorylation rate $k_b$; we then repeat this
procedure by varying $k_{\rm f}$. The result is shown in
\fref{PPN}B. Clearly, there exists an optimal value of $k_{\rm b}$
that maximizes $I(p;t)$. Moreover, the optimal value $k_{\rm b}^{\rm
  opt}$ becomes indepdendent of $k_{\rm f}$ when $k_{\rm f}$ becomes
so small that the system enters the regime in which it responds
linearly to changes in the mean input $\bar{s}$. We then fixed the
phosphorylation rate to $k_{\rm f}=0.01/{\rm h}$, and compute $I(p;t)$
as a function of $k_{\rm b}$ for different levels of the input-noise
strength, see \fref{PPN}C. It is seen that the optimal
dephosphorylation rate $k_{\rm b}^{\rm opt}$ is essentially
independent of the input noise strength $\sigma^2_s$.  In the
simulations corresponding to Fig. 2 of the main text, we therefore
kept $k_{\rm b}$ constant at $k_{\rm b}^{\rm opt}=0.3/{\rm h}$ and
$k_{\rm f}$ constant at $k_{\rm f}=0.01/{\rm h}$ when we varied
$\sigma^2_s$.

The observation that $k_{\rm b}^{\rm opt}$ is independent of $k_{\rm
  f}$ and $\sigma^2_s$ can be understood by noting that to maximize
information transmission, the system should operate in the
linear-response regime in which the mean output $\bar{x}$ responds
linearly to changes in the mean input $\bar{s}$. This regime tends to
enhance information because it ensures that in the presence of a
sinusoidal input, the output $x_p(t)$ will not be distorted and be
sinusoidal too. In this linear-response regime, the system can be
analyzed analytically, see \eref{muopt} in section \ref{sec:ANA_PPN}
below. This equation, which accurately predicts the optimum seen in
\fref{PPN}B and \fref{PPN}C, reveals that the optimal
dephosphorylation rate depends on the frequency of the driving signal,
$\omega$, and the correlation time of the noise, $\tau_c$, but not on
the noise strength $\sigma^2_s$ and the coupling $\rho$ to the input
signal, given by $\rho=k_{\rm f} x_T$. Increasing the gain $\rho$
amplifies not only the true signal, but also the noise in that signal
(see also \eref{kfs}), such that the signal-to-noise ratio is
unaltered. Indeed, increasing the gain only helps in the presence of
internal noise, which \b{here and the main text}, however, is
zero. 

\b{In sections \ref{sec:CompIntNoise} and
  \ref{sec:AnaIntNoise} we discuss the role of internal noise. As
  \fref{OptCouplingIntExtNoise} shows, in the presence of not only
  input noise but also internal noise, there exists an
  optimal, non-zero, coupling strength, which arises as a trade-off
  between lifting the amplitude of the output above the internal noise
  (which necessitates a sufficiently large coupling strength, see
  \eref{SNR_HO_IntNoise}) and minimizing the distortions of the shape
  of the output signal. However, for biologically relevant copy
    numbers the internal noise is small, while signal distortions only
    kick in at large coupling strengths. Consequently, the optimum is
    broad (\fref{OptCouplingIntExtNoise}).  The chosen coupling strength here is in the plateau regime
    in which the mutual information is maximized in the presence of
    both internal and input noise.}



\subsection{Uncoupled-hexamer model: Kai system of {\it Prochlorococcus}}
{\bf Background} The uncoupled-hexamer model (UHM) presented in the main text is a
minimal model of the Kai system of the cyanobacterium {\it
  Proclorococcus} and, possibly, the purple bacterium {\it
  Rhodopseudomonas palustris}. The well characterized clock of the
cyanobacterium {\it S. elongatus} consists of three proteins, KaiA,
KaiB and KaiC, which are all essential for sustaining free-running
oscillations \cite{Ishiura:1998vc}. And, indeed, many cyanobacteria
possess at least one copy of each {\it kai} gene. One exception is
{\it Proclororoccus}, which contains {\it kaiB} and {\it kaiC}, but
misses a (functional) {\it kaiA} gene. Interestingly, in daily
(12h:12h) light-dark (LD) cycles, the expression of many genes,
including {\it kaiB} and {\it kaiC}, is rhythmic, but in constant
conditions these rhythms damp very rapidly
\cite{Holtzendorff:2008dj,Zinser:2009js}. Similar behavior is observed for
the purple bacterium {\it R. palustris}, which possesses homologs of
the {\it kaiB} and {\it kaiC} genes
\cite{Ma:2016ca}: under LD conditions, the KaiC homolog appears to be
phosphorylated in a circadian fashion, but under constant conditions,
the oscillations decay very rapidly; physiological activities, such as
the nitrogen fixation rates, follow a similar pattern
\cite{Ma:2016ca}. Of particular interest is the observation that under
LD conditions but not under LL conditions, the growth rate is
significantly reduced in the strain in which the {\it kaiC} homolog
was knocked out \cite{Ma:2016ca}. This strongly suggests that the
(homologous) Kai system plays a role as a timekeeping mechanism, which
relies, however, on oscillatory driving.

{\bf Model} Our model is inspired by the models that in recent years
have been developed for {\it S. elongatus}
\cite{VanZon2007,Rust2007,Zwicker2010,Lin2014,Paijmans:2017gx}. These
models share a number of characteristics that are essential for
generating oscillations and entrainment (see also next section). The
central clock component is KaiC, a hexamer, that can switch between an
active state in which the phosphorylation level tends to rise and an
inactive one in which it tends to fall. The model lacks KaiA because
{\it Proclororoccus} and {\it R. palustris} miss a functional {\it
  kaiA} gene \cite{Holtzendorff:2008dj,Zinser:2009js,Ma:2016ca}. In
{\it S. elongatus}, KaiB does not directly affect the rates of
phosphorylation and dephosphorylation, but mainly serves to stabilize
the inactive state and mediate KaiA binding by inactive KaiC
\cite{Lin2014,Paijmans:2017gx}. KaiB is therefore not modelled
explicitly \cite{Lin2014,Paijmans:2017gx}. The main entrainment signal
for {\it S. elongatus} is the ratio of ATP to ADP levels, which
depends on the light intensity, and predominantly couples to KaiC in
its active conformation
\cite{Rust2011,Pattanayak:2015jm,Paijmans:2017gx,Paijmans:2017gp}. These
observations give rise to the following chemical rate equations \b{of
  our deterministic model}:
\begin{align}
\dot{c}_0 &= k_{\rm s} \tilde{c}_0 - k_{\rm f} s(t) c_0 \elabel{UHM_F}\\
\dot{c}_i &= k_{\rm f} s(t) (c_{i-1} - c_i) &i \in (1,\dots,5)\\
\dot{c}_6 &= k_{\rm f} s(t) c_{5} - k_{\rm s} c_6\\
\dot{\tilde{c}}_6 &=  k_{\rm s} c_6 - k_{\rm f} \tilde{c}_6\\
\dot{\tilde{c}}_i &=  k_{\rm b} (\tilde{c}_{i+1} - \tilde{c}_i)& i \in (1,\dots,5)\\
\dot{\tilde{c}}_0 &=  k_{\rm b} \tilde{c}_1 - k_{\rm s} \tilde{c}_0 \elabel{UHM_L}\end{align}
Here, $c_i$, with $i=0,\dots,6$, is the concentration of active
$i$-fold phosphorylated KaiC in its active conformation, while
$\tilde{c}_i$ is the concentration of inactive $i$-fold phosphorylated
KaiC. The quantity $k_{\rm s}$ is the conformational switching rate,
$k_{\rm b}$ is
the dephosphorylation rate of inactive KaiC, and $k_{\rm f} s(t)$ is the
phosphorylation rate of active KaiC, $k_{\rm f}$, times the input
signal $s(t)$.

\begin{figure*}[t]
\includegraphics[width=2\columnwidth]{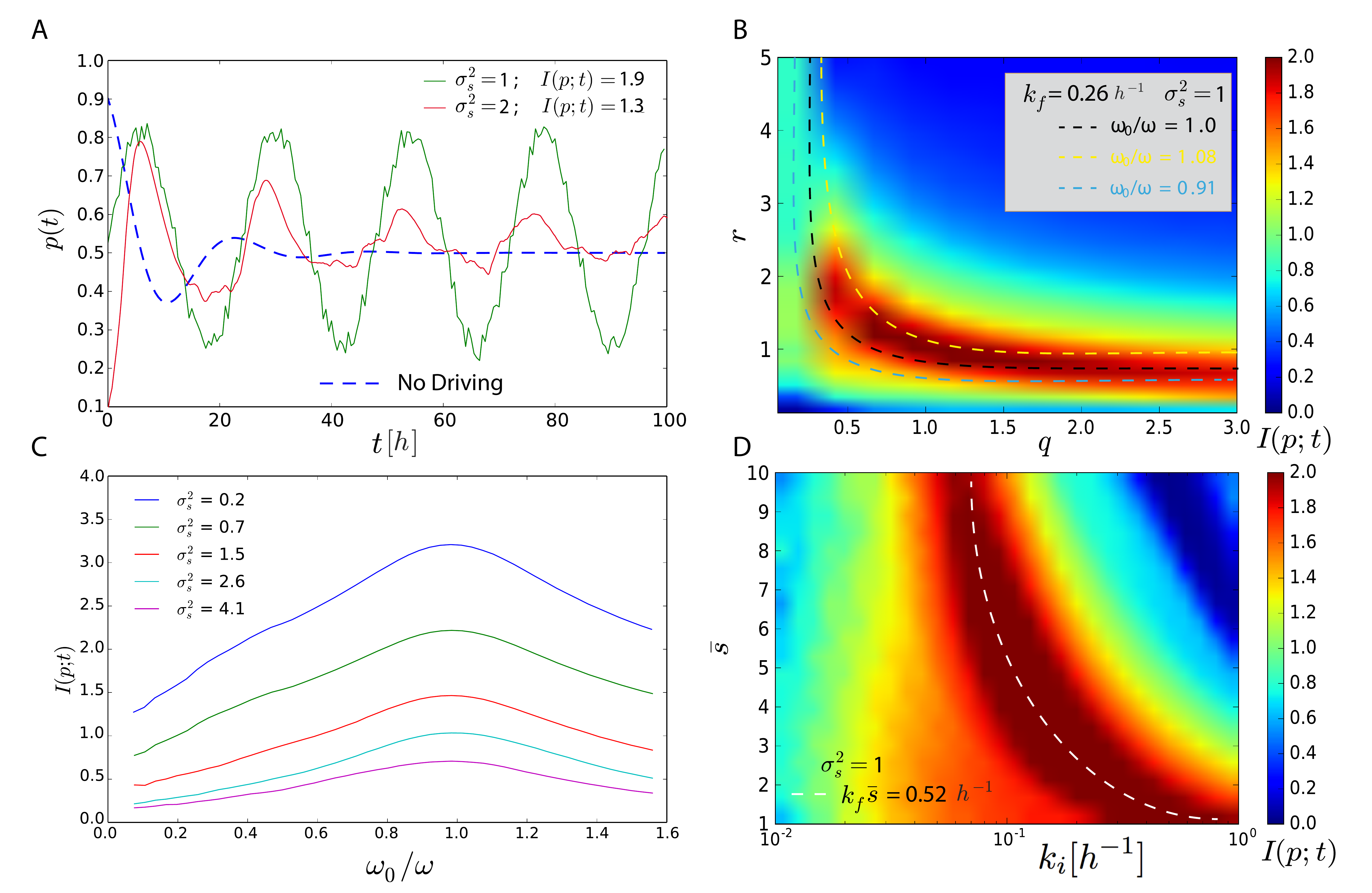}
\caption{The \b{deterministic} uncoupled-hexamer model. (A) Time
  traces of $p(t)$ in the absence of driving (dashed line) and in the
  presence of driving (solid lines), for two different values of the
  input-noise strength $\sigma^2_s$; the corresponding values of the mutual
  information $I(p;t)$ are also shown. Note that in the absence of
  driving, the system relaxes in an oscillatory fashion to a stable
  fixed point. (B) Heatmap of the mutual information $I(p;t)$ as a
  function of the scaling factor ${q}$ that scales both the
  dephosphorylation rate $k_{\rm b}$ and the the mean phosphorylation
  rate $k_{\rm f}\bar{s}$ (see \eref{kfs}) and the ratio ${r}=k_{\rm
    b}/(k_{\rm f}\bar{s})$ of these quantities.  The mean
  phosphorylation rate $k_{\rm f}\bar{s}$ is changed by varying
  $k_{\rm f}$ while keeping $\bar{s}=2$ constant.  Superimposed are
  contour lines of constant $\omega_0 = \omega_0 (q,r)$ (see
  \eref{UHMT0}). It is seen that in the regime where $I(p;t)$ is high,
  $I(p,t)$ is almost constant along these contour lines, showing that
  $I(p;t)$ predominantly depends on $k_{\rm f}$ and $k_{\rm b}$ via
  $\omega_0$. (C) The mutual information $I(p;t)$ as a function of
  $\omega_0$, which was varied by scaling $k_{\rm f}$ and $k_{\rm b}$
  keeping $r=k_{\rm b}/(k_{\rm f}\bar{s}) = 1$ and $\bar{s}=2$, for
  different values of the input-noise strength $\sigma^2_s$. It is
  seen that there exists an optimal intrinsic frequency $\omega_0^{\rm
    opt}$ that maximizes $I(p;t)$. Moreover, $\omega_0^{\rm opt}$ is
  nearly independent of $\sigma^2_s$, corresponding to an intrinsic
  period $T_0=2\pi / \omega_0^{\rm opt} \approx 23.1 {\rm h}$. (D) The
  mutual information $I(p;t)$ as a function of $k_{\rm f}$ and
  $\bar{s}$, keeping $k_{\rm b}=0.52/{\rm h}$ constant. Superimposed
  is the line along which $k_{\rm f}\bar{s}=k_{\rm b}=0.52/{\rm h}$ is
  constant and hence the intrinsic period $T_0$ is constant (see
  \eref{UHMT0}) and equal to $T_0=23.1{\rm h}$. Along this line also
  $I(p;t)$ is essentially constant, meaning that the strength of the
  forcing, set by $k_{\rm f}$, is not very critical. This mirrors the
  behavior seen for the push-pull network (see \fref{PPN}). It is due
  to the fact that increasing the forcing raises not only the
  amplitude but also the noise, keeping the signal-to-noise ratio and
  hence the mutual information essentially unchanged. The noise
  correlation time $\tau_c = 0.5 {\rm h}$.  \flabel{UHM}}
\end{figure*}

The output is the phosphorylation fraction of KaiC proteins
(monomers), given by \cite{VanZon2007,Zwicker2010,Paijmans:2017gx}
\begin{align}
p(t) = \frac{1}{6}\frac{\sum_{i=0}^6 i (c_i +
  \tilde{c}_i)}{\sum_{i=0}^6 (c_i + \tilde{c}_i)}.
\elabel{pdef}
\end{align}
\fref{UHM}A shows a time trace of the phosphorylation level $p(t)$ of
both a driven and a non-driven uncoupled-hexamer model. 

{\bf Intrinsic frequency} Because the cycles of the different hexamers
are not coupled via KaiA as in the coupled-hexamer model and in {\it
  S. elnogatus}, the system cannot sustain free-running
oscillations. In this respect, the system is similar to the push-pull
network in the sense that a perturbation of the non-driven system will
relax to a stable fixed point. However, this model differs from the
push-pull network in that it has a characteristic frequency
$\omega_0=2\pi / T_0$ with intrinsic period $T_0$, arising from the
phosphorylation cycle of the KaiC hexamers. Consequently, while a
perturbed (non-driven) push-pull network will relax exponentially to
its stable fixed point, the uncoupled-hexamer model will, when not
driven, relax in an oscillatory fashion to its stable fixed point with
an intrinsic frequency $\omega_0$ (see \fref{UHM}A). To predict the
latter, we note that the dynamics of \erefsrange{UHM_F}{UHM_L} can be
written in the form $\dot{\bf x} = {\bf A} {\bf x}$, and when all rate
constants are equal, $k_{\rm f} \bar{s}= k_{\rm b} = k_{\rm s}$, the
eigenvalues and eigenvectors of ${\bf A}$ can be computed
analytically. The eigenvectors are complex exponentials. For a cycle
with $N$ sites with hopping rate $k$, the frequency associated with
the lowest-lying eigenvalue is $k \sin(2 \pi/N)$, which to leading
order is $2\pi k / N$, corresponding to a period $T_0 = N / k$. Please
note that this is also the period of a single multimer with $N$
(cyclic) sites with $N$ equal rates of hopping from one site to the
next. We therefore expect that, to a good approximation, the intrinsic
frequency $\omega_0=2\pi/T_0$ of an ensemble of hexamers corresponds
to the intrinsic period of a single hexamer:
\begin{align}
T_0 \simeq \frac{2}{k_{\rm s}} + \frac{6}{k_{\rm f}\bar{s}} + \frac{6}{k_{\rm b}}
\simeq \frac{6}{k_{\rm f}\bar{s}}+\frac{6}{k_{\rm b}}\elabel{UHMT0},
\end{align}
where we recall that in the non-driven system the phosphorylation
rate is $k_{\rm f} \bar{s}$.
We verfied that this approximation is very accurate by fitting the
  relaxation of $p(t)$ of the UHM to a function of the form
  $e^{-\gamma t} \sin(\omega_0 t)$, with $\omega_0 = 2\pi / T_0$. The
  intrinsic period $T_0$ obtained in this way is to an excellent
  approximation given by \eref{UHMT0}.

\noindent {\bf Setting the parameters}\\
  The parameters were set as follows: the conformational switching
  rate $k_{\rm s}$ was set to be larger than the (de)phosphorylation
  rates $k_{\rm s}\gg \{k_{\rm f},k_{\rm b}\}$, as in the original
  models \cite{VanZon2007,Zwicker2010,Paijmans:2017gx}. This leaves
  for a given input noise $\eta_s$, three parameters to be optimized:
  the phosphorylation rate $k_{\rm f}$, the dephosphorylation rate
  $k_{\rm b}$, and the mean input signal $\bar{s}$. The product
  $k_{\rm f} \bar{s}$ determines the mean phosphorylation rate, while
  $k_{\rm f}$ separately determines the strength of the forcing,
  i.e. the amplitude of the oscillations in the phosphoryation rate
  (see \eref{kfs}). The quantities $k_{\rm f}\bar{s}$ and $k_{\rm b}$
  together determine the intrinsic frequency $\omega_0=2\pi / T_0$
  (see \eref{UHMT0}) and the symmetry of the phosphorylation cycle,
  set by the ratio $r\equiv k_{\rm b} / (k_{\rm f}\bar{s})$.  

  {\bf Optimal intrinsic frequency} We therefore first computed for
  different input-noise strengths $\sigma^2_s$, the mutual information
  $I(p;t)$ as a function of the ratio ${r}=k_{\rm b}/(k_{\rm
    f}\bar{s})$ and a scaling factor $q$ that scales both $k_{\rm f}$
  and $k_{\rm b}$, keeping $\bar{s}=2$. \fref{UHM}B shows the heatmap
  of $I(p;t) = I(r,q)$ for $\sigma^2_s=1$, but qualitatively similar
  results were obtained for other values of $\sigma^2_{s}$ (as
  discussed below). Since the intrinsic frequency $\omega_0$ depends
  on both ${r}$ and $q$ (see \eref{UHMT0}), we have superimposed
  contourlines of constant $\omega_0$. Interestingly, the figure shows
  that in the relevant regime of high mutual information, $I(p;t)$
  follows the contourlines of constant $\omega_0$. This shows that
  $I(p;t)$ depends on ${r}$ and ${q}$ predominantly through $\omega_0
  ({r},{q})$, $I (p;t) \approx I(\omega_0(r,q))$. It demonstrates that
  the mutual information is primarly determined by the intrinsic
  period $T_0$---the time to complete a single cycle---and not by the
  evenness of the pace around the cycle set by $r$.

  To reveal the dependence of $I(\omega_0)$ on $\sigma^2_s$, we show
  in panel C for different values of $\sigma^2_s$, $I(p;t)$ as a
  function of $\omega_0$, which was varied by scaling $k_{\rm f}$ and
  $k_{\rm b}$ via the scaling factor $q$, keeping the ratio of $k_{\rm
    f}\bar{s}$ and $k_{\rm b}$ constant at ${r}=1$ (while also keeping
  $\bar{s}=2$). Clearly, there is an optimal frequency $\omega_0^{\rm
    opt} \approx 1.04 \omega$ corresponding to an optimal $k=k_{\rm
    f}\bar{s}=k_{\rm b}=0.52/{\rm h}$, that maximizes the mutual
  information which is essentially independent of $\sigma^2_s$. In
  Fig. 2 of the main text, when we vary $\sigma^2_s$, we thus kept
  $k=k_{\rm f}\bar{s} = k_{\rm b} = 0.52/{\rm h}$ constant, with
  $k_{\rm f}=0.26/{\rm h}$ and $\bar{s}=2$.

Interestingly, the optimal intrinsic frequency $\omega_0^{\rm opt}$ is
not equal to the driving frequency $\omega$: $\omega_0^{\rm opt}>
\omega$, yielding an intrinsic period $T_0^{\rm opt}\approx 23.1 {\rm
  h}$ that is smaller than 24 hrs. This can be understood by analyzing
the simplest model that mimics the uncoupled-hexamer model: the
(damped) harmonic oscillator, which, like the uncoupled-hexamer model,
is a linear system with a characteristic frequency. As described in
\ref{sec:ANA_HO}, we expect generically for such a system that the
optimal intrinsic frequency is larger than the driving frequency:
$\omega_0^{\rm opt} > \omega$. This is because while the amplitude $A$
of the output (the ``signal'') is maximal at resonance, $\omega_0 =
\omega$ (see \eref{A_HO}), input-noise averaging is maximized
(i.e. output noise $\sigma_x$ minimized) for large $\omega_0$ (see
\eref{noiseHOCol}), such that the signal-to-noise ratio $A/\sigma_x$
is maximal for $\omega_0^{\rm opt} > \omega$.

{\bf Mutual information is less sensitive to coupling strength}
Lastly, while $k_{\rm f}\bar{s}$ and $k_{\rm b}$ are vital by setting
the intrinsic period $T_0$ (\eref{UHMT0}) that maximizes the mutual
information (panels B and C of \fref{UHM}), we now address the
importance of the coupling strength, which is set by $k_{\rm f}$
separately (see \eref{kfs}). To this end, we computed the mutual
information $I(p;t)$ as a function of $k_{\rm f}$ and $\bar{s}$,
keeping the dephosphorylation rate constant at $k_{\rm b}=0.52/{\rm
  h}$. \fref{UHM}D shows the result. It is seen that there is, as in
panel B, a band along which the mutual information is highest. This
band coincides with the superimposed dashed white line along which
$k_{\rm f}\bar{s}=0.52 / {\rm h}$ and hence $T_{\rm 0}$ are constant
(see \eref{UHMT0}). This shows that the mutual information $I(p;t)$ is
predominantly determined by the intrinsic period $T_0$: as the
parameters are changed in a direction perpendiular to this line (and
$T_0$ changes most strongly), then $I(p;t)$ falls dramatically. In
contrast, along the dashed white line of constant $T_0$, $I(p;t)$ is
nearly constant. It shows that the precise strength of the forcing,
set by $k_{\rm f}$, is not critical for the mutual information. This
behavior mirrors that observed for the push-pull network. While
increasing $k_{\rm f}$ increases the amplitude of the oscillations in
$p(t)$, it also increases the noise, such that the signal-to-noise
ratio and hence the mutual information are essentially unchanged. The
same behavior is observed for the minimal model of this system, the
harmonic oscillator, described in \ref{sec:ANA_HO}. 

\b{Yet, as for the push-pull network, in the presence of {\em internal}
  noise there exists an optimal coupling strength, as shown in
  \fref{OptCouplingIntExtNoise}B and discussed in section
  \ref{sec:CompIntNoise}.  However, as for the push-pull
  network, the optimium is broad: the signal needs to be lifted above
  the internal noise, yet for larger coupling the effective input
  noise (which scales with the coupling) dominates over the
  internal noise, leading to a regime in which the mutual information
  remains essentially unchanged; the chosen coupling strength here is
  in this regime (\fref{OptCouplingIntExtNoise}).}

To sum up, in the simulations corresponding to Fig. 2 of the main
text, we kept $k_{\rm b}=k_{\rm f}\bar{s} = 0.52 / {\rm h}$, with
$\bar{s}=2$ and $k_{\rm f}=0.26/{\rm h}$.

\subsection{Coupled-hexamer model: Kai system of {\it S. elongatus}}
{\bf Backgroud} In contrast to the cyanobacterium {\it
  Prochlorococcus} and the purple bacterium {\it R. palustris}, the
cyanobacterium {\it S. elongatus} harbors all three Kai proteins,
KaiA, KaiB, and KaiC, and can (therefore) exhibit self-sustained,
limit-cycle oscillations \cite{Ishiura:1998vc}. The circadian system
combines a transcription-translation cycle (TTC)
\cite{Xu2000,Nakahira2004,Nishiwaki2004} with a protein
phosphorylation cycle (PPC) of KaiC \cite{Tomita:2005uv}, and in 2005 the
latter was reconstituted in the test tube \cite{Nakajima2005}. The
dominant pacemaker appears to be the protein phosphorylation cycle
\cite{Zwicker2010,Teng:2013cf}, although at higher growth rates the
transcription-translation cycle is important for maintaining robust
oscillations \cite{Zwicker2010,Teng:2013cf}. Changes in light
intensity induce a phase shift of the in-vivo clock and cause a change
in the ratio of ATP to ADP levels \cite{Rust2011}. Moreover, when
these changes in ATP/ADP levels were experimentally simulated in the
test tube, they induced a phase shift of the protein phosphorylation
cycle which is similar to that of the wild-type clock
\cite{Rust2011}. These experiments indicate that the phosphorylation
cycle is not only the dominant pacemaker, but also the cycle that
couples the circadian system to the light input. We therefore focused
on the protein phosphorylation cycle.

Due to the wealth of experimental data, the in-vitro protein
phosphorylation cycle of {\it S. elongatus} has been modeled
extensively in the past decade
\cite{VanZon2007,Rust2007,Clodong2007,Mori2007a,Zwicker2010,Lin2014,Paijmans:2017gx}. In
\cite{Paijmans:2017gx} we presented a very detailed thermodynamically
consistent statistical-mechanical model, which is based on earlier
models \cite{VanZon2007,Zwicker2010,Lin2014} and can explain most of
the experimental observations. The coupled-hexamer model (CHM)
presented here is a minimal version of these models.  It contains the
necessary ingredients for describing the autonomous
protein-phosphorylation oscillations and the coupling to the light
input, i.e. the ATP/ADP ratio.

The model is similar to the uncoupled-hexamer model described in the
previous section, with KaiC switching between an active state in which
the phosphorylation level tends to rise and an inactive in which it
tends to fall. The key difference between the two systems is that the
CHM also harbors KaiA, which synchronizes the oscillations of the
individual hexamers via the mechanism of differential affinity
\cite{VanZon2007,Rust2007}, allowing for self-sustained
oscillations. Specifically, KaiA is needed to stimulate
phosphorylation of active KaiC, yet inactive KaiC can bind KaiA
too. Consequently, inactive hexamers that are in the dephosphoryation
phase of the phosphorylation cycle---the laggards---can take away KaiA
from those KaiC hexamers that have already finished their
phosphorylation cycle---the front runners. These front runners are
ready for a next round of phosphorylation, but need to bind KaiA for
this. By strongly binding and sequestering KaiA, the laggards can thus
take away KaiA from the front runners, thereby forcing them to slow
down. This narrows the distribution of phosphoforms, and effectively
synchronizes the phosphorylation cycles of the individual hexamers
\cite{VanZon2007}. The mechanism appears to be active not only during
the inactive phase, but also during the active phase: KaiA has a
higher binding affinity for less phosphorylated KaiC
\cite{VanZon2007,Lin2014}. Since KaiB serves to mainly stabilize the
inactive state and mediate the sequestration of KaiA by inactive KaiC,
KaiB is, as in the UHM and following \cite{Lin2014,Paijmans:2017gx},
only modelled implicitly.

\begin{figure*}[t]
\includegraphics[width=2\columnwidth]{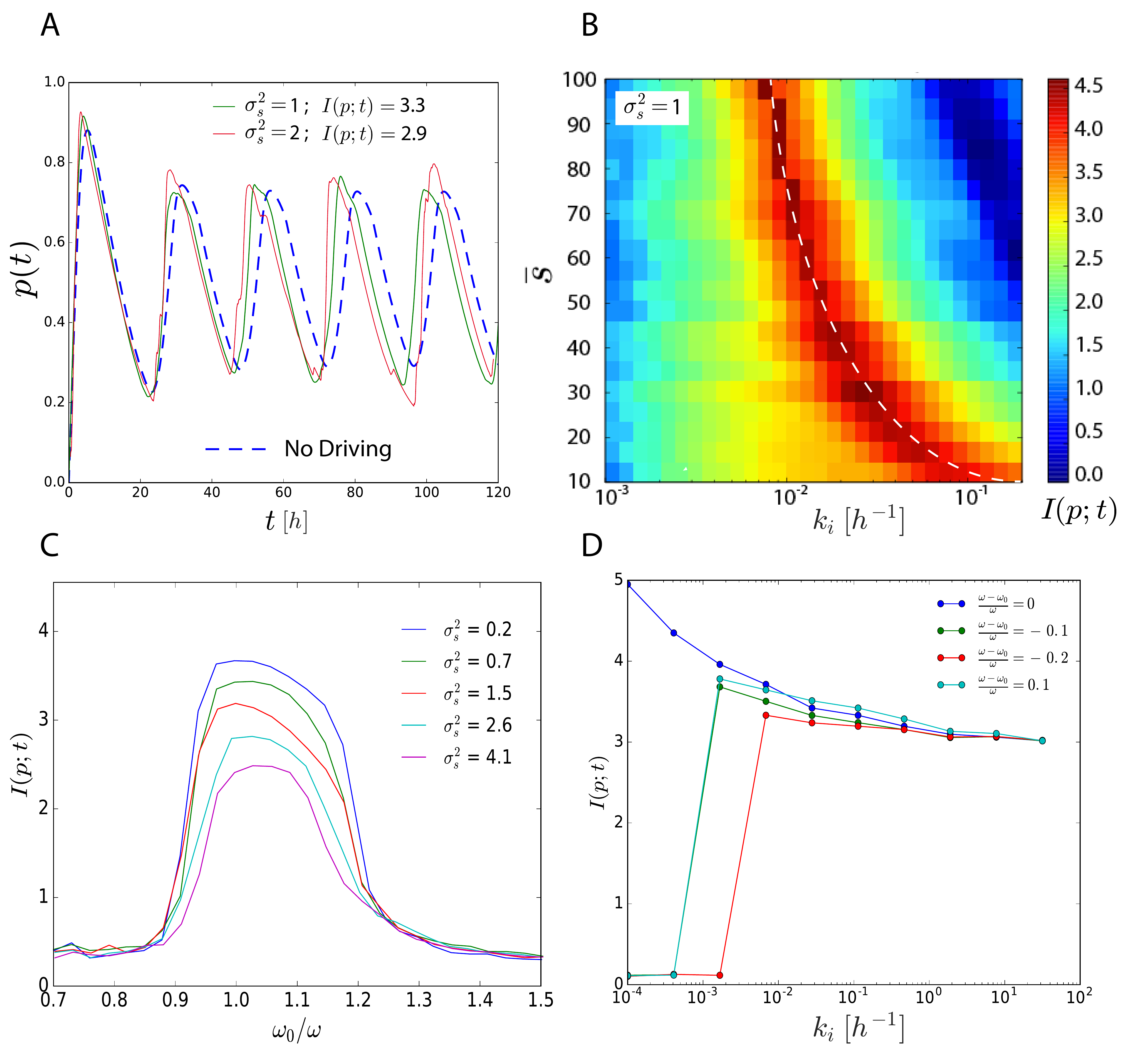}
\caption{The \b{deterministic} coupled-hexamer model. (A) Time traces
  of $p(t)$ in the abence of driving (dashed line) and in the presence
  of driving (solid lines), for two different values of the
  input-noise strength $\sigma^2_s$; the corresponding values of the
  mutual information $I(p;t)$ are also shown. In the absence of
  driving, the system exhibits stable, limit-cycle
  oscillations. \b{(B) The mutual information $I(p;t)$ as a function
    of $\bar{s}$ and $k_i = k_1=\dots=k_5$, for $\sigma^2_s = 1$;
    $k_{\rm b}=0.1875/{\rm h}$ is kept constant while $k_{\rm ps}$ is
    scaled with the same factor as $k_i$.  While the intrinsic
    frequency $\omega_0$ is mainly determined by the product $k_i
    \bar{s}$ (the average phosphorylation rate, see \eref{kfs}), the
    coupling strength is set by $k_i$. Superimposed the white-dashed
    line along which $k_i\bar{s}$ and hence the intrinsic frequency
    $\omega_0$ is constant and equal to the driving frequency
    $\omega_0=\omega=2\pi / 24\, {\rm h^{-1}}$.  Clearly, the mutual
    information decreases rapidly as the intrinsic frequency is
    altered significantly away from the driving frequency $\omega$, as
    shown more clearly in panel C. Along the (dashed white) line of
    constant intrinsic frequency, the mutual information decreases as
    the coupling is increased; this is more clearly illustrated in
    panel D. (C) The mutual information $I(p;t)$ as a function of the
    intrinsic frequency $\omega_0$, which was varied by scaling all
    phosphorylation rates $\{k_{\rm ps}, k_i, k_{\rm b}\}$ by a factor
    $q$, while keeping $\bar{s}=2$.  It is seen that there exists an
    optimal (de)phosphorylation rate that maximizes $I(p;t)$, which
    weakly depends on $\sigma^2_s$. It corresponds to an intrinsic
    period $T_0=25.1 {\rm h}$ of the free-running clock.  (D) The
    mutual information as a function of $k_i$ for different detuning
    strengths $(\omega - \omega_0) / \omega$, all for $\sigma^2_s =
    1$; $k_{\rm ps}$ is scaled by the same factor as $k_i$; $k_{\rm
      b}=0.1875/{\rm h}$ is kept constant and $\bar{s}$ is changed
    such that $k_i \bar{s}$ and hence the intrinsic frequency
    $\omega_0$ remains constant along each curve. It is seen that for
    zero detuning, $\omega_0 = \omega$ (corresponding also to the
    white dashed line in panel B), the mutual information increases
    continuously as the coupling strength is decreased; this is
    because decreasing the coupling makes it possible to minimize
    input-noise propagation. However, for finite detuning, the mutual
    information first rises as $k_i$ is lowered (because that
    minimizes input-noise propagation), but then suddenly drops to
    zero when the system leaves the Arnold tongue: for non-zero
    detuning, a minimal coupling is necessary to phase-lock the system
    to the driving signal~\cite{Monti:2018hs}.  The noise correlation
    time $\tau_c = 0.5 {\rm h}$.}  \flabel{CHM}}
\end{figure*}

{\bf Model} Since computing the mutual information accurately requires
very long simulations, we sought to develop a minimal version of the
PPC model presented in \cite{VanZon2007,Zwicker2010,Paijmans:2016fd},
which can describe a wealth of data including the concentration
dependence of the self-sustained oscillations and the coupling to
ATP/ADP \cite{VanZon2007,Paijmans:2016fd,Paijmans:2017gp}.  This model
\b{is deterministic and} described by the following chemical rate equations:
\begin{align}
  \dot{c}_0 =& k_{\rm s} \tilde{c}_0 - s(t) c_0 \left[k_0\frac{A}{A+K_0} +
  k_{\rm ps} \frac{K_0}{A+K_0}\right] \elabel{CHM_F}\\
  \dot{c}_i =& s(t) c_{i-1} \left[k_{i-1} \frac{A}{A+K_{i-1}} + k_{\rm
      ps} \frac{K_{i-1}}{A+K_{i-1}}\right] \nonumber\\
&-  s(t) c_i\left[
    k_i \frac{A}{A+K_i} +
   k_{\rm ps} \frac{K_i}{A+K_i}\right] \hspace*{0cm} i\in (1,\dots,5)\\
  \dot{c}_6 =&  s(t) c_{5}\left[k_{5} \frac{A}{A+K_{5}} + k_{\rm ps}\frac{K_5}{A+K_5}\right] - k_{\rm s} c_6\\
  \dot{\tilde{c}}_6 =&  k_{\rm s} c_6 - k_{\rm b} \tilde{c}_6\\
  \dot{\tilde{c}}_i =&  k_{\rm b} (\tilde{c}_{i+1} - \tilde{c}_i)
  \hspace*{1cm} i \in (1,\dots,5)\\
  \dot{\tilde{c}}_0 =&  k_{\rm b} \tilde{c}_1 - k_{\rm s} \tilde{c}_0 \\
  A = &A_{\rm T} - \sum_{j=0}^{5} c_j \frac{A}{A+K_j} - \sum_{j=0}^6
  b_j \tilde{c}_j\frac{A^{b_j}}{A^{b_j}+\tilde{K}_j^{b_j}}  \elabel{Afree}
\end{align}
Here, $c_i$ and $\tilde{c}_i$ are the concentrations of active and
inactive $i$-fold phosphorylated KaiC, $A$ is the concentration of
free KaiA. The rates $k_i$ are the rates of KaiA-stimulated
phosphorylation of active KaiC and $k_{\rm ps}$ is the spontaneous
phosphorylation rate of active KaiC when KaiA is not bound. Please
note that both rates are multiplied by the input signal $s(t)$, since
both rates depend on the ATP/ADP ratio \cite{Paijmans:2017gx}. The
dephosphorylation rate $k_{\rm b}$ is independent of the ATP/ADP ratio
\cite{Lin2014,Paijmans:2017gx} and hence $k_{\rm b}$ is not multiplied
with $s(t)$. As in the UHM, $k_{\rm s}$ is the conformational
switching rate. The last equation, \eref{Afree}, gives the
concentration $A$ of free KaiA under the quasi-equilibrium assumption
of rapid KaiA (un)binding by active KaiC with affinity $K_i$
(second term right-hand side) and rapid binding of KaiA by inactive
KaiC, where each $i$-fold
phosphorylated inactive KaiC hexamer can bind $b_i$ KaiA dimers (last
term right-hand side \eref{Afree}).  The mechanism of differential
affinity is implemented via two ingredients: 1) the dissociation
constant of KaiA binding to active KaiC, $K_i$, depends on the
phosphorylation level $i$, with less phosphorylated KaiC having a
higher binding affinity: $K_i < K_{i+1}$
\cite{VanZon2007,Lin2014,Paijmans:2017gx}; 2) inactive KaiC can
strongly bind and sequester KaiA
\cite{VanZon2007,Lin2014,Paijmans:2017gx}; this is modeled by the last
term in \eref{Afree}.

{\bf Autonomous oscillations} \fref{CHM}A shows a time trace of $p(t)$
(\eref{pdef}) for both a driven and a non-driven
coupled-hexamer model. Clearly, in contrast to the push-pull network
and the uncoupled-hexamer model, this system exhibits free running
simulations. Note also that the autonomous oscillations are slightly
asymmetric as observed experimentally, and as shown also by the
detailed models on which this minimal model is based
\cite{VanZon2007,Zwicker2010}. Lastly, while the driving signal is
sinusoidal, the output signal of the driven system remains
non-sinusoidal. This is because this system is non-linear; this
behavior is indeed in marked contrast to the behavior seen for the
linear UHM (see \fref{UHM}) and that of the PPN (\fref{PPN}) which
operates in the linear regime.  The slight asymmetry in the
oscillations also explains why in the regime of very low noise, this
system has a slightly lower mutual information than that of push-pull
network or the uncoupled-hexamer model, as seen in Fig. 1 of the main
text.

\noindent {\bf Setting the parameters}\\
{\b {\bf Free-running oscillator}} We first set the parameters to get
autonomous oscillations, keeping $s(t)=\bar{s}=2$. These parameters
were inspired by the parameters of the model upon which the current
model is built \cite{VanZon2007}. Specifically, the KaiA binding
affinity of active KaiC, given by $K_i$, was chosen such that it obeys
differential affinity, $K_0 < K_1 < K_2 < K_3 < K_4 < K_5$ , as in the
PPC model of \cite{VanZon2007,Zwicker2010,Paijmans:2016fd}.  In
addition, in our model, $b_i=2$ for $i=1,2,3,4$ and $b_i=0$ for
$i=0,5,6$, meaning that $i=1-4$ fold phosphorylated inactive KaiC
hexamers can each bind two KaiA dimers with strong affinity
$\tilde{K}_i=\tilde{K}$.  The conformational switching rate $k_{\rm
  s}$ was set to be higher than all the (de)phosphorylation rates,
$k_{\rm s} >> \{k_i, k_{\rm ps}, k_b\}$ and the values of $k_i, k_{\rm
  ps}, k_{\rm b}$ were, again apart from a scaling factor to set the
optimal intrinsic frequency as described below, identical to those of
the PPC model of \cite{VanZon2007,Lin2014,Paijmans:2016fd}. These
parameter values allowed for robust free-running oscillations (see
\fref{CHM}A) in near quantitative agreement with the oscillations of
the more detailed PPC model of
\cite{VanZon2007,Lin2014,Paijmans:2016fd}.

{\bf Driven oscillator: Optimal intrinsic frequency} We then studied
the driven system. We computed the mutual information $I(p;t)$ as a
function of the mean signal $\bar{s}$ and the phosphorylation rates
$k_i=k_1=\dots=k_5$, see \fref{CHM}B. While the intrinsic frequency is
primarily determined by the mean phosphorylation rate $k_i \bar{s}$,
as illustrated by the dashed-white line of constant intrinsic
frequency $\omega_0$, the coupling strength is (for a given mean $k_i
\bar{s}$) set by the amplitude $k_i$ (see \eref{kfs}).  Panel B shows
that the mutual information changes markedly in the direction
perpendicular to the white line, indicating that $I(p;t)$ strongly
depends on $\omega_0$. To illustrate this further, we varied the
intrinsic frequency $\omega_0$ of the autonomous oscillations by
varying all (de) phosphorylation rates $\{k_i, k_{\rm ps}, k_b\}$ by a
constant factor and computed the mutual information $I(p;t)$ as a
function of this factor and hence $\omega_0$. The result is shown in
\fref{CHM}C.  Clearly, as for the uncoupled-hexamer model, there
exists an optimal intrinsic frequency $\omega_0^{\rm opt}$ that
maximizes $I(p;t)$. The optimal intrinic
frequency depends on the input-noise strength: for low input noise,
$\omega_0^{\rm opt}<\omega$, but then $\omega_0^{\rm opt}$ increases
with $\sigma^2_s$ to become similar to $\omega$ in the high noise
regime.  We also see, however, that the dependence of $\omega_0^{\rm
  opt}$ on $\sigma^2_s$ is rather weak (\fref{CHM}B). We therefore
kept the parameters in the simulations corresponding to Fig. 2 of the
main text, constant.

\b{{\bf Driven oscillator: mutual information increases with
    decreasing coupling strength as long as the system remains inside
    the Arnold tongue. } Along the white dashed line of panel B
  (corresponding to the blue line in panel D), $\omega_0 = \omega$, and the
  mutual information $I(p;t)$ decreases as the coupling strength $k_i$
  is increased. Indeed, when there is no detuning ($\omega_0 =
  \omega$) and no internal noise, $I(p;t)$ is maximized when the
  coupling strength goes to zero. This can be understood by noting
  that a) the limit-cycle oscillator has, in stark contrast to the
  push-pull network and the uncoupled-hexamer system, an intrinsic
  robust amplitude, which does not rely on driving by the input
  signal; b) decreasing the coupling reduces the propagation of the
  input fluctuations. In section \ref{sec:LCO_HO} we prove
  analytically that concerning the robustness to input noise: a) the
  optimal regime is that of weak coupling; b) in this regime, systems
  based on a limit-cycle attractor, such as the CHM, are superior to
  those based on a fixed-point attractor, such as the PPN and the
  UHM.}

\b{\bf Driven oscillator: With non-zero detuning, coupling is
  necessary to keep the system inside the Arnold tongue.}
\b{Importantly, there will always be a finite amount of
  internal noise. In addition, the intrisic period will never be
  exacly $24{\rm h}$. In both cases, coupling is essential to keep the
  system in phase with the driving signal.  In the next section we
  discuss the role of internal noise, but in panel D of \fref{CHM} we
  show for the deterministic CHM the importance of coupling when there
  is a finite amount of detuning $(\omega - \omega_0) /
  \omega$. Clearly, for non-zero detuning, the mutual information
  first rises as the coupling strength is decreased (because that
  minimizes input-noise propagation), but then suddenly drops as the
  system moves out of the Arnod tongue: when the intrinsic period does
  not match the period of the driving signal, a minimal coupling is
  essential to firmly lock the oscillations to the input signal (keeping the system inside the Arnold tongue); indeed, as panel D
  shows, the required coupling strength increases with the amount of
  detuning \cite{Monti:2018hs}.}

\begin{figure*}[t]
\includegraphics[width=2\columnwidth]{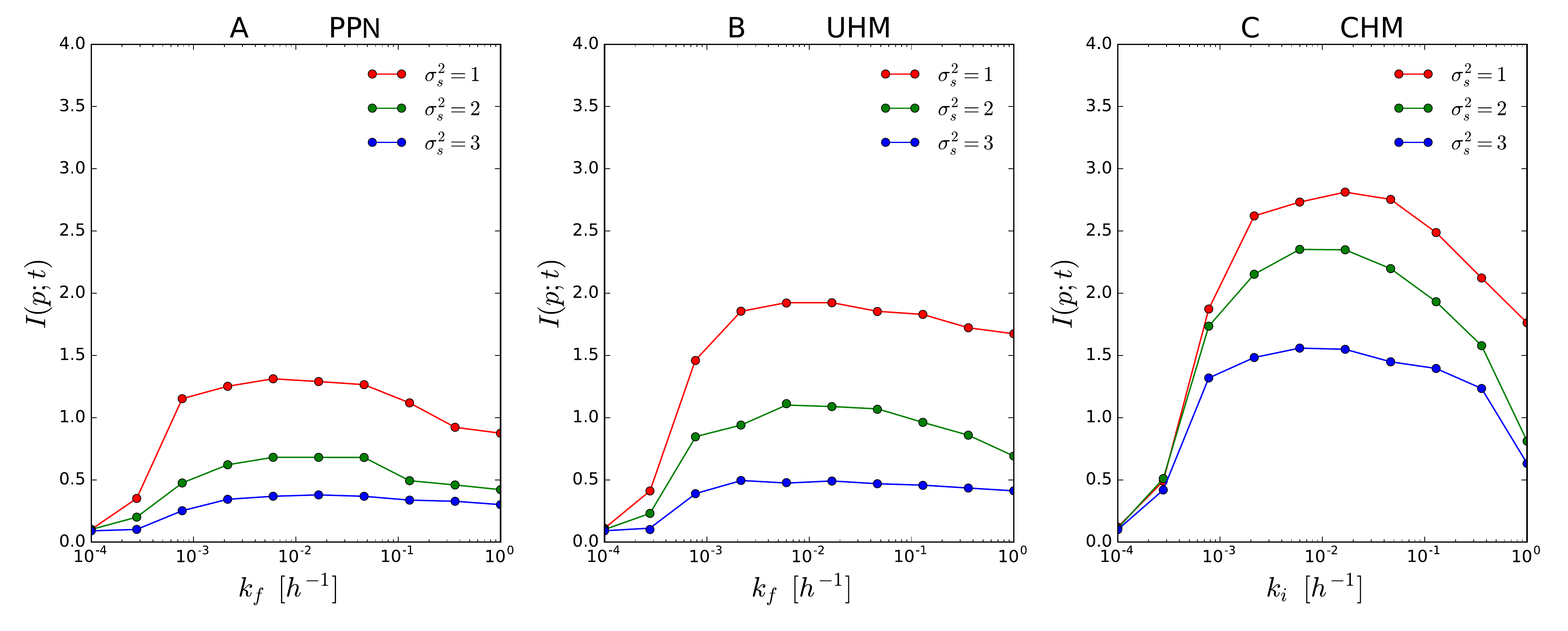}
\caption{\b{Optimal coupling strength in the three different
    computational models in the presence of input noise and internal
    noise. To isolate the role of internal noise, the detuning for
    both the uncoupled and coupled hexamer model was set to zero.
    Shown is the mutual information $I(p;t)$ as a function of the
    coupling strength $k_{\rm f}$ and $k_i$ for different input-noise
    levels $\sigma^2_s$, for: (A) The push-pull network (PPN); (B)
    Uncoupled hexamer model (UHM); (C) Coupled hexamer model (CHM). In
    all three models, the internal noise was kept constant, by keeping
    the copy number of the central clock protein at $N=1000$; this
    number is comparable to the number of KaiC hexamers as measured
    for the cyanobacterium {\it S. elongatus} in vivo
    \cite{Kitayama:2003un}. The results were obtained by performing
    stochastic Gillespie simulations. For the push-pull network (panel
    A), the de-phosphorylation rate $k_{\rm b}$ is set to the optimal
    one as predicted by \eref{muopt}. The stochastic models of the UHM
    and CHM are based of the PPC of \cite{Zwicker2010}.  It is seen
    that for all three models there exists an optimal coupling
    constant that maximizes the mutual information. For the PPN and
    UHM, the optimum is broad: for low coupling, the internal noise
    dominates, and coupling is necessary to lift the signal (amplitude
    oscillations) above the internal noise; for higher coupling, the
    input noise dominates over the internal noise, and the
    signal-to-noise ratio (and hence the mutual information) becomes
    independent of the coupling strength; for even larger coupling
    strength, the mutual information goes down because of signal
    distortion. For the CHM, the optimum is sharper, arising from a
    pronounced trade-off between minimizing input-noise propagation
    and maximizing internal-noise suppression. Parameters: PPN:
    $k_{\rm b}=0.3/{\rm h}$; UHM: $\bar{s}$ is scaled with $k_{\rm f}$
    such that $k_{\rm f}\bar{s}=k_{\rm b}=0.5/{\rm h}$ and
    $\omega_0=\omega$ (see \eref{UHMT0}); CHM: $\bar{s}$ is scaled
    with $k_i$ such that $\omega_0=\omega$; $k_{\rm ps}$ is scaled
    with $k_i$; $k_{\rm b}=0.1875/{\rm h}$ is kept constant; other
    parameters, see \ref{tab:Models}.}
  \flabel{OptCouplingIntExtNoise}}
\end{figure*}

\b{{\bf Setting the coupling strength and the other parameters} The
  fact that the mutual information depends on the amount of detuning
  (\fref{CHM}D) and also internal noise, as shown in the next section
  (\fref{OptCouplingIntExtNoise}), raises the question what is the
  natural procedure to set its value. We have decided to set the
  relative coupling strength to a value that is comparable to the
  coupling strength of the PPC of {\it S. elongatus}. Specifically,
  Fig. 3B of Phong {\it et al.} \cite{Phong:2013fr} shows that the
  kinase rate of the CII domain increases from $0.1 / {\rm h}$ at an
  ATP fraction of 25\% to $0.42 / {\rm h}$ at an ATP fraction of
  100\%. Assuming the ATP fraction oscillates between these levels
  inside the cell \cite{Rust2011}, the amplitude over the mean of the
  oscillations of the kinase rate is around $0.6$. This should be
  compared to $k_i / (k_i \bar{s}) = 1 / \bar{s}$ in our model (see
  \eref{kfs}). With $\bar{s}=2$, the coupling strength is indeed
  comparable to that of the PPC of {\it S. elongatus}.  We thus kept
  $\bar{s}=2$ fixed and then optimized over the intrinsic frequency by
  scaling the (de)phosphorylation rates $k_i, k_{\rm ps}, k_{\rm b}$,
  as shown in \fref{CHM}C. This yielded $\omega_0^{\rm opt}=0.96
  \omega$, corresponding to an intrinsic period $T_0 = 25.1 {\rm h}$.
  Table \ref{tab:Models} gives an overview of all the
  parameters. Finally, we emphasize that the chosen coupling strength
  is a conservative estimate: if the ATP fraction oscillates from 0.2
  to 0.6 inside the cell \cite{Rust2011}, then the in vivo coupling
  strength will be lower; as panel D shows, the performance of the
  CHM, regarding robustness to input noise,
  will then even be higher. In fact,
as  \fref{OptCouplingIntExtNoiseDetuning}A shows, the optimal
    coupling strength that maximizes the mutual information for the
    CHM in the presence of both detuning and internal noise at
    biologically relevant strengths, is even lower than that
    corresponding to Fig. 2 of the main text. In comparing the CHM
  against the UHM and PPN, we thus consider a ``worst-case'' scenario
  for the CHM. Indeed, even for this scenario, the CHM is much more
  robust to input noise than the PPN and UHM, as Fig. 2 of the main
  text shows.}

\subsection{Robustness to internal noise}
\label{sec:CompIntNoise}

\b{The computational models of the readout systems considered in the
  main text and above are deterministic; only the input signal is
  stochastic. In this section, we address the question how robust the
  results on our computational models are to the presence of internal
  noise that arises from the inherent stochasticity of chemical
  reactions.  To isolate the effect of internal noise, we first zoom
  in on the interplay between internal and input noise in the absence
  of any detuning for the UHM and CHM (\fref{OptCouplingIntExtNoise}), and
  then we study the biologically relevant regime with a finite
  amount of detuning
  (\fref{OptCouplingIntExtNoiseDetuning}). \fref{OptCouplingIntExtNoise}
  shows that in the presence of both sources of noise, all
  computational models exhibit an optimal coupling strength that
  maximizes information
  transmission. \fref{OptCouplingIntExtNoiseDetuning} then
  demonstrates that in the biologically relevant regime, at least for
  cyanobacteria: 1) the optimal coupling is weak because the input noise
  dominates over the internal noise; 2) the coupled-hexamer
  model is more robust to input noise than the push-pull
  network and the uncoupled-hexamer model. We elucidate these results
  using our analytical models in sections \ref{sec:LCO_HO} and \ref{sec:AnaIntNoise}.}

\b{{\bf Stochastic simulations} To investigate the role of internal
  noise, we have performed stochastic Gillespie simulations
  \cite{Gillespie:1977dc} of all three computational models. These
  simulations take into account the inherent stochasticity of the
  chemical reactions, yet do assume that the system remains
  well-stirred at all times.  We keep the magnitude of the internal
  noise fixed by keeping the copy number $N$ of the central clock
  component, $X$ in the PPN and the KaiC hexamer in the UHM and CHM,
  constant at $N=1000$; this number is comparable to the number of
  KaiC hexamers in the cyanobacterium {\it S. elongatus}
  \cite{Kitayama:2003un}. The stochastic model of the PPN and the UHM
    are the stochastic versions of the deterministic models studied
    above and in the main text, taking into account the stochastic
    phosphorylation and dephosphorylation of $X$ and KaiC,
    respectively. For the stochastic model of the CHM, we have adopted
    the stochastic PPC model, including its parameter values
  \cite{Zwicker2010}; here, KaiA and KaiB binding is modeled
    explicitly, but since these reactions are much faster than the
    (de)phosphorylation reactions, this is not important---to an
    excellent approximation, this model is the stochastic
    equivalent of the deterministic CHM studied in the main text and
    above.}

\subsubsection{The interplay between input and internal noise with no detuning}

\begin{figure*}[t]
\includegraphics[width=2\columnwidth]{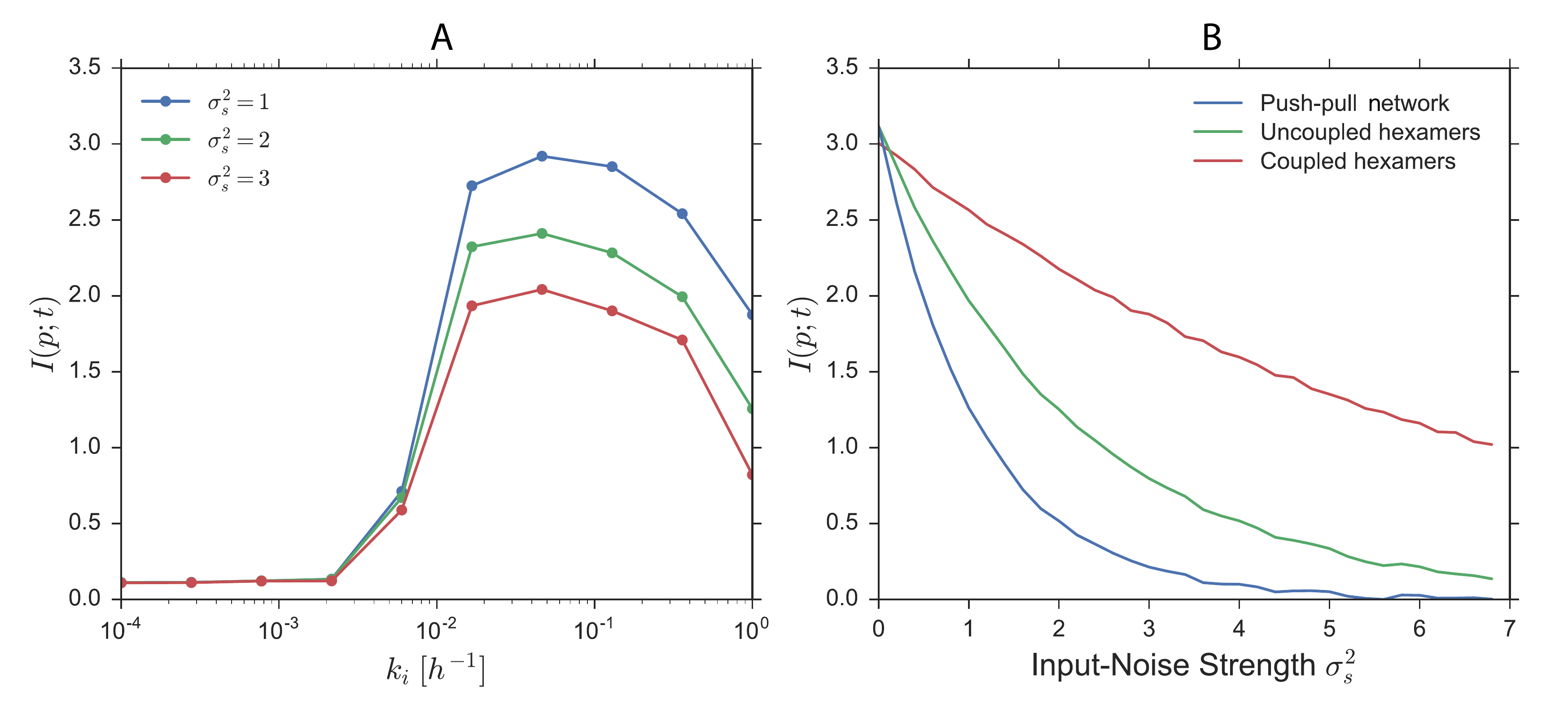}
\caption{\b{Comparing the coupled-hexamer model (CHM) with
    biologically relevant levels of internal noise and detuning
    against the optimal push-pull network (PPN) and optimal
    uncoupled-hexamer model (UHM). For all models, the internal noise
    was kept constant by keeping the copy number of the central clock
    component at $N=1000$, which is comparable to the number of KaiC
    hexamers as measured in vivo for the cyanobacterium {\it
      S. elongatus} \cite{Kitayama:2003un}. For the CHM, the amount of
    detuning was set to $(\omega-\omega_0) / \omega = - 0.1$, which
    corresponds to that measured for the reconstitued PPC of {\it
      S. elongatus} \cite{Nakajima2005}.  Weather data
    \cite{Gu:2001vh} indicates that the average input noise strength
    is $\sigma^2_s \approx 1-2$, but since there will be variations in
    the fluctuations in the light intensity from day to day, we also
    consider higher input noise strengths (see section
    \ref{sec:Input}) (A) The mutual information $I(p;t)$ of the CHM as
    a function of coupling strength $k_i$ for 3 different input-noise
    strengths; $k_{\rm b}=0.1875/{\rm h}$ and $k_{\rm ps}$ is scaled
    by the same factor as $k_i$; $\bar{s}$ is changed such that the
    intrinsic frequency and hence amount of detuning is constant along
    each curve.  It is seen that the mutual information is initially
    low but then sharply rises with $k_i$ as the system enters the
    Arnold tongue where the CHM becomes firmly locked to the driving
    signal (compare with \fref{CHM}D). When the coupling strength is
    raised further, the mutual information goes through a maximum,
    which arises from a trade-off between minimizing input-noise
    propagation and maximizing internal-noise
    suppression. Importantly, the optimal coupling strength is low,
    indicating that the input noise dominates over the internal
    noise. Please also note that the maximum is broader than that in
    \fref{CHM}D, due to the internal noise. (B) The mutual information
    $I(p;t)$ as a function of input-noise strength for the three
    different computational models. For the CHM, all parameters (but
    most notably $k_i, k_{\rm ps}, k_{\rm b}$) have the baseline
    parameter values corresponding to Fig. 2 of the main text and
    shown in Table \ref{tab:Models}, except $\bar{s}$, which was
    changed such that the detuning is $(\omega-\omega_0) / \omega = -
    0.1$. For the UHM and PPN, all parameters (including $\bar{s}$)
    have the baseline parameter values corresponding to Fig. 2 and
    shown in table \ref{tab:Models}; these parameter values maximize
    the mutual information in the absence of internal noise (see
    \fref{PPN}C and \fref{UHM}C).  The figure shows that the principal
    finding of our manuscript, shown in Fig. 2 of the main text, is
    robust to the presence of internal noise: in the limit of low
    input-noise, the mutual information is similar for all
    systems. Yet, in the regime of high input noise, the CHM has the
    highest mutual information.
    \flabel{OptCouplingIntExtNoiseDetuning} The results were obtained
    by performing stochastic Gillespie simulations
    \cite{Gillespie:1977dc}}.}
\end{figure*}

\b{In the previous sections, we have seen that for the deterministic
  push-pull network and the deterministic uncoupled-hexamer model, the
  mutual information is essentially independent of the coupling
  strength in the weak-coupling regime, because increasing the
  coupling strength increases both the amplitude of the output (the
  gain) and the amplification of the input noise, leaving the
  signal-to-noise ratio unchanged. In contrast, for the CHM, when the
  intrinsic clock period is not equal to that of the driving signal, a
  minimal amount of coupling is necessary to phase-lock the clock to
  the driving and put the system inside the Arnold tongue
  (\fref{CHM}D). Yet, once the system is inside the Arnold tongue the
  coupling should be as low as possible to minimize input-noise
  propagation.}

\b{However, for all three systems, we expect that in the presence of
  {\em internal} noise there is a positive effect of increasing the
  coupling strength, although, interestingly, the origin of the effect
  is different for the three respective systems: for the fixed-point
  attractors (PPN and UHM), increasing the coupling helps to raise the
  the amplitude of the oscillations (the signal) above the internal
  noise, while for the limit-cycle attractor (CHM) increasing the
  coupling increases the restoring force that  contains the effect of the
  internal noise. Section \ref{sec:AnaIntNoise} discusses these effects
  in more detail.}

\b{In \fref{OptCouplingIntExtNoise} we show for all three models
  separately, the mutual information $I(p;t)$ as a function of the
  coupling strength, for different strengths of the input noise,
  keeping the internal noise constant. We see that in all cases there
  exists an optimal coupling strength that maximizes the mutual
  information, as predicted by the analytical models discussed in
  section \ref{sec:AnaIntNoise}. For the fixed-point attractors, the
  PPN and the UHM, the optimum is broad: a minimal coupling is
  required to raise the signal above the internal noise, but for
  larger coupling strengths the effect of the input noise, which
  increases with the coupling, dominates over the internal noise, and
  in this regime the signal-to-noise ratio is essentially constant;
  for even larger coupling, however, the signal will saturate (because
  $p(t)$ is bounded by zero and unity), and this will lead to
  non-sinusoidal oscillations, causing the mutual information to go
  down. For the limit-cycle attractor (the CHM), the optimum is more
  pronounced, arising from a sharp trade-off between minimizing
  input-noise propagation (which favors weak coupling) and maximizing
  internal noise suppression (which favors strong coupling).  Indeed,
  panel C shows that the optimal coupling strength decreases as the
  input noise is increased, precisely as this argument predicts.  }

\subsubsection{Interplay between internal and input noise with detuning}
\b{In vivo, not only a finite amount of internal noise is inevitable, but
also a non-zero amount of detuning. In this section, we compare the
three computational models in the presence of both internal noise and
detuning at biologically relevant levels.} 

\b{Panel A of \fref{OptCouplingIntExtNoiseDetuning} shows for the CHM
  the mutual information $I(p;t)$ as a function of the coupling
  strength $k_i$, for three different input-noise levels, in the
  presence of internal noise and detuning at biologicallly relevant
  levels. As above, the internal noise is set by the copy number
  $N=1000$ corresponding to the number of KaiC hexamers in {\it
    S. elongatus} \cite{Kitayama:2003un}, while the detuning is $(\omega
  - \omega_0) / \omega=-0.1$ as measured experimentally for the
  reconstituted PPC of {\it S. elongatus} \cite{Nakajima2005}. Panel A
  exhibits a mixture of the behavior of \fref{CHM}D corresponding to
  the CHM with finite detuning and no internal noise, and that of
  \fref{OptCouplingIntExtNoise}C corresponding to no detuning but with
  internal noise present: to increase the mutual information, the
  coupling strength first has to rise to bring the system inside the
  Arnold tongue (compare with \fref{CHM}D). Yet once inside the Arnold
  tongue, $I(p;t)$ features an optimum arising from the
  interplay between minimizing input-noise propagation and maximizing
  internal noise suppression. We also see that the optimal
    coupling strength, for all input-noise levels, is lower than that
    of the CHM of Fig. 2 of the main text; with such a weaker
    coupling, the robustness of the CHM to input noise would be even higher.}

\b{In \fref{OptCouplingIntExtNoiseDetuning} we compare the performance of the
  three computational models as a function of input-noise strength,
  in the presence of both internal noise and detuning at biologically
  relevant levels.  Clearly, as observed
  for the deterministic systems corresponding to Fig. 2 of the main
  text, for low input noise, the performance of the three systems is
  very similar. Yet, for high input noise, the CHM is far superior. We
  thus conclude that the principal result of the main text, namely
  that a limit-cycle oscillator such as the CHM is more robust to
  input noise than a damped oscillator such as the PPN or UHM, is
  robust to the presence of internal noise.}

\b{We can understand this result by noting that in the presence of
  biologically relevant amounts of internal noise and input noise, the
  optimal coupling is weak because the input noise dominates over the
  internal noise.  In fact, experiments have revealed that the clock of
  {\it S. elongatus} has a strong temporal stability with a
  correlation time of several months, indicating that the internal
  noise is indeed small \cite{Mihalcescu:2004ch}.  As we prove analytically
  in \ref{sec:LCO_HO}, in the input-noise dominated regime a
  limit-cycle oscillator, such as the CHM, is generically more
  resilient to input noise than a system with a fixed point
  attractor, such as the PPN and UHM. Reducing the coupling minimizes
  the amplification of the input noise in all systems, but only the
  limit-cycle oscillator (CHM) can still sustain robust
  large-amplitude oscillations in this regime.}

\b{For larger internal noise strengths than that
considered here, thus outside the biological realm, it might be
beneficial to increase the coupling further. Strong coupling makes it
possible to exploit the fact that the output $p(t)$ is naturally
bounded between zero and unity; the noise can thus be tamed by
continually pushing $p(t)$ against either zero and unity. This
generates, however, strongly non-sinusoidal, square-wave like
oscillations, which are not experimentally observed
\cite{Rust2007}. We thus leave the regime of strong coupling for
future work.}

\begin{figure*}[t]
\includegraphics[width=2\columnwidth]{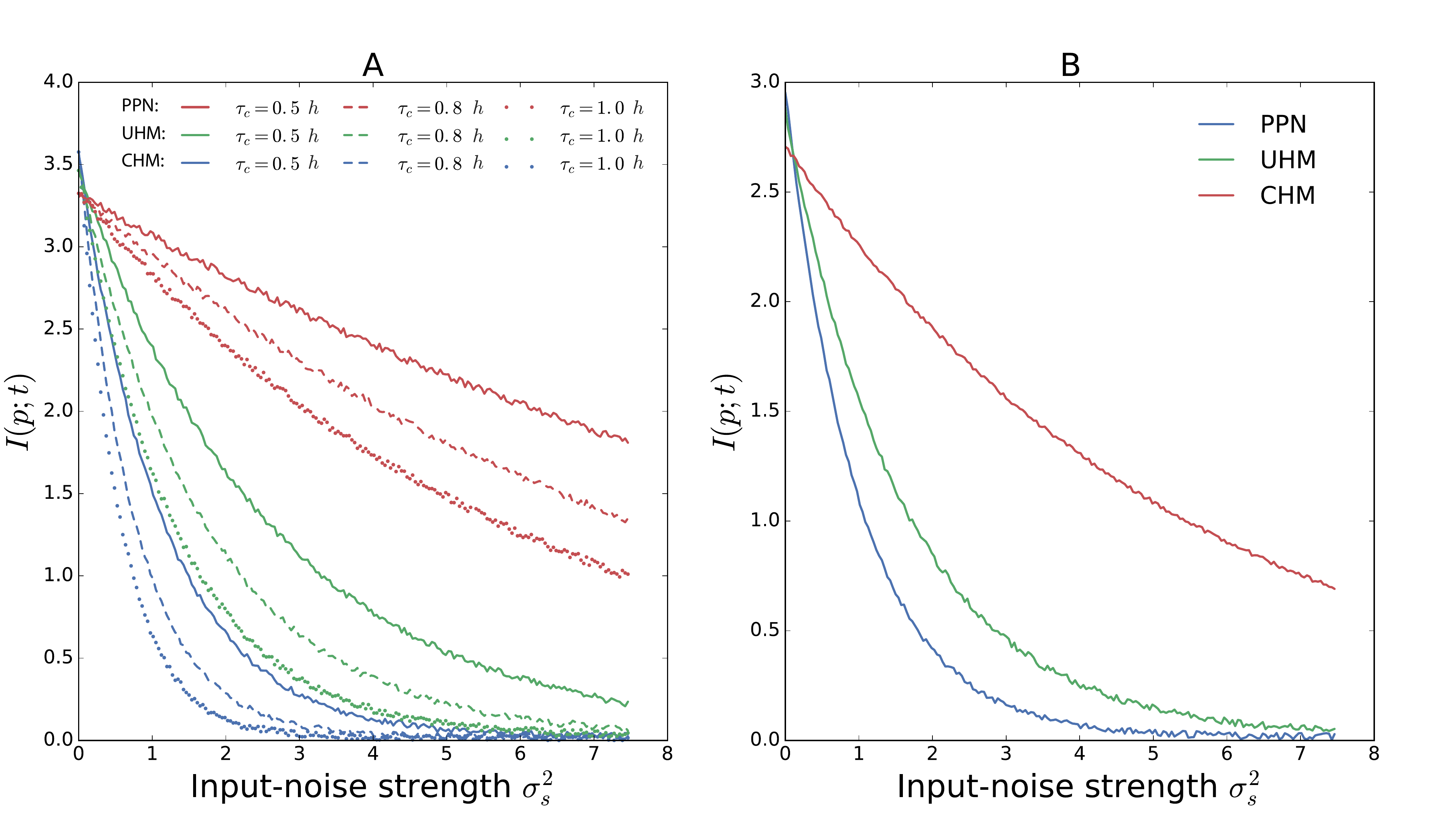}
\caption{Robustness of the pricipal resut of our paper, Fig. 2 of the
  main text, to the shape and correlation time of the input signal. (A)
  Robustness to correlation time of the input noise. It is seen that
  increasing the correlation time $\tau_c$ of the input noise lowers
  the mutual information $I(p;t)$. This is because a higher
  correlation time impedes noise averaging
  \cite{Paulsson:2004dh,TanaseNicola:2006bh,Govern:2014ez}. Yet, for all
  values of $\tau_c$ the result of Fig. 2 of the main text is
  recapitulated: when the input-noise strength $\sigma^2_s$ is low,
  all readout systems are equally accurate; yet, in the high noise
  regime, the coupled-hexamer model is superior. (B) Robustness to the
  shape of the input signal. Here, the input is a truncated sinusoidal
  signal so that during the night $s(t)=0$, while during the day
  $s(t)$ is a half sinusoid (see \eref{TruncInput}). As expected,
  shutting off the driving during the night lowers the mutual
  information (compare with panel A). More strikingly, in the regime
  of low input noise, all readout systems are again equally
  informative on time. Clearly, the push-pull network and
  uncoupled-hexamer model do not need to be driven constantly; it is
  sufficient that the light drives the phosphorylation of the readout
  proteins during the day, so that they can dephosphorylate
  spontaneously during the night. In the regime of high input-noise,
  the coupled-hexamer system is again optimal. In panel B, the noise
  correlation time $\tau_c = 0.5 {\rm h}$. Other parameters are in
  Table \ref{tab:Models}.  \flabel{Robustness}}
\end{figure*}

\subsection{Robustness to shape of input signal}
\label{sec:Robustness}
We have tested the robustness of our principal result, shown in
  Fig. 2 of the main text, by varying a number of key parameters. We
  first varied the correlation time $\tau_c$ of the noise, see
\fref{Robustness}A. Clearly, the main result is robust to
  variations in the value of $\tau_c$: in the limit of small
  input-noise $\sigma^2_s$ all three time-keeping systems are equally
  accurate, while for large input noise the bonafide clock is far
  superior. We have also varied the nature of the input
  signal. Specifically, instead of a sinusoidal signal we have also
  studied a truncated sinusoidal signal $s(t)$, which drops to zero
  for 12 hours during the night but is a half-sinusoid for 12 hours
  during the day:
\begin{align}
s(t) = h(t) \left\{\sin(\omega t) + \eta_s(t)\right\},
\elabel{TruncInput}
\end{align}
where $h(t)=0$ for $0<t<12$ and $h(t)=1$ for $12 < t< 24$. The result
is shown in \fref{Robustness}B. It is seen that the principal result
of Fig. 2 of the main text is also insensitive to the precise choice
of the input signal.

The robustness of our principal observations indicate they are
universal and should be observable in minimal generic models. These
are described in the next sections.

\subsection{Computing the mutual information}
\b{The mutual information is given by
\begin{align}
I(p;t) = \int_0^t dp \int_0^T dt P(p,t) \log_2
\frac{P(p,t)}{P(p)P(t)},
\elabel{MI1}
\end{align}
where $P(p,t)$ is the joint probability distribution of the
phosphorylation level $p$ and time $t$ and $P(p)$ and $P(t)$ are the
marginal probability distribution functions of $p$ and $t$,
respectively. When $p$ and $t$ are statistically independent,
$P(p,t)=P(p)P(t)$ and the mutual information $I(p;t)$ is indeed
zero. More generally, $2^{I(p;t)}$ corresponds the number of time
points $t$ that can be inferred uniquely from the phosphorylation
level $p$; it thus corresponds to the number of distinguishable
mappings between $t$ and $p$ \cite{Walczak:1324157}.  The mutual
information depends on the entropy of the input distribution $H(t)$
and the accuracy of signal transmission, which can be seen by
rewriting \eref{MI1} as
\begin{align}
I(p;t) = H(t) - \avg{H(t|p)}_p,\elabel{MI2}
\end{align}
where 
\begin{align}
H(t) = - \int_0^T dt P(t)\log_2 P(t)
\end{align}
 is the entropy of the
input distribution $P(t) = 1 / T$ and 
\begin{align}
\avg{H(t|p)}_p = -\int_0^1 dp P(p) \int_0^T dt P(t|p) \log_2 P(t|p)
\end{align}
is the average of the entropy of the conditional distribution of $t$
given $p$, $P(t|p)$. The input entropy $H(t)$ quantifies the a priori
uncertainty on the input, while $\avg{H(t|p)}_p$ quantifies the
uncertainty on the input $t$ after the output $p$ has been
measured. \eref{MI3}  shows that the mutual information can be
interpreted as the reduction in the uncertainty on the input $t$, by
measuring the output $p$. The conditional entropy $\avg{H(t|p)}_p$
depends on the reliability of signal transmission, and goes to zero
when the signal is transduced perfectly. Indeed, since the input
distribution $P(t)$ is continuous, the mutual information diverges
when there is no input noise (and no internal noise). The highest
mutual information reported in Fig. 2 of the main text thus
corresponds to the smallest input-noise level studied. For a more
detailed discussion of the mutual information, we refer to
\cite{Walczak:1324157}.}

\b{The mutual information is symmetric with respect to its arguments,
  and \eref{MI1} can also be rewritten as}
\begin{align}
I(p;t) = H(p) - \avg{H(p|t)}_t.\elabel{MI3}
\end{align}
where 
\begin{align}
H(p) = -\int_0^1 dp P(p) \log_2 P(p)
\end{align}
is the entropy of the output distribution $P(p)$  and
\begin{align}
\avg{H(p|t)}_t = -\frac{1}{T} \int_0^T dt \int_0^1 dp P(p|t) \log_2 P(p|t)
\end{align}
is the average of the conditional entropy of $P(p|t)$, with $P(p|t)$
the conditional distribution of $p$ given $t$.
\b{We have used this form to compute $I(p;t)$. In numerically
computing the mutual information, we have verified that the results
are independent of the bin size of the distribution of $p$, following
the approach of} \cite{Cheong:2011jp}. 

\section{Analytical models}
\label{sec:ANA}
\subsection{Push-pull network}
\label{sec:ANA_PPN}
The equation for the push-pull network is
\begin{align}
\dot{x}_p &= k_{\rm f} s(t) (x_T - x_p(t)) - k_{\rm b} x_p\\
&\simeq k_{\rm f}s(t) x_T - k_{\rm b} x_p,
\end{align}
where in the last equation we have assumed that $x_T \gg x_p$, which
is the case when $k_{\rm f} s(t) \ll k_{\rm b}$. In this regime, the
push-pull network operates in the linear regime, leading to sinusoidal
oscillations, which tend to enhance information transmission
\cite{Monti:2016bp}. In what follows, we write, to facilitate
comparison with other studies on noise transmission
\cite{Tostevin:2010bo,Monti:2016bp} $\rho \equiv k_{\rm f} x_{\rm T}$,
$\mu = k_{\rm b}$
and, for notational convenience, $x_p = x$. We thus study
\begin{align}
\dot{x} = \rho s(t) - \mu x(t).
\end{align}
The equation can be solved analytically to yield
\begin{align}
x(t) = \int_{-\infty}^t dt^\prime \chi(t-t^\prime)s(t),
\end{align}
with $\chi(t-t^\prime) = \rho e^{-\mu(t-t^\prime)}$. With the input signal
given by
\begin{align}
 s(t) = \sin
  (\omega t)+\bar{s} + \eta_s (t),
\end{align}
the output is
\begin{align}
x(t) =  A \sin(\omega t - \phi) +
\bar{x} + \eta_x (t)
\end{align}
where the amplitude is 
\begin{align}
A = \frac{\rho}{\sqrt{\mu^2+\omega^2}},
\end{align}
the phase difference of the output with the input is
\begin{align}
\phi =
\arctan (\omega / \mu), 
\end{align}
the mean is 
\begin{align}
\bar{x} = \rho \bar{s} / \mu
\end{align}
and the noise is
\begin{align}
\eta_x = \rho \int_{-\infty}^t dt^\prime e^{-\mu (t-t^\prime)} \eta_s
  (t^\prime).
\end{align}
The variance of the output, assuming the system is in steady state, is then
\begin{align}
\sigma^2_x &= \avg{(x(0) - \bar{x}(0))^2} \\
&=\rho^2\int_{-\infty}^0 \int_{-\infty}^0 dt dt^\prime e^{\mu (t+t^\prime)}
\avg{\eta_s (t) \eta_s(t^\prime)}.
\end{align}
Assuming that the input noise has variance $\sigma^2_s$ and decays
exponentially with correlation time $\tau_c=\lambda^{-1}$, meaning that
$\avg{\eta_s(t) \eta_s(t^\prime)} = \sigma^2_s e^{-\lambda
  |t-t^\prime|}$, the variance of the output is
\begin{align}
\sigma^2_x &= \rho^2 \sigma^2_s  \left[\int_{-\infty}^0\int_{-\infty}^t dt dt^\prime e^{\mu
  (t+t^\prime)} e^{-\lambda (t- t^\prime)} + \right.\\
& \left.\int_{-\infty}^0 \int_{t}^0 dt dt^\prime e^{\mu
  (t+t^\prime)} e^{+\lambda (t- t^\prime)}\right]\\
&=g^2 \frac{\mu}{\mu+\lambda} \sigma^2_s,
\end{align}
with the gain given by $g\equiv \rho / \mu$. 

The signal-to-noise ratio $A/\sigma_x$ is then
\begin{align}
\frac{A}{\sigma_x}=\sqrt{\frac{\mu(\mu+\lambda)}{\mu^2+\omega^2}}\frac{1}{\sigma_s},
\end{align}
which has a maximum at the optimal relaxation rate \cite{Monti:2016bp}
\begin{align}
\mu^{\rm opt} &= \frac{\omega^2}{\lambda} \left(1+\sqrt{1+\left(\lambda/\omega\right)^2}\right).\elabel{muopt}
\end{align}
This optimum arises from a trade-off between the amplitude, which
increases as $\mu$ increases, and input-noise averaging, which
improves as $\mu$ decreases. Another point to note is that the optimal
signal-to-noise ratio does not depend on $\rho = k_{\rm f} x_{\rm T}$,
and hence not on $k_{\rm f}$ and $x_{\rm T}$: while increasing $\rho$
increases the amplitude of the signal, it also amplifies the noise in
the input signal. Increasing the gain $\rho$ (via $x_{\rm T}$ and/or
$k_{\rm f}$) only helps in the presence of intrinsic noise, because
increasing the amplitude of the signal helps to raise the signal above
the intrinsic noise \cite{Monti:2016bp}, \b{as discussed in sections
  \ref{sec:CompIntNoise} and \ref{sec:AnaIntNoise}}. However, in
the deterministic models considered in this study, the intrinsic noise
is zero.

\subsection{The harmonic oscillator and the uncoupled-hexamer model}
\label{sec:ANA_HO}
The uncoupled-hexamer model (UHM) is linear. Moreover, because each
hexamer has a phosphorylation cycle with a characteristic oscillatino
frequency $\omega_0$, this system is akin to the harmonic
oscillator. Indeed, when not driven, both the UHM and the harmonic
oscillator relax in an oscillatory fashion to a stable fixed point. To
develop intuition on the behavior of the UHM, we therefore here
analyze the behavior of a harmonic oscillator driven by a noisy
sinusoidal signal.

The equation of motion of the driven harmonic oscillator is
\begin{align}
\ddot{x} + \omega^2_0 x + \gamma \dot{x} = \rho s(t), \elabel{HO}
\end{align}
where $\omega_0$ is the characteristic frequency, $\gamma$ is the
friction and $\rho$ describes the strength of the coupling to the
input signal $s(t)$. We assume that $s(t)= \sin(\omega
t)+\eta_s(t)$. We note that while the undriven harmonic oscillator is
isomorphic to the undriven UHM, their coupling to the input is different: in
the UHM, the hexamers are, motivated by the Kai system
\cite{Rust2011,Pattanayak:2015jm}, only coupled to the input during
their active phosphorylation phase, while the harmonic oscillator is
coupled continuously; moreover, in the harmonic oscillator the noise
is additive, while in the UHM the signal multiplies the
phosphorylation rate, leading to multiplicative noise. Yet, the
behavior of the two models is qualitatively similar, as discussed below.

Solving \eref{HO} in Fourier space yields $\tilde{x}(\omega) =
\tilde{\chi}(\omega) \tilde{s}(\omega)$, with
\begin{align}
\tilde{\chi}(\omega) = \frac{\rho}{\omega_0^2 - \omega^2 - i \omega
  \gamma}.
\end{align}
Hence, the time evolution of $x(t)$ is
\begin{align}
x(t) &= \frac{1}{2\pi} \int_{-\infty}^{\infty} d\omega 
e^{-i\omega t}
\tilde{\chi}(\omega) s(\omega)\\
&= \frac{\rho}{2\pi} \int_{-\infty}^{\infty} d\omega
\int_{-\infty}^{\infty} dt^\prime
\frac{e^{i \omega(t^\prime - t)} s(t^\prime)}{ \omega^2_0 - \omega^2 -
  i \omega \gamma}.
\end{align}
We do the integral over $\omega$ first. The integrand has poles at
\begin{align}
\omega = \frac{-i \gamma}{2} \pm
\sqrt{\omega_0^2-\frac{\gamma^2}{4}} \equiv \frac{-i
  \gamma}{2} \pm \omega_1.
\end{align}
This yields
\begin{align}
x(t) &= \frac{\rho}{2\pi} \int_{-\infty}^{\infty} s(t^\prime)
\theta(t-t^\prime) (2\pi i) \times \\
& \left[\frac{e^{i(-i\frac{\gamma}{2} + \omega_1)(t^\prime -
    t)}}{2\omega_1}-\frac{e^{i(-i\frac{\gamma}{2} - \omega_1)(t^\prime -
    t)}}{2\omega_1}\right]\\
&=\frac{\rho}{\omega_1} \int_{-\infty}^t dt^\prime e^{-\frac{\gamma}{2} (t-t^\prime)}
  \sin (\omega_1 (t-t^\prime)) s(t^\prime).
\end{align}
With $s(t) = \sin(\omega t)$, this yields
\begin{align}
x(t) 
&=\frac{-\gamma \omega \cos[\omega t] + (-\omega^2 + \omega_0^2) \sin[\omega t]}{
\gamma^2 \omega^2 + (\omega^2 - \omega_0^2)^2}\elabel{xHO}
\end{align} 
This can also be rewritten as
\begin{align}
x(t) = A \sin (\omega t + \phi),
\end{align}
with the amplitude given by
\begin{align}
A 
&=\frac{\rho}{\sqrt{\gamma^2 \omega^2 + (\omega^2 - \omega_0^2)^2}}\elabel{A_HO}
\end{align}
and the phase given by
\begin{align}
\phi = \arctan \left[\frac{- 4 \gamma \omega}{\gamma^2 + 4(\omega_1^2 -
\omega^2)}\right].
\end{align}
\eref{A_HO} shows that the amplitude increases as the friction
decreases and that the amplitude is maximal when the intrinsic
frequency equals the driving frequency; in fact, when $\gamma\to 0$ and
$\omega_0 = \omega$, the amplitude diverges.

\begin{figure*}[t]
{\sf (A)} \hspace*{-0.2cm}
\includegraphics[width=5cm]{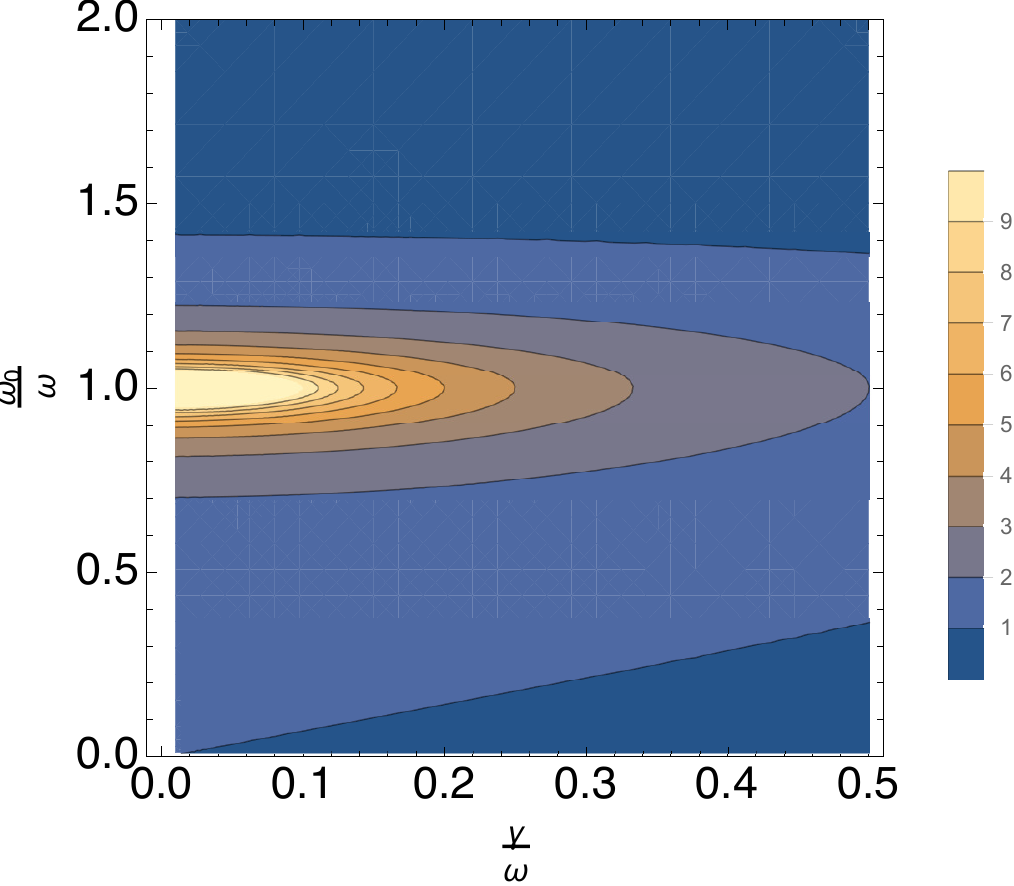}
\hspace*{0.2cm}
{\sf (B)} \hspace*{-0.2cm}
\includegraphics[width=5cm]{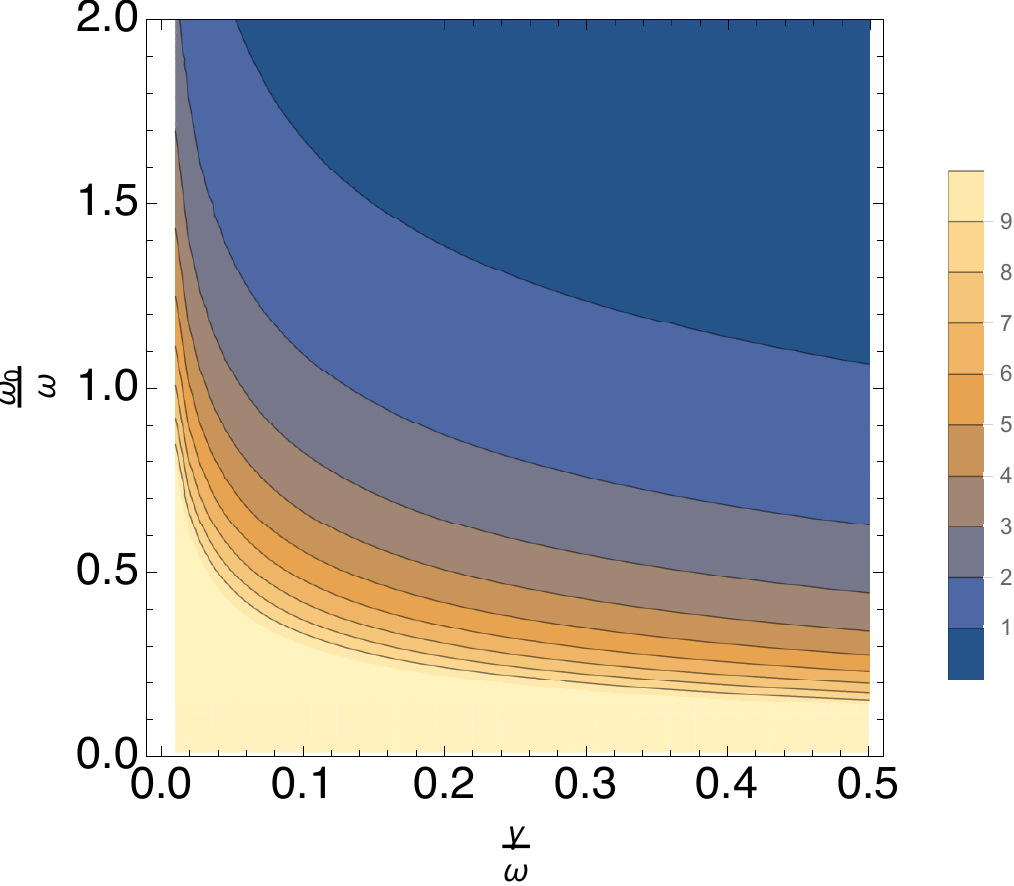}
\hspace*{0.2cm}
{\sf (C)} \hspace*{-0.2cm}
\includegraphics[width=5cm]{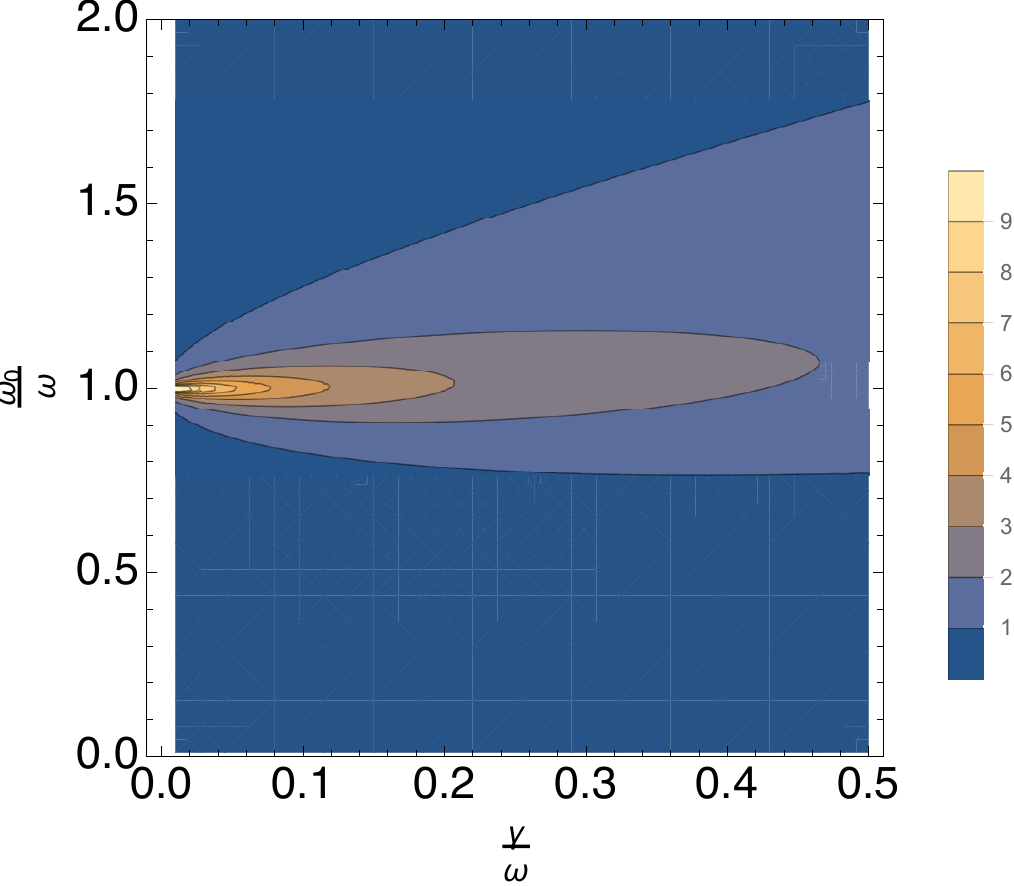}
\caption{The amplitude (A), standard deviation $\sigma_x$ (B), and
  signal-to-noise ratio $A/\sigma_x$ (C) as a function of the the
  intrinsic frequency $\omega_0$ and friction $\gamma$ for the
  harmonic oscillator. It is seen that the amplitude peaks when
  $\gamma=0$ and the intrinsic frequency equals the driving frequency,
  $\omega_0 = \omega$ (A).  The noise peaks at $\gamma=0$ and at
  $\omega_0=0$ (B).  Because the amplitude peaks at $\omega_0=\omega$,
  while the noise peaks at $\omega_0=0$, there is an optimal intrinsic
  frequency $\omega_0^{\rm opt}>\omega$ that maximizes the signal-to-noise ratio (C).
See also \fref{HO_SNR_Omega0_Gamma}.
  \flabel{HO_Amp_Noise}}
\end{figure*}

With an input noise with variance $\sigma^2_s$ and decay rate $\lambda$,
the noise in the output, $\sigma^2_x = \avg{\delta x^2(0)}$, is given by
\begin{align}
\sigma^2_x &= \frac{\rho^2}{\omega_1^2} \int_{-\infty}^0 dt
\int_{-\infty}^0 dt^\prime e^{\frac{\gamma}{2} (t+t^\prime)} \sin(\omega_1 t)
  \sin(\omega_1 t^\prime) \avg{\eta_s (t) \eta_s (t^\prime)}\\
&=\frac{\rho^2 \sigma^2_s}{\omega_1^2} \left[ \int_{-\infty}^0 dt
  \int_{-\infty}^t dt^\prime
e^{\frac{\gamma}{2} (t+t^\prime)} \sin(\omega_1 t)\sin(\omega_1 t) e^{-\lambda(t-t^\prime)} \right.\nonumber\\
&+\left.\int_{-\infty}^0dt \int_t^0dt^\prime e^{\frac{\gamma}{2} (t+t^\prime)} \sin(\omega_1 t)
  \sin(\omega_1 t^\prime) e^{-\lambda(t^\prime - t)}\right]\\
&= \rho^2\sigma^2_s \frac{16(\gamma + \lambda)}{\gamma[(\gamma+2\lambda)^2 + 4\omega_1^2](\gamma^2 +
    4  \omega_1^2)}\\
&= \rho^2\sigma^2_s \frac{(\gamma + \lambda)}{\gamma \omega_0^2[\lambda(\gamma+\lambda)+\omega_0^2]}\elabel{noiseHOCol}
\end{align}
This expression shows that the noise diverges for all frequencies when
the friction $\gamma\to 0$. It also shows that the noise diverges for
$\omega_0 \to 0$ for all values of $\gamma$, or, conversely, that it goes
to zero for $\omega_0 \to \infty$. This can be understood by
  imagining a particle with mass $m=1$ in a harmonic potential well
  with spring constant $k$, giving a resonance frequency $\omega_0^2
  = k/m = k$, which is buffeted by stochastic forces: its variance
  decreases as the spring constant $k$ and intrinsic frequency
  $\omega_0$ increase.

\begin{figure}[b]
\includegraphics[width=\columnwidth]{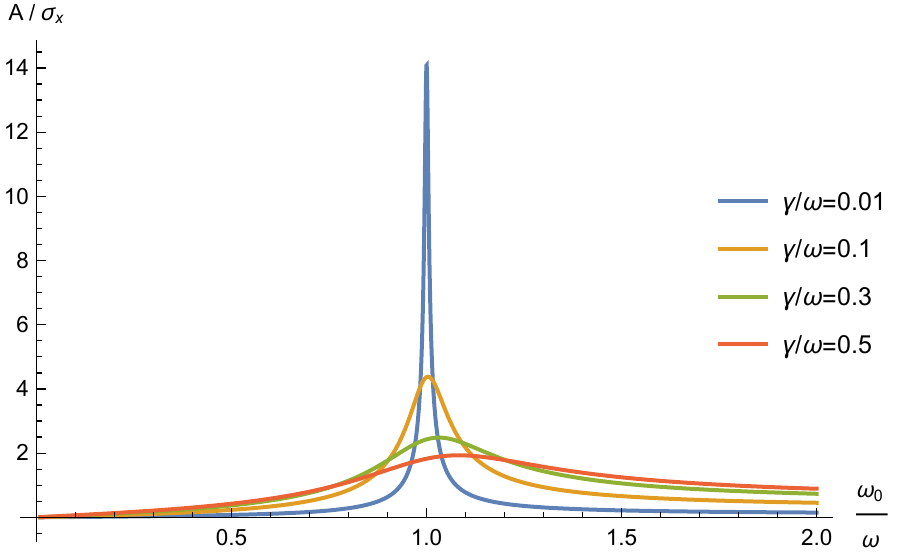}
\caption{The signal-to-noise $A/\sigma_x$ of the harmonic oscillation as a function of
  $\omega_0$ for different values of $\gamma$. Because the amplitude
  $A$ exhibits a strong maximum at $\omega_0 = \omega$, the SNR peaks
  around $\omega_0 = \omega$. However, the maximum is not precisely
  at $\omega_0=\omega$, because the noise $\sigma_x$ peaks at
  $\omega_0=0$ and not at $\omega_0=\omega$. Depending on the friction, there
  thus exists an optimal intrinsic frequency $\omega_0^{\rm opt}>\omega$. Note also that when $\omega\neq
  \omega_0$, it is actually beneficial to have friction, $\gamma\neq
  0$. 
  \flabel{HO_SNR_Omega0_Gamma}}
\end{figure}

\frefstwo{HO_Amp_Noise}{HO_SNR_Omega0_Gamma} show the amplitude $A$,
noise $\sigma^2_x$, and signal-to-noise ratio $A/\sigma_x$ for the
harmonic oscillator. Clearly, the amplitude is maximal at resonance,
diverging when $\gamma \to 0$ (\fref{HO_Amp_Noise}A). The noise is
maximal at $\omega_0\to 0$, and also diverges for all frequencies when
$\gamma \to 0$ (\fref{HO_Amp_Noise}B). However, the amplitude rises
more rapidly as $\gamma \to 0$ than the noise does, leading to a
global optimum of the signal-to-noise ratio for $\omega_0 = \omega$
and $\gamma \to 0$ (\fref{HO_Amp_Noise}C). However, biochemical
networks have, in general, a finite friction, and then the optimal
intrinsic frequency is off resonance, as most clearly seen in
\fref{HO_SNR_Omega0_Gamma}. In fact, since the noise is minimized for
$\omega_0 \to \infty$ while the amplitude is maximized at resonance,
$\omega_0 = \omega$, the optimal frequency $\omega_0^{\rm opt}$ that
maximizes the signal-to-noise ratio is in general $\omega_0^{\rm opt}
> \omega$, as indeed also observed for the uncoupled hexamer model
(see \fref{UHM}B).

Because noise is commonly modeled as Gaussian white noise, as in our
Stuart-Landau model below, rather than colored noise as assumed here,
we also give, for completeness, the expression for $\sigma^2_x$ when
the input noise is Gaussian and white, $\avg{\eta_s(t)
  \eta_s(t^\prime)} = \sigma^2_{s, \rm white} \delta (t-t^\prime)$. It
is
\begin{align}
\sigma^2_x &= \frac{\rho^2\sigma^2_{s, \rm white}}{2\gamma
  \omega_0^2}.
\elabel{sigmaxsqHOwhite}
\end{align}
This is consistent with \eref{noiseHOCol}, by noting that the
integrated noise strength of the colored noise is $2 \int_0^\infty dt
\sigma^2_{s} e^{-\lambda
t}=2\sigma^2_{s}/\lambda$, while the integrated noise
strength of the white noise case is $\sigma^2_{s,\rm white}$. Indeed,
with this identification, \eref{noiseHOCol} in the limit of large
$\lambda$ reduces to the above expression for the white noise
case.

\subsection{Comparison between push-pull network and harmonic
  oscillator in the high friction limit}
\label{sec:COMP_UHM_PP}
Intuitively, one would expect that in the high-friction limit the
harmonic oscillator peforms similarly to the push-pull network. The
signal-to-noise ratio ${\rm SNR}=A/\sigma_x$ indeed becomes the same
in this limit. However, the amplitude and the noise separately scale
differently, because the friction in the harmonic oscillator also
reduces the strenght of the signal and the noise: in the
high-friction limit, the equation of motion of the harmonic oscillator
becomes $\dot{x}_{\rm HO} = \rho s(t) /
\gamma - \omega_0^2 / \gamma x(t) + \rho \eta_s (t) / \gamma$, showing that the
friction renormalizes both the signal and the noise. However, such a
renormalization of both the signal and the noise should not affect the
signal-to-noise ratio. Moreover, we now see that in this high-friction
limit the harmonic oscillator relaxes with a rate $\omega_0^2 /
\gamma$, which is to be compared with $\mu$ of the push-pull network,
for which $\dot{x}_{\rm PP} = \rho s(t) - \mu x(t) + \rho \eta_s(t)$. From this we can
anticipate that while the amplitude and the noise will be different,
the signal-to-noise ratio will be the same. Concretely,
in the high-friction limit the amplitude, the noise and the
signal-to-noise ratio of the harmonic oscillator become
\begin{align}
A^{\rm HO} &= \frac{\rho}{\gamma \omega}\\
\sigma_x^{\rm HO}&=\frac{\rho\sigma_s}{\omega_0\sqrt{\gamma \lambda}}\\
{\rm SNR}^{\rm HO}
&=\sqrt{\frac{\omega_0^2}{\gamma}}\frac{\sqrt{\lambda}}{\omega}=\frac{\sqrt{\mu \lambda}}{\omega},
\end{align}
where in the last line we have made the identification $\mu =
\omega_0^2 / \gamma$. For the push-pull network, the corresponding
quantities, in the limit that $\mu \to 0$, are
\begin{align}
A^{\rm PP} &=\frac{\rho}{\mu}\\
\sigma_x^{\rm PP}&=\frac{\rho \sigma_s}{\mu \lambda}\\
{\rm SNR}^{\rm PP}&=\frac{\sqrt{\mu \lambda}}{\omega}.
\end{align}
Clearly, the signal-to-noise ratio of the two models are the
same in the limit of high friction. 

\fref{SNR_HO_PP} compares the behavior of the harmonic oscillator
against that of the push-pull system. Clearly, for small $\gamma$, the
signal-to-noise ratio SNR of the harmonic oscillator is larger than
that of the push-pull network, showing that building an oscillatory
tendency with a resonance frequency into a readout system can enhance
the signal-to-noise ratio.  However, in the large-friction limit, the
SNR is the same of both models, as expected.

\begin{figure}[t]
\includegraphics[width=\columnwidth]{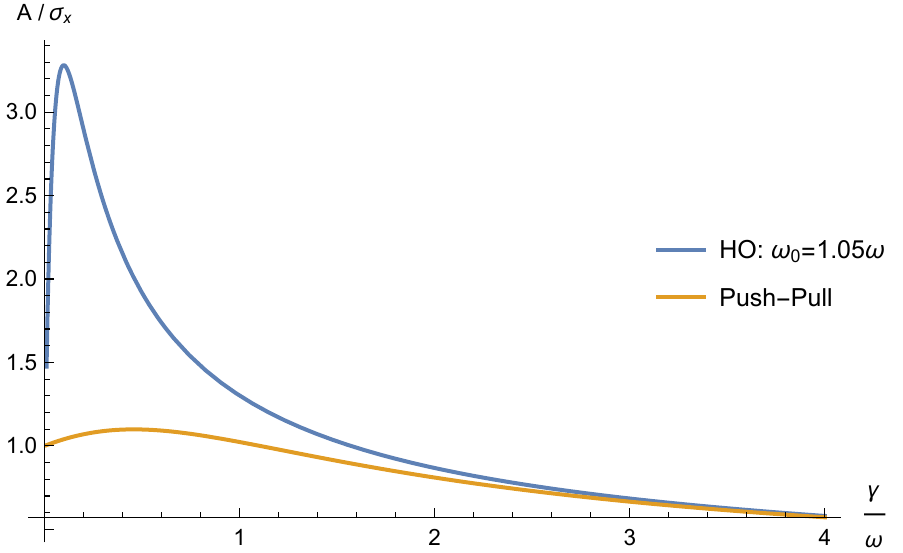}
\caption{The signal-to-noise $A/\sigma_x$ as a function of $\gamma$
  for the harmonic oscillator and the push-pull network. For the
  harmonic oscillator, the friction is varied, while $\omega_0$ is
  kept constant; for the push-pull network $\mu$ is varied according
  to $\mu = \omega_0^2 / \gamma$. It is seen that for low and
  intermediate friction the harmonic oscillator outperforms the
  push-pull network, but that in the high-friction limit they perform similarly. \flabel{SNR_HO_PP}}
\end{figure}

\subsection{Weakly non-linear oscillator and the coupled-hexamer
  model}
\label{sec:SL}
The coupled-hexamer model (CHM) is a non-linear oscillator that can
sustain autonomous limit-cycle oscillations in the absence of any
driving. Here, we describe the Stuart-Landau model, which provides a
universal description of a weakly non-linear system near the Hopf
bifurcation where a limit cycle appears. We use it to analyze the
time-keeping properties of a system as it is altered from essentially
a damped linear oscillator to a weakly non-linear oscillator, see
Fig. 3 of the main text. Our treatment follows largely that of
Pikovsky {\it et al.} \cite{Pikovsky2003}.

\subsubsection{The amplitude equation}
We consider the weakly non-linear oscillator \cite{Pikovsky2003}:
\begin{align}
\ddot{x} + \omega_0^2 x = f(x,\dot{x}) + \rho s(t),
\elabel{nlo}
\end{align}
with $s(t) = \sin(\omega t) + \bar{s} + \eta_s$ being the driving
signal as before. The quantity $f(x,\dot{x})$ describes the
non-linearity of the autonomous oscillator and the parameter
$\rho$ controls the strength of the forcing. The description
presented below is valid in the regime where the non-linearity
$f(x,\dot{x})$ is small and the strength of the driving, quantified by
$\rho$, is small. We begin by developing the formalism in the
deterministic limit $\eta_s = 0$, in which $s(t)$ is periodic with
period $T = 2\pi/\omega$, before returning to the effects of noisy
driving.  In contrast to previous sections, our discussion here is
limited to input noise that is not only Gaussian but white,
$\avg{\eta_s(t)} = 0$ and $\avg{\eta_s(t)\eta_s(t')} = \sigma_s^2
  \delta(t - t')$.

\eref{nlo} is close to that of a linear oscillator. We therefore
expect that its solution has a nearly sinusoidal form. Moreover, we
expect at least over some parameter range the frequency of the system
is entrained by that of the driving signal. We therefore write the solution as
\begin{align}
x(t) = {\rm Re} \left[A(t) e^{i \omega t}\right] = \frac{1}{2}\left( A(t) e^{i
  \omega t} + {\rm c.c.}\right),
\end{align}
where ${\rm c.c.}$ denotes complex conjugate.  The above equation has
the form of an harmonic oscillation with frequency $\omega$, but with
a time-dependent complex amplitude $A(t)$. We emphasize that the
observed frequency may deviate from $\omega$, when the amplitude
$A(t)$ rotates in the complex plane.

The above equation determines only the real part of the complex number
$A(t)e^{i \omega t}$. To fully specify $A(t)$, we also need to set the
imaginary part of $A(t)e^{i\omega t}$, which we choose to do via
\begin{align}
y(t) &= - \omega {\rm Im} \left[A(t) e^{i \omega t}\right]=\frac{1}{2}
\left(i \omega A(t) e^{i\omega t} + {\rm c.c.}\right)\\
&=\dot{x}.
\end{align}
The relation $y(t) = \dot{x}$ thus specifies the imaginary part of the
amplitude $A(t)$. Hence, the complex amplitude can be written as
\begin{align}
\elabel{Axy}
A(t) e^{i \omega t} = x(t) - i y(t) / \omega.
\end{align}

Writing $A(t) = R(t) e^{i\phi(t)}$, it can be verified that
\begin{align}
x(t) &= R(t) \cos (\phi(t) + \omega t) \elabel{x}\\
y(t) &= - \omega R(t) \sin(\phi(t) + \omega t) \elabel{y}\\
R^2(t) &= x^2(t) + y^2(t) / \omega^2,
\elabel{Rsq}
\end{align}
and that the specification $\dot{x}(t) = y(t)$ implies that 
\begin{align}
\elabel{dRdphi}
\frac{\dot{R}(t)}{R(t)} = \dot{\phi}(t) \tan (\phi(t) + \omega t).
\end{align}

\eref{y} shows that the time derivative of $y(t)$ is
\begin{align}
\dot{y} &=-\omega^2 x \nonumber \\
&- \omega
\left[\dot{R}(t) \sin(\phi(t) + \omega t) + R(t) \dot{\phi}(t)\cos(\phi(t) +
  \omega t)\right]\elabel{ydot}
\end{align}
On the other hand, we know that
\begin{align}
i \omega \dot{A} e^{i\omega t} &=- \omega
\left[\dot{R}(t) \sin(\phi(t) + \omega t) + R(t) \dot{\phi}(t)\cos(\phi(t) +
  \omega t)\right]\nonumber\\
&+ i \omega\left[\dot{R}(t) \cos(\phi(t) + \omega t) - R(t)
  \dot{\phi}(t) \sin (\phi(t) + \omega t)\right] \elabel{ImAdot}\\
&=\dot{y} + \omega^2 x.\elabel{ioAdt}
\end{align}
where in \eref{ImAdot} we have exploited that the imaginary part is
zero because of \eref{dRdphi}. Combing the above equation with
\eref{nlo}, noting that $\dot{y} = \ddot{x}$, yields the following
equation for the time evolution of the amplitude:
\begin{align}
\dot{A} = \frac{e^{-i \omega t}}{i \omega} \left[(\omega^2 -
    \omega_0^2) x + f(x,y) + \rho s(t)\right].
\elabel{Adot}
\end{align}

\subsubsection{Averaging}
The above transformation is exact. To make progress, we will use the
method of averaging \cite{Guckenheimer:1983up}. Specifically, we will
time average \eref{Adot} over one period $T$
\cite{Guckenheimer:1983up,Pikovsky2003}. Averaging the driving $e^{-i
  \omega t} s(t) / (i \omega)$ yields the complex constant $E/(2\omega)$. The
second term of \eref{Adot} can be expanded in polynomials of
$x(t)=(1/2) {\rm Re} A(t) e^{i \omega t}$ and $y(t)=(1/2) {\rm Im}A(t)
e^{i \omega t}$, yielding powers of the type $(A(t) e^{i \omega t})^n
(A^*(t) e^{-i \omega t})^m$. After multiplying with $e^{-i \omega t}$
and averaging over one period $T$, only the terms with $m=n-1$ do not
vanish. Consequently, the terms that remain after averaging have the
form $g(|A|^2) A$, with an arbitrary function $g$. For small
amplitudes only the linear term proportional to $A$ and the first
non-linear term, $\propto |A|^2 A$ term are important. Finally,
averaging the first term of \eref{Adot} yields a term linear in $A$.

Summing it up, the time evolution of the amplitude of the system with
deterministic driving ($\eta_s=0$) is given by \cite{Pikovsky2003}
\begin{align}
\dot{A} &= -i \frac{\omega^2 - \omega_0^2}{2\omega} A + \alpha A - (\beta
+ i \kappa) |A|^2 A -  \frac{\rho}{2\omega} E
\end{align}
The parameters have a clear interpretation. The parameters $\alpha$
and $\beta$ describe, respectively, the linear and non-linear growth
or decay of oscillations. To have stable oscillations, both in the
presence and absence of driving, large amplitude oscillations
dominated by the nonlinear term need to decay, which means that
$\beta$ must be positive, $\beta>0$; this parameter is fixed in all
our calculations. The parameter that allows us to alter the system
from one that shows damped oscillations in the absence of driving to
one that can generate autonomous oscillations which do not rely on
forcing, is $\alpha$.  For the system to sustain free-running
oscillations, small amplitude oscillations, dominated by the linear
term, must grow, meaning that $\alpha$ must be positive,
$\alpha>0$. The case with $\alpha>0$ thus describes a system that can
perform stable limit cycle oscillations, making it a bonafide clock.
The case $\alpha<0$ describes a system that in the absence of any
driving, $E=0$, relaxes in an oscillatory fashion to a stable fixed
point with $A=0$. In the presence of weak driving, the amplitude
  $A$ at the fixed point will be non-zero but small, making the effect
  of the non-linearity weak. The
case $\alpha<0$ thus describes a system that is effectively a damped
harmonic oscillator, which only dispays sustained oscillations when
forced by an oscillatory signal. This system mimics the
uncoupled-hexamer model.

The parameter $\kappa$ describes the non-linear dependence of the
oscillation frequency on the amplitude. For the isochronous scenario
in which the phase moves with a constant velocity,
$\kappa=0$, which is what we will assume henceforth. 

Defining the parameter $\nu \equiv (\omega^2 - \omega_0^2)/(2\omega)$
and the parameter $\epsilon \equiv \rho/(2\omega)$, we can then
rewrite the above equation as
\begin{align}
\dot{A} = -i \nu A + \alpha A - \beta |A|^2 A -  \epsilon E,
\elabel{AmpEq}
\end{align}
where $A$ is the complex time-dependent amplitude, $E$ is a complex
constant, and $\nu$, $\alpha$, and $\beta$ are real
constants. \eref{AmpEq} is Eq. 2 of the main text. It provides a
universal description of a driven weakly nonlinear system near the
Hopf bifurcation where the limit cycle appears \cite{Pikovsky2003}.

To model the input noise we will add the noise term to \eref{AmpEq}:
\begin{align}
\dot{A} = -i \nu A + \alpha A - \beta |A|^2 A - \epsilon E + \rho\bar{\eta}_s(t),
\elabel{AmpEqNoise}
\end{align}
where $\bar{\eta}_s(t)$ is the noise $\eta_s(t)$ averaged
over one period of the driving:
\begin{align}
\bar{\eta}_s(t) \equiv \frac{1}{T} \int_{t-T/2}^{t+T/2} dt^\prime
\frac{e^{-i\omega t^\prime}}{i\omega} \eta_s(t^\prime).
\elabel{etasbar}
\end{align}
Since $\eta_s(t)$ is real but its prefactor $e^{-i\omega
  t}/{i\omega}$ is complex, $\bar{s}(t)$ is, in general,
complex.
Below we will describe the characteristics of the noise $\bar{\eta}_s$.

\subsubsection{Linear-Noise Approximation} {\bf Scenarios} By varying
$\alpha$ we will interpolate between two scenarios: the damped
oscillator, modeling the UHM, with $\alpha < 0$, and the weakly
non-linear oscillator that can sustain free-running oscillations,
modeling the CHM, with $\alpha > 0$. For the system with $\alpha<0$,
the amplitude of $x(t)$ when not driven is $A=0$: the system comes to
a standstill. When the system is driven, the amplitude will be
nonzero, but constant since the system is essentially linear as
described above. For the system with $\alpha>0$, $A(t)$ can exhibit
distinct types of dynamics, depending on the strength of driving and
the frequency mismatch characterized by $\nu$
\cite{Pikovsky2003}. However, here we do not consider the regimes that
$A(t)$ rotates in the complex plane; we will limit ourselves to the
scenario that $A(t) = A$ is constant, meaning that $\nu$ cannot be too
large \cite{Pikovsky2003}.  

{\bf Overview} Before we discuss the linear-noise approximation in
detail, we first give an overview. The central observation is that
both for the driven damped oscillator with $\alpha < 0$ and the driven
limit-cycle oscillator with $\alpha>0$, the complex amplitude $A$ is constant,
corresponding to a stable fixed point of the amplitude equation,
\eref{AmpEq}. In the spirit of the linear-noise approximation used to
calculate noise in biochemical networks, we then expand around the
fixed point to linear order, and evaluate the noise at the fixed
point. This approach thus assumes that the distribution of the
variables of interest is Gaussian, centered at the fixed point. More
concretely, we first expand $A(t)$ to linear order around its stable
fixed point, which is obtained by setting $\dot{A}$ in \eref{AmpEq} to
zero. This makes it possible to compute the variance of
$A$. Importantly, this variance is that of a Gaussian distribution in
the frame that co-rotates with the driving, as can be seen from
\erefstwo{x}{y}. To obtain the variance of $x$ and $y$ in the original
frame, we then transform this distribution back to original frame of
$x$ and $y$. If we can make this transformation linear, then it is
guaranteed that the distribution of $x$ and $y$ will also be
Gaussian. As we will see, the transformation can be made linear by writing $A$ as $A=u
+ i v$, where $u$ and $v$ are the real and imgainary parts of $A$,
respectively.

{\bf Expanding $A$ around its fixed point} We write $A(t) = u(t)
+ i v(t)$. \eref{AmpEqNoise} then yields for the real and imaginary part of $a(t)$:
\begin{align}
\dot{u} &= \nu v + \alpha u - \beta (u^2+v^2) u - \epsilon e_u +
\rho \bar{\eta}_u \elabel{dotu}\\
\dot{v} &=-\nu u + \alpha v - \beta (u^2+v^2) v - \epsilon e_v +
\rho \bar{\eta}_v \elabel{dotv}
\end{align}
Here, $\bar{\eta}_u$ and $\bar{\eta}_v$ are the real and imaginary
parts of the averaged noise $\bar{\eta}_s$, given by \eref{etasbar};
they are discussed below. The quantities $e_u$ and $e_v$ are the real
and imaginary parts of the driving $E$. Their respective values depend
on the phase of the driving, which is arbitrary and can be chosen
freely. For example, when the driving is $s(t) = \sin (\omega t)$,
then $e_u = 1$ and $e_v=0$, while if the signal is $s(t)
= \cos(\omega t)$, then $e_u=0$ and $e_v=1$.

We now expand $u(t)$ and $v(t)$ around their steady-state values,
$u^\ast$ and ${v^\ast}$, respectively.  Inserting this in the above
equations and expanding up to linear order yields
\begin{align}
\dot{\delta u}&= c_1 \delta u + c_2 \delta v + \rho \bar{\eta}_u \elabel{dotdu}\\
\dot{\delta v}&= c_3 \delta u + c_4 \delta v + \rho \bar{\eta}_v, \elabel{dotdv}
\end{align}
with
\begin{align}
c_1 &= \alpha - \beta (3 {u^\ast}^2 + {v^\ast}^2) \elabel{c1}\\
c_2 &=\nu - \beta 2 {u^\ast} {v^\ast}\\
c_3 &=-\nu - \beta 2 {u^\ast}{v^\ast}\\
c_4 &= \alpha - \beta ({u^\ast}^2 + 3 {v^\ast}^2).\elabel{c4}
\end{align}
The fixed points ${u^\ast}$ and ${v^\ast}$ are obtained by solving the
cubic equations \erefstwo{dotu}{dotv} in steady state.

{\bf Noise characteristics} We next have to specify the noise
characteristics of $\bar{\eta}_u(t)$ and $\bar{\eta}_v(t)$.
\eref{etasbar} reveals that the noise terms are given by 
\begin{align}
\bar{\eta}_u(t) &=-\frac{1}{\omega T}\int_{t-T/2}^{t+T/2}dt^\prime
\sin(\omega t^\prime) \eta_s(t^\prime) \\
\bar{\eta}_v(t)&=-\frac{1}{\omega T}\int_{t-T/2}^{t+T/2} dt^\prime
\cos(\omega t^\prime) \eta_s(t^\prime).
\end{align}
The method of averaging \cite{Anishchenko:2007tf} reveals that to
leading order the statistics of these quantities can be approximated by
\begin{align}
\avg{\bar{\eta}_u(t)
  \bar{\eta}_u(t^\prime)}&=\avg{\bar{\eta}_v(t)
  \bar{\eta}_v(t^\prime)}
=\frac{\sigma_s^2}{2\omega^2} \delta (t-t^\prime)\\
\avg{\bar{\eta}_u(t) \bar{\eta}_v(t^\prime)} &=0.
\end{align}

{\bf Variance-co-variance} 
From here, there are (at least) three ways to obtain the variance and
co-variance matrix of $u$ and $v$. Since the system is linear, it can be directly
solved in the time domain. Another route is via the power
spectra \cite{TanaseNicola:2006bh,Warren:2006ky}. Here, we obtain it from \cite{Gardiner85}
\begin{align}
{\bf A}{\bf C}_{uv} + {\bf C}_{uv}{\bf A^{\rm T}} = - {\bf D}_{uv}.
\elabel{ACCA}
\end{align}
The matrix ${\bf C}_{uv}$ is the variance-covariane matrix with
elements $\sigma^2_{uu}, \sigma^2_{uv}, \sigma^2_{vu}, \sigma^2_{vv}$
and ${\bf A}$ is the Jacobian of \erefstwo{dotdu}{dotdv} with elements
$A_{11} = c_1, A_{12} = c_2, A_{21} = c_3, A_{22} = c_4$.  The matrix
${\bf D}_{uv}$ is the noise matrix of $\avg{\bar{\eta}^2_{u/v}}$,
where we absorb the coupling strength $\rho = 2 \omega
\epsilon$ (cf. \eref{AmpEq}) in the noise strength:
\begin{align}
{\bf D}_{uv} &= \elabel{Duv}\left(
\begin{array}{cc}
2\epsilon^2\sigma^2_s & 0\\
0&2\epsilon^2\sigma^2_s
\end{array}
\right).
\end{align}

{\bf Transforming back} The variance-covariance matrix ${\bf C}_{uv}$,
with elements $\sigma^2_{uu}, \sigma^2_{uv}, \sigma^2_{vu},
\sigma^2_{vv}$, characterizes a Gaussian distribution in the complex
plane
\begin{align}
P(u,v) = \frac{1}{2\pi\sqrt{|{\bf C}_{uv}|}}e^{-\frac{1}{2}{\bf a}^{\rm T} {\bf C}_{uv}^{-1} {\bf a}},
\end{align}
where $|{\bf C}_{uv}|$ is the determinant of the variance-covariance
matrix ${\bf C}_{uv}$ and ${\bf C}_{uv}^{-1}$ is the inverse of ${\bf
  C}_{uv}$, and ${\bf a}$ is a vector with elements $\delta u,\delta
v$ (the deviations of the real and imaginary parts of $A=a$ from their
respective fixed points ${u^\ast}$ and ${v^\ast}$) with ${\bf a}^{\rm
  T}$ its transpose. This distribution $P(u,v)$ defines a distribution
in the co-rotating frame of the oscillator in the complex plane. To
obtain $P(x,y)$ in the original non-co-rotating frame, we need to
rotate this distribution.  \eref{Axy} shows that the corresponding
rotation is described by
\begin{align}
x(t) &= u \cos(\omega t)- v \sin (\omega t) \elabel{xt}\\
y(t) &= - \omega u \sin(\omega t) - \omega v \cos (\omega t) \elabel{yt},
\end{align}
which defines the rotation matrix 
\begin{align}
{\bf Q} = \left(
\begin{array}{cc}
\cos(\omega t) & - \sin(\omega t) \\
-\omega \sin(\omega t) &-\omega \cos(\omega t)
\end{array}
\right)
\end{align}
such that ${\bf z} = {\bf Q} {\bf a}$, with ${\bf z}$ the vector with
elements $\delta x(t) = x(t) -x^\ast(t), \delta y(t) = y(t) -
y^\ast(t)$, where $x^\ast, y^\ast$ are the rotating ``fixed'' points
of $x(t)$ and $y(t)$, i.e. their time-dependent mean values, given by
\erefstwo{xt}{yt} with $u=u^\ast$ and $v=v^\ast$. Hence, the
distribution of interest is given by
\begin{align}
P(x,y|t)=\frac{1}{2\pi \sqrt{|{\bf C}_{xy}|}} e^{-\frac{1}{2}{\bf
  z}^{\rm T}{\bf C}_{xy}^{-1}
    {\bf z}},\elabel{Pxygt}
\end{align}
where 
\begin{align}
{\bf C}_{xy}^{-1} = [{\bf Q}^{-1}]^{\rm T} {\bf C}_{uv}^{-1} {\bf
  Q}^{-1}
\elabel{CxyInv}
\end{align}
 and its inverse ${\bf C}_{xy}$ is the variance-covariance
matrix for $x,y$, with elements
$\sigma^2_{xx}(t),\sigma^2_{xy}(t),\sigma^2_{yx}(t),\sigma^2_{yy}(t)$, which
depend on time because ${\bf Q}$ depends on time.

{\bf Mutual information $I(p;t)$} Lastly, the oscillations in the
phosphorylation $p(t)$ of the hexamer models correspond to the
oscillations in $x(t)$ in the Stuart-Landau model. We therefore need
to compute the mutual information $I(x;t)$, not
$I(x,y;t)$. Specifically, we calculate the mutual information from
\begin{align}
I(x,t) = H(x) - \avg{H(x|t)}_t,
\end{align}
where the entropy $H(x) = -\int dx P(x) \log P(x)$ with $P(x) = 1/T
\int_0^T dt P(x|t)$ and the conditional entropy $H(x|t) = -
1/T\int_0^T dt\int dx P(x|t) \log P(x|t)$, with $P(x|t) =
1/\sqrt{2\pi\sigma^2_{xx}(t)} e^{-(x(t)-x^\ast
  (t))^2/(2\sigma^2_{xx}(t))}$. We emphasize that both the variance
$\sigma^2_{xx}(t)$ and the average $x^\ast(t)$ depend on time.

{\bf Summing up Approach and Parameters Fig. 3 main text} To sum up
the procedure, to compute the noise in $A=a$ we first need to obtain
the steady state values of its real and imaginary part, $\bar{u}$ and
$\bar{v}$ (see \erefsrange{c1}{c4}). These are obtained from setting
the time derivatives of $u(t)$ and $v(t)$ in
\erefstwo{dotu}{dotv} to zero;  this involves solving a cubic equation,
which we do numerically. We then compute the variance-covariance
matrix ${\bf C}_{uv}$ via \eref{ACCA}, where the elements of the Jacobian ${\bf A}$ are
given by \erefsrange{c1}{c4} and the noise matrix ${\bf D}_{uv}$
is given by \eref{Duv}.  After having obtained ${\bf C}_{uv}$, we
find the variance-covariance matrix for $x$ and $y$, ${\bf C}_{xy}$,
from \eref{CxyInv}. For Fig. 3 of the main text, $\nu = 0$,
  $\beta = \omega$, $\epsilon = 0.5\omega$. 

\subsubsection{Comparing limit cycle oscillator with damped oscillator}
Fig. 3 of the main text shows that the mutual information $I(x;t)$
increases with $\alpha$, especially when the input noise is large. To
elucidate this further, we show in \fref{SLPxyNoise} for two different
values of $\alpha$ and for two levels of the input noise, the dynamics
of the system in the plane of $x$ and $y$.  The panels not only show
the mean trajectory, indicated by the dashed line, but also samples
$(x,y)$ from $P(x,y|t_i)$ for evenly spaced time points $t_i$; 
$P(x,y|t)$ is given by \eref{Pxygt} and samples from the same time
point $t_i$ have the same color. It is seen that when the input noise
is low (left two panels), the respective distributions (``blobs'') are
well separated, both for $\alpha=-\omega$, when the system is a damped
oscillator (D.O.) (top row), and for $\alpha=3\omega$ (bottom row), when the
system is a limit-cycle oscillator (L.C.O.). However, when the input
noise is large (right column), the blobs of the damped oscillator
become mixed, while the distributions $P(x,y|t)$ of the limit-cycle
oscillator are still fairly well separated.

\begin{figure}[t]
\includegraphics[width=\columnwidth]{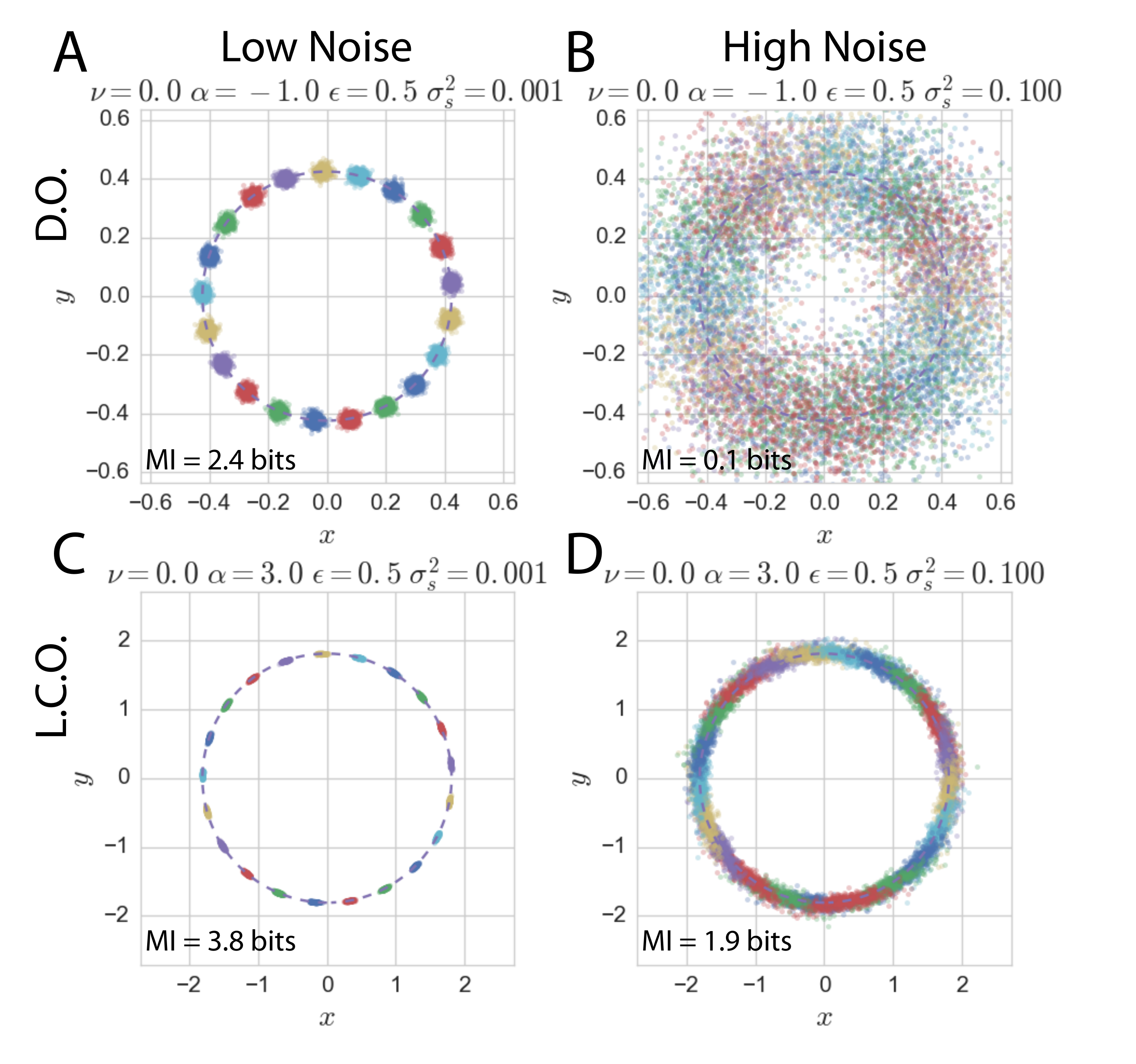}
\caption{The dynamics of the Stuart-Landau model when
  $\alpha=-\omega$, corresponding to a damped oscillator (D.O., top
  row), and when $\alpha=3\omega$, corresponding to a limit-cycle
  oscillator (L.C.O.,bottom row), both for low input noise,
  $\sigma^2_s=0.001\omega$ (left column) and high input noise,
  $\sigma^2_s=0.1\omega$ (right column). Dashed line denotes the mean
  trajectory of $(x,y)$, and the points are samples of $(x,y)$ from
  the distribution $P(x,y|t_i)$ for evenly spaced time points $t_i$;
  $P(x,y|t)$ is given by \eref{Pxygt} and points belonging to the same
  time have the same color. It is seen that when the input noise is
  low, the distributions corresponding to the different times are
  still well separated, both for the limit-cycle oscillator and the
  damped oscillator. Yet, for high noise, only for the L.C.O. are the
  distributions still reasonably separated, leading to a mutual
  information that is still close to 2 bits. In contrast, for the
  D.O., the distributions are mixed, leading to a low mutual
  information close to zero. \flabel{SLPxyNoise}}
\end{figure}

To interpret this further, we note that the mutual information $I(x;t)
= H(t) - H(t|x)$. Here, $H(t)$ is the entropy of the input signal,
which is constant, i.e. does not depend on the design of the
system. The dependence of $I(x;t)$ on the design of the system is thus
governed by the conditional entropy, given by $H(t|x) =
\avg{\avg{-\log P(t|x)}_{P(t|x)}}_{P(x)}$. The quantity $\avg{-\log
  P(t|x)}_{P(t|x)}$ quantifies the uncertainty in estimating the time
$t$ from a given output $x$; the average $\avg{\dots}_{P(x)}$
indicates that this uncertainty should be averaged over all output
values $x$ weighted by $P(x)$. The conditional entropy $H(t|x)$ is low
and $I(x;t)$ is high when, averaged over $x$, the distribution
$P(t|x)$ of times $t$ for a given $x$ is narrow. We can now interpret
\fref{SLPxyNoise}: The smaller the number of blobs that intersect the
line $x$, the higher the mutual information. Or, concomitantly, the
more the distributions are separated, the higher the mutual
information---information transmission is indeed a packing
problem. Clearly, when the input noise is low, the time can be
inferred reliably from the output even with a damped oscillator (top
left panel). For high input noise, however, the mutual information of
the damped oscillator falls dramatically because the blobs now overlap
strongly. In contrast, the distributions of the limit-cycle oscillator
are still reasonably separated and $I(x;t)$ is still
almost close to 2 bits.

\begin{figure}[t]
\includegraphics[width=\columnwidth]{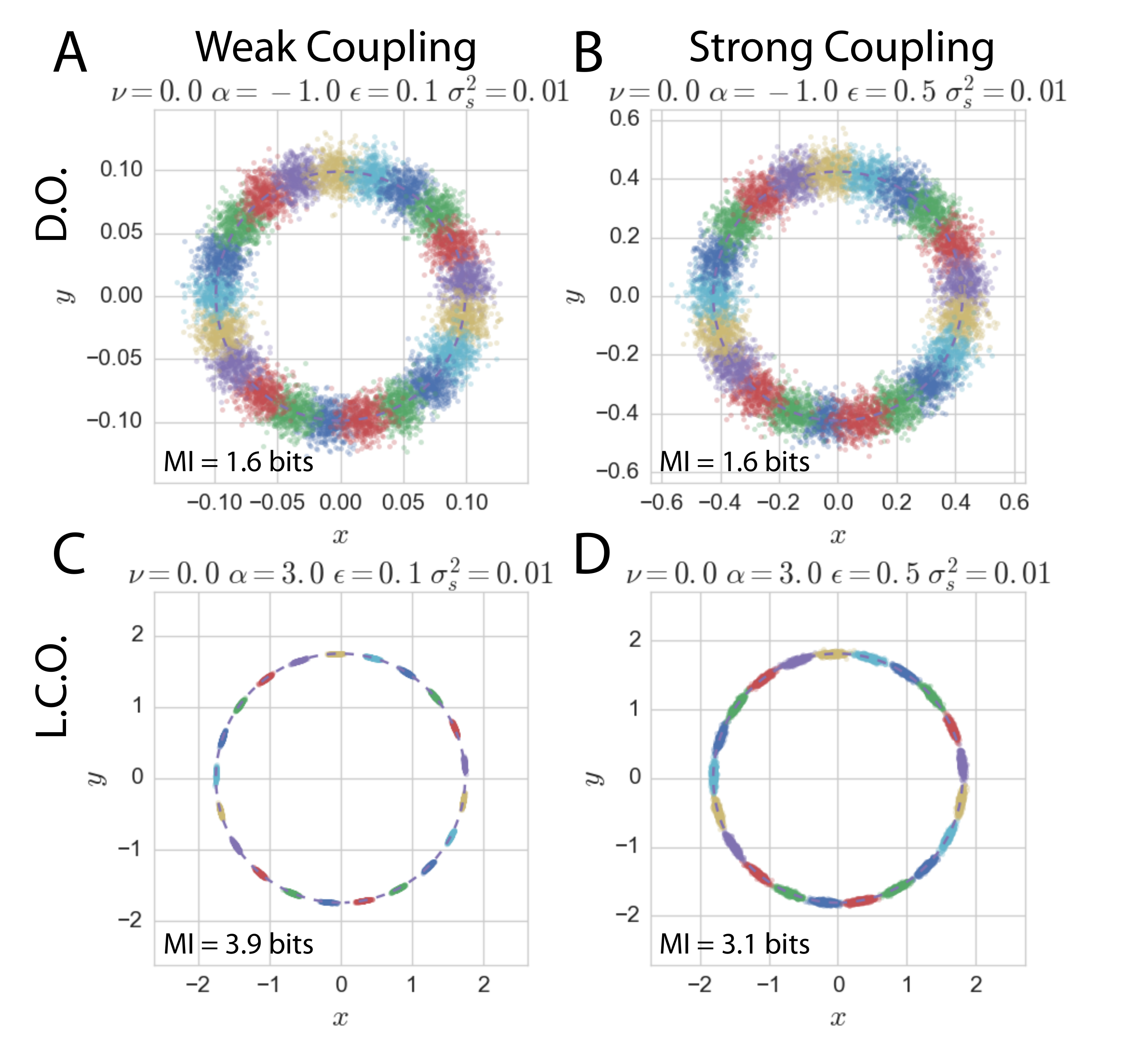}
\caption{The dynamics of the Stuart-Landau model when
  $\alpha=-\omega$, corresponding to a damped oscillator (D.O., top
  row), and when $\alpha=3\omega$, corresponding to a limit-cycle
  oscillator (L.C.O., bottom row), both for weak coupling,
  $\epsilon=0.1\omega$ (left column) and strong coupling,
  $\epsilon=0.5\omega$ (right column); the input noise is set to a low
  value, $\sigma^2_s = 0.01\omega$. Dashed line denotes the mean
  trajectory of $(x,y)$, and the points are samples of $(x,y)$ from
  the distribution $P(x,y|t_i)$ for evenly spaced times $t_i$;
  $P(x,y|t)$ is given by \eref{Pxygt} and points belonging to the same
  time have the same color. It is seen that for the D.O. the amplitude
  and the noise are small when the coupling is small (top left panel;
  note the scale on the x- and y-axis). Increasing the coupling,
  however, not only raises the amplitude (the gain), but also amplifies the
  noise, leaving the mutual information unchanged: a damped oscillator
  cannot lift the trade-off between gain and noise.  In contrast, the
  limit-cycle oscillator already exhibits large amplitude oscillations
  even for weak coupling; \b{at the same time, lowering the coupling
    strength does reduce input-noise propagation. The limit-cycle
    oscillator can thus lift the trade-off between gain and input:
    lowering the coupling raises the mutual
    information. Section \ref{sec:LCO_HO} makes these arguments
    quantitative.} It is also interesting to note that especially the
  fluctuations in the radial direction, the amplitude fluctuations,
  are strongly reduced in the L.C.O., due to the non-linearity of the
  system. \flabel{SLPxyCoupling}}
\end{figure}

\fref{SLPxyNoise} also nicely illustrates that the mutual information
would be increased if the system could estimate the time not from $x$
only, but instead from $x$ {\em and} $y$: this removes the degeneracy
in estimating $t$ for a given $x$ associated with sinusoidal
oscillations \cite{Monti:2016bp}. One mechanism to remove the
degeneracy is to have a readout system that not only reads out the
amplitude of the clock signal, but also its derivative, for example
via incoherent feedback loops \cite{Becker:2015iu}. Another
possibility is that the clock signal is read out by 2 (or more)
proteins that are out of phase with each other, as shown in
\cite{Monti:2016bp}. Indeed, while we have computed the instantaneous
mutual information between time and the output at a given time, the
trajectory of the clock signal provides more information about time,
which could in principle be extracted by appropriate readout systens
\cite{Monti:2016bp}.

Lastly, we show in \fref{SLPxyCoupling} the dynamics for two different
values of $\alpha$ and for two different values of the coupling
strength $\epsilon$. The top left panel shows that when $\epsilon$ is
small, the amplitude of the damped oscillator is very weak---note the
scale on the x- and y-axis. To increase the amplitude of the output,
the coupling strength must be increased. However, this amplifies the
input noise as well, such that the mutual information remains
unchanged (top right panel): the damped oscillator faces a fundamental
trade-off between gain and input noise that cannot be lifted. In
contrast, the limit-cycle oscillator (bottom row) already exhibits
strong amplitude oscillations even when the coupling strength
$\epsilon$ is small: the amplitude of the cycle---a bonafide
limit-cycle---is determined by the properties of the system, and is
only very weakly affected by the strength of the forcing. \b{At the
  same time, weakening the coupling does reduce the propagation of
  input noise. These two observations together explain why for the
  limit-cycle oscillator the mutual information increases as the
  coupling is reduced. In \ref{sec:LCO_HO} we elucidate these
  arguments further, and show that concerning the robustness to input
  noise, the weak-coupling regime is the optimal regime that maximizes
  the mutual information, and that in this regime a limit-cycle
  oscillator is superior over a damped oscillator.}

\begin{figure}[t]
\includegraphics[width=\columnwidth]{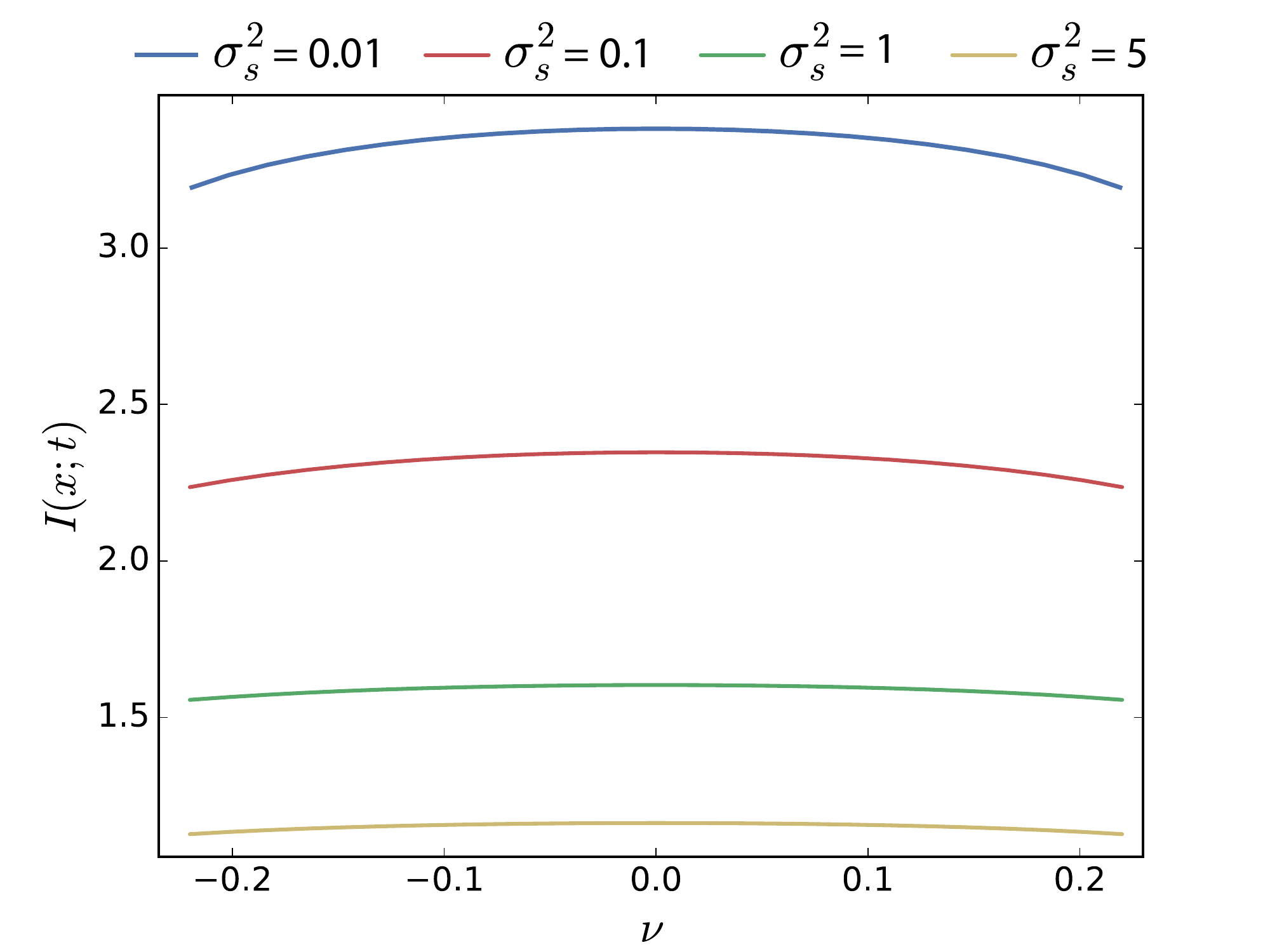}
\caption{The mutual information $I(x;t)$ in the Stuart-Landau model as
  a function of $\nu = (\omega^2 - \omega_0^2)/(2\omega)$, for
  different input-noise strengths $\sigma^2_s$. It is seen that the
  mutual information is maximized at $\nu=0$ (corresponding to
  $\omega_0=\omega$) for all input noise levels.  $\beta = 1.0
  \omega$; $\epsilon = 0.5\omega$; $\alpha=3\omega$; $\sigma^2_s$ in
  units of $\omega$. \flabel{SLnu}}
\end{figure}

\subsubsection{Optimal intrinsic frequency}
\fref{CHM}B shows that the optimal intrinsic frequency $\omega_0^{\rm
  opt}$ that maximizes the mutual information $I(p;t)$ for the
coupled-hexamer model (CHM) depends, albeit very weakly, on the
input-noise strength $\sigma^2_s$. Here we wondered whether the
Stuart-Landau model could reproduce this feature. \fref{SLnu} shows
the result. The figure shows the mutual information $I(x;t)$ as a
function of $\nu = (\omega^2 - \omega_0^2)/(2\omega)$ for different
values of $\sigma^2_s$. It is seen that the dependence of $I(x;t)$ on
$\nu$ is rather weak, yielding a broad maximum that peaks at $\nu=0$
(corresponding to $\omega_0 = \omega$) for all noise strengths. This
suggests that the optimal $\omega_0^{\rm opt}<\omega$ observed for low
input noise in the CHM arises from a stronger non-linearity in that
system than captured by the Stuart-Landau model, which describes
weakly non-linear oscillators.

\subsection{Why limit cycle oscillators are generically more
  robust to input noise than damped oscillators in the weak-coupling regime}
\label{sec:LCO_HO}
\b{The principal result of our manuscript, illustrated in Fig. 2 of
  the main text, is that a limit-cycle oscillator is more robust to
  input noise than a damped oscillator. We now address the question
  how generic this observation is, and whether it can explained from a
  simple scaling argument. To answer these questions, we will
  investigate the analytical models discussed in the previous
  sections, which are valid in the regime of weak coupling.  We will
  analyse the harmonic oscillator described in \ref{sec:ANA_HO}, which
  applies not only to the uncoupled hexamer model (UHM), but also, in
  the high-friction limit, to the push-pull network (PPN), as
  described in \ref{sec:COMP_UHM_PP}. For the coupled-hexamer model,
  we will analyse not only the Stuart-Landau model described in
  \ref{sec:SL}, but also a phase-oscillator model within the
  phase-averaging approximation \cite{Monti:2018hs}. While the
  Stuart-Landau model gives a universal description of weakly
  non-linear oscillators near the Hopf bifurcation, the
  phase-oscillator model within the phase-averaging approximation
  gives a general description of (potentially highly) non-linear
  oscillators in the weak coupling regime \cite{Monti:2018hs};
  importantly, both descriptions give the same scaling argument,
  strongly suggesting it applies to most, if not all, limit-cycle
  oscillators. The principal finding of our analysis of these models
  is that damped oscillators such as the UHM and PPN cannot lift the
  trade-off between the amplification of the output signal (the gain)
  and the propagation of input noise, while limit-cycle oscillators
  can because their oscillations have an inherent robust amplitude,
  which does not rely on external driving. Before we derive the
  principal result in detail in the paragraphs below, we first give an
  overview of the main arguments, for the case where there is no internal
  noise. In the next section (\ref{sec:AnaIntNoise}), we then discuss
  the role of internal noise and how the optimal design of the readout
  system depends on the relative amounts of internal and external
  noise.}

\b{{\bf Overview} To understand why limit-cycle oscillators (CHM) are
  generically more robust to input noise than damped oscillators (UHM
  and PPN), the role of the coupling strength $\rho$ is key. For a
  damped oscillator, the amplitude $A$ of the output oscillations (the
  signal) scales linearly with the coupling strength, $A\sim
  \rho$. However, increasing the coupling not only amplifies the
  signal, but also the input noise. Moreover, it does so by the same
  amount: the standard deviation of the output signal, $\sigma_x$ also
  scales linearly with $\rho$, $\sigma_x \sim \rho$. Consequently, the
  number of distinct time points that can be resolved, the
  signal-to-noise ratio $A / \sigma_x$, is independent of $\rho$:
  damped oscillators cannot lift the trade-off between gain and input
  noise by optimizing the coupling strength, as can also be seen in
  panels A and B of \fref{SLPxyCoupling}.} 

\b{This is in marked contrast
  to the behavior of a limit-cycle oscillator. A limit-cycle
  oscillator has an intrinsic amplitude $A$, which does not rely on
  external driving, as the amplitude of a damped oscillator does. Its
  amplitude is thus essentially independent of $\rho$, and, more
  specifically, it goes to a non-zero value as $\rho \to 0$. Moreover,
  while the amplitude remains finite, the propagation of input noise
  does go to zero as $\rho \to 0$, because, as we will show, $\sigma_x
  \sim \sqrt{\rho}$. Hence, the signal-to-noise ratio $A / \sigma_x
  \sim 1/\sqrt{\rho}$ rises as the coupling is decreased, as
  \fref{SLPxyCoupling}C/D illustrate. Although this scaling law
  naively suggests that the optimal coupling strength is $\rho
  \rightarrow 0$, we will show below that, in real systems, internal
  noise and detuning between the driving and intrinsic oscillator
  frequencies always cut off the divergence at small but finite
  $\rho$.} 

\b{Importantly, we find exactly the same scaling relation
  $A/\sigma_x \sim 1/\sqrt{\rho}$ for
  both the Stuart-Landau model and the phase-oscillator model within
  the phase-averaging approximation, which is the natural description
  of non-linear limit-cycle oscillators in the weak-coupling regime
  \cite{Pikovsky2003,Monti:2018hs}. Our analysis thus shows that
  concerning the robustness to input noise: 1) the optimal regime that
  maximizes the signal-to-noise ratio is the weak-coupling regime; 2)
  in this regime, limit-cycle oscillators are generically more robust
  than damped oscillators. We emphasize that the weak-coupling regime
  is precisely the regime where our analysis applies, indicating that
  our principal result applies to a very broad class of
  oscillators. Moreover, this result can be understood intuitively:
  while both a damped and a limit-cycle oscillator can reduce the
  propagation of input noise by lowering the coupling strength, only
  the limit-cycle oscillator still exhibits a robust amplitude in the
  weak-coupling regime, raising the signal-to-noise ratio (see
  \fref{SLPxyCoupling}). In the next paragraphs, we derive and
  elucidate the scaling of $A$ and $\sigma_x$ with $\rho$ for both
  oscillator models. The role of internal noise is discussed in the next
  section. }

\b{{\bf Damped oscillators} We will first reiterate the main findings for
the harmonic oscillator (the uncoupled hexamer model), described in
\ref{sec:ANA_HO}; these findings also apply to the push-pull network,
which corresponds to the high-friction limit of the harmonic
oscillator (see section \ref{sec:COMP_UHM_PP}).  The amplitude of the
harmonic oscillator is given by \eref{A_HO} and repeated here for
completeness:
\begin{align}
A 
&=\frac{\rho}{\sqrt{\gamma^2 \omega^2 + (\omega^2 -
    \omega_0^2)^2}} \sim \rho\elabel{A_HO2}.
\end{align}
Importantly, the amplitude increases linearly with the coupling
strength $\rho$. This result can be understood by noting that the
driving force $\rho s(t)$ scales with $\rho$ while the restoring force
$-\omega_0^2 x$ is independent of $\rho$ (see \eref{HO}). The variance
$\sigma^2_x$ of the output is, for Gaussian white input noise of
strength $\sigma^2_s$ (see
\eref{sigmaxsqHOwhite}):
\begin{align}
\sigma^2_x &= \frac{\rho^2\sigma^2_{s}}{2\gamma
  \omega_0^2} \sim \rho^2.
\elabel{sigmaxsqHOwhite2}
\end{align}
Clearly, the noise in the output $\sigma_x$ scales with the coupling
strength $\rho$. This is because increasing the coupling strength not only
amplifies the true signal $\sin(\omega t)$ but also the noise in the input
signal, $\eta_s$ (see \eref{HO}). Because both the amplitude $A$ and
the noise $\sigma_x$ scale with the couplgin strength $\rho$, the
signal-to-noise ratio is independent of the coupling strength $\rho$:
\begin{align}
\frac{A}{\sigma_x} &=\frac{\sqrt{2\gamma} \omega_0}{\sigma_s\sqrt{\gamma^2 \omega^2 + (\omega^2 -
    \omega_0^2)^2}}\sim \rho^0.\elabel{SNR_HO}
\end{align}
Indeed, these systems cannot lift the trade-off between gain and
noise: amplifying the signal inevitably also amplifies the noise in
the input. This is in marked contrast to the
limit-cycle oscillators, as we show next.
}

\b{ {\bf Limit-cycle oscillator: Stuart-Landau model}  To develop
    our argument, we consider the case that the frequency mismatch
    $\nu=(\omega^2-\omega_0^2)/(2\omega)=0$. Moreover, we choose the
    phase of the driving signal such that $e_v=0$, as a result of
    which $v^\ast=0$ (see~\eref{dotv}). With $v^\ast=0$, the
    steady-state value of the phase is $\phi^\ast=0$, while the mean
    amplitude of the limit cycle becomes
    $R^\ast=|u^\ast|$. Importantly, this amplitude, which can be
    obtained by solving the cubic equation for $u$ (\eref{dotu}) is
    very insensitive to the coupling strength $\rho$---this is indeed
    a hallmark of a limit-cycle oscillator. As a result, even for the
    weakest coupling strengths $\rho$, the system exhibits a robust
    amplitude $A=R^\ast$, as illustrated in \fref{SLPxyCoupling}C/D.
  } \b{ Since with $v^\ast=0$ the amplitude is $R^\ast=|u^\ast|$, its
    variance is $\sigma^2_R = \sigma^2_u$. Moreover, the variance in
    the phase is $\sigma^2_\phi = \sigma^2_v / {R^\ast}^2$. With
    $\nu=0$ and $v^\ast=0$, $c_2$ and $c_3$ in \erefsrange{c1}{c4} are
    both zero, which then yields the following expressions for the
    variance in $u$ and $v$ (using that $\epsilon \equiv \rho /
    (2\omega)$):
\begin{align}
\sigma^2_u &= \frac{\rho^2 \sigma^2_s}{4 (-\alpha + \beta 3
  {u^\ast}^2)\omega^2}\elabel{sigu}\\
\sigma^2_v &=\frac{\rho^2 \sigma^2_s}{4 (-\alpha + \beta {u^\ast}^2)
  \omega^2}.\elabel{sigv}
\end{align}
Before we discuss the signal-to-noise ratio in the limit-cycle
oscillator, we note that for a harmonic oscillator with $\beta=0$, the
method of averaging yields $\alpha = -\gamma / 2$, showing that the
result above indeed reduces to that for a harmonic oscillator with
$\omega_0=\omega$ (see \eref{sigmaxsqHOwhite2}).
}
\b{
We now analyze the numerator and denominator of \erefstwo{sigu}{sigv}
for the limit-cycle oscillator with $\beta>0$. The numerator increases
with the coupling strength $\rho$, as observed for the harmonic
oscillator; this reflects the fact that also in the limit-cycle
oscillator, the input fluctuations are amplified by the gain
$\rho$. This numerator is the same for both $\sigma^2_u$ and
$\sigma^2_v$. The denominator, however, is
larger for $\sigma^2_u$ than for $\sigma^2_v$. Indeed, the restoring
force for amplitude fluctuations, corresponding to
$\sigma^2_u=\sigma^2_R$, is larger than that for the phase
fluctuations, $\sigma^2_\phi=\sigma^2_v/{R^\ast}^2$. This is the
remnant of the fact that limit-cycle oscillators, in the absence of
any driving, exhibit a neutral mode in the direction along the limit
cycle; even with the coupling, this thus remains the soft mode. It is
predominantly these fluctuations, $\sigma^2_v$, that limit the
precision in estimating the time.  Interestingly, since we have chosen
the phase of the input such that $v^\ast=0$ and $R^\ast=|u^\ast|$, an
inspection of \eref{dotu} shows that $-\alpha + \beta
{u^\ast}^2=\epsilon/R^\ast=\rho/(2\omega R^\ast)$. Hence, we find that
\begin{align}
\sigma^2_v = \frac{\rho^2 \sigma^2_sR^\ast}{2\rho \omega} \sim \rho.
\elabel{varLCO}
\end{align}
The expression shows that the coupling not only amplifies the input
noise (the numerator), but also that it generates a restoring force
that tames these fluctuations (the denominator). The latter is in
marked contrast to the harmonic oscillator, which lacks this restoring
force (see \eref{sigmaxsqHOwhite2}). Consequently, while the output
noise $\sigma^2_x$ of the
harmonic oscillator scales as $\rho^2$ (\eref{sigmaxsqHOwhite2}),
that of the limit-cycle oscillator scales as $\rho$. We also
note that the restoring force decreases with the amplitude $R^\ast$ of
the limit cycle.  }

\b{
\eref{varLCO} shows that the signal-to-noise ratio $A/\sigma_v =
R^\ast/\sigma_v$ is given by
\begin{align}
\frac{A}{\sigma_{v}} =\frac{1}{\sigma_s}\sqrt{\frac{2\omega R^\ast
    }{\rho}}\sim \frac{1}{\sqrt{\rho}},
\elabel{SNR_SL}
\end{align}
where we have used that for small $\rho$ the amplitude $R^\ast$ has a
finite value.  Clearly, in the weak coupling limit, the
signal-to-noise ratio of the limit-cycle oscillator increases as
$\rho$ decreases, in contrast to the signal-to-noise ratio of the
damped oscillator, which is independent of $\rho$ (\eref{SNR_HO}). As
a result, for sufficiently weak coupling, a limit-cycle oscillator
will inevitably become superior to a damped oscillator.
Fundamentally, the reason is that the limit-cycle oscillator has an
intrinsic amplitude which does not rely on external driving, while the
damped oscillator does not: in both systems the input
fluctuations are only weakly amplified in the weak-coupling regime, but
only the limit-cycle oscillator has in this regime still a strong amplitude
that raises the signal above the noise.}

\b{We can also obtain a signal-to-noise ratio by dividing the amplitude
of the limit-cycle $A=2\pi$ by the standard deviation of the phase,
$\sigma_\phi=\sigma_v/R^\ast$:
\begin{align}
\frac{A}{\sigma_\phi} = \frac{2\pi}{\sigma_s}\sqrt{\frac{2\omega R^\ast}{\rho}}
\end{align}
This indeed gives the same scaling with the coupling constant $\rho$ and the
radius of the limit cycle $R^\ast$.
}

\b{ {\bf Limit-cycle oscillator: Phase-averaging method} The
  Stuart-Landau model describes a weakly non-linear system near the
  Hopf bifurcation. Yet, the coupled-hexamer model exhibits
  large-amplitude oscillations. We therefore also investigate a
  phase-oscillator model, which describes non-linear oscillators with
  a robust limit cycle. We analyze this model via the phase-averaging
  method, which applies in the regime that the intrinsic frequency
  $\omega_0$ is close to the driving frequency $\omega$ and the
  coupling $\rho$ is weak \cite{Pikovsky2003,Monti:2018hs}. This
  framework provides a description of the dynamics of the phase
  difference $\psi \equiv \phi - \omega t$ between the phase of the
  clock, $\phi$, and that of the external signal $\omega t$:
\begin{align}
\dot{\psi} = \nu + \rho_\psi Q(\psi) + \rho_\psi \eta_s,
\end{align}
where, as before, $\nu = (\omega^2 - \omega^2_0)/(2 \omega)$,
$\eta_\psi$ is a Gaussian white noise source
$\avg{\eta_\psi(t)\eta_\psi(t^\prime)} = \sigma^2_{s} \delta
(t-t^\prime)$, $\rho_\psi$ is the coupling strength, and
$Q(\psi)=\int_0^T dt^\prime Z(\psi + \omega t^\prime) s(t^\prime)$ is
the force acting on $\psi$, given by the convolution of instantaneous
phase-response curve $Z(\phi)$ and the driving signal $s(t)$
\cite{Pikovsky2003,Monti:2018hs}. In the phase-locked regime, the
deterministic equation $\dot{\psi} = \nu + \rho_\psi Q(\psi)$ always
has a stable fixed point $\psi^*$. Linearizing about this fixed point,
we find: 
\begin{align}
\dot{\delta \psi} = - \rho_\psi \zeta \delta \psi + \rho \eta_\psi,
\elabel{dotdpsi}
\end{align}
where $\zeta$ is the linearization of the force $Q(\psi)$ around the
fixed point $\psi^*$.  From this we obtain for the variance
\begin{align}
\sigma^2_\psi = \frac{\rho_\psi^2 \sigma^2_{s}}{2\rho_\psi \zeta}\sim \rho_\psi.
\elabel{sigma2psi}
\end{align}
We note that, as in the Stuart-Landau description (see \eref{varLCO}), the numerator
scales with $\rho_\psi^2$, because of the amplification of the input
fluctuations. The denominator scales, as in the
Stuart-Landau model, with $\rho_\psi$, reflecting the fact that the
restoring force that tames fluctuations in $\psi$ increases with the
coupling strength $\rho_\psi$. In fact, not only the scaling with
$\rho$ is the same in the Stuart-Landau model and the phase-averaging
method, but also the scaling with $R^\ast$; this can be understood by
noting that $\rho_\psi = \rho / R^\ast$, which comes from the factor
$\partial \phi / \partial x$ that arises in reducing the dynamics of
$x$ to that of $\phi$ and $\psi$ see
\cite{Pikovsky2003,Monti:2018hs}.}

\b{
The amplitude of the limit cycle $A=2\pi$ is constant. This
means that in this description, the signal-to-noise ratio---the
number of time points that can be inferred from the phase $\psi$---scales as
\begin{align}
\frac{A}{\sigma_\psi} \sim \frac{1}{\sqrt{\rho}}.
\end{align}
Hence, as found for the Stuart-Landau model (\eref{SNR_SL}), also in this description the signal-to-noise ratio of a
limit-cycle oscillator increases as the coupling strength decreases, in
contrast to that of a damped oscillator for which the signal-to-noise
ratio is independent of coupling strength.
}

\b{{\bf Role of detuning} Lastly, while a finite detuning $\nu\neq 0$
  necessitates a minimal coupling strength $\rho$ to bring the system
  inside the Arnold tongue, as illustrated in \fref{CHM}D,
  \eref{dotdpsi} indicates that inside the Arnold tongue the scaling
  of the signal-to-noise ratio $A/\sigma_\psi$ with $\rho$ does not
  depend on the amount of detuning $\nu$---the detuning generates a
  constant force which affects the fixed point $\psi^\ast$, but it
  does not affect the restoring force for fluctuations around $\psi^\ast$.}

\subsection{Role of internal noise}
\label{sec:AnaIntNoise}
\b{In the above sections we studied the robustness of the three
  different systems to input noise. We now address the role of
  internal noise, which arises from the intrinsic stochasticity of
  chemical reactions. First, in the next section, we study the
  signal-to-noise ratio of these systems in the presence of internal
  noise only. In the subsequent section, we then address their
  performance in the presence of both internal and input noise.  The
  coupled-hexamer model is again described by the Stuart-Landau model
  and the phase-averaging method
  of the previous section, while the push-pull network and the
  uncoupled-hexamer model are described by the damped
  oscillator of section \ref{sec:ANA_HO}; the latter system describes not
  only the uncoupled-hexamer model, but also, in the high-friction
  limit, the push-pull network of \ref{sec:ANA_PPN} (see also
  \ref{sec:COMP_UHM_PP}).}

\subsubsection{Robustness to internal noise}
\label{sec:AnaIntNoiseOnly}
\b{The derivation of the signal-to-noise ratio of the respective
  systems in the presence of internal noise closely follows that on
  input noise: the principal difference
  concerns the scaling of the noise with the coupling strength.}

\b{{\bf Damped oscillator} To study the role of internal noise, we can
  add an intrinsic noise term to \eref{HO}. This will yield the same
  expression for $\dot{x}$ as that in the presence of external noise,
  except that the external noise term scales with the coupling
  strength $\rho$, while the internal noise term does not. Hence, we
  find for the variance of the output $\sigma^2_x$ in the presence of internal
  Gaussian white noise of strength $\sigma^2_{\rm int}$:
\begin{align}
\sigma^2_x &= \frac{\sigma^2_{\rm int}}{2\gamma
  \omega_0^2}.
\elabel{sigmaxsqHOIntNoise}
\end{align}
Note that the noise $\sigma^2_x$ is independent of the coupling strength.}

\b{The expression for the amplitude is still given by \eref{A_HO}
  (\eref{A_HO2}). Combining this expression with
  \eref{sigmaxsqHOIntNoise} then yields the following expression for
  the signal-to-noise ratio for the damped oscillator with internal
  noise only:
\begin{align}
\frac{A}{\sigma_x} &=\frac{\sqrt{2\gamma} \omega_0 \rho}{\sigma_{\rm int}\sqrt{\gamma^2 \omega^2 + (\omega^2 -
    \omega_0^2)^2}}\sim \rho.
\elabel{SNR_HO_IntNoise}
\end{align}
Clearly, the signal-to-noise ratio now increases with the coupling
strength $\rho$. Whereas with input noise both the noise $\sigma_x$
and the amplitude $A$ scale with $\rho$ such that the signal-to-noise
ratio is independent of $\rho$, with internal noise the amplitude $A$
scales with $\rho$ but the noise $\sigma_x$ does not; increasing the
coupling thus makes it possible to raise the output signal above the
internal noise.  }

\b{{\bf Limit-cycle oscillator: Stuart-Landau model} Also for the
  Stuart-Landau model, the principal difference between the internal
  and input noise is that the former does not scale with the coupling
  strength $\rho$ while the latter does. Following the steps from
  \eref{sigu} to \eref{varLCO}, but with the effective input noise
  $\rho^2 \sigma^2_s$ replaced by the internal noise $\sigma^2_{\rm
    int}$, we find that in the presence of internal noise, the output
  noise is given by
\begin{align}
\sigma^2_v = \frac{\sigma^2_{\rm int}R^\ast}{2\rho \omega}.
\elabel{varLCOIntNoise}
\end{align}
Importantly, $\sigma^2_v$ decreases as the coupling $\rho$ is
increased. As we have seen above for the case of input noise,
\eref{varLCO}, for the limit-cycle oscillator the coupling to the
input yields a restoring force that increases with $\rho$.}

\b{With the amplitude $A=R^\ast$, we then obtain the following
  signal-to-noise ratio:
\begin{align}
\frac{A}{\sigma_{v}} =\frac{\sqrt{2\rho \omega
    R^\ast}}{\sigma_{\rm int}}\sim \sqrt{\rho}.
\elabel{SNR_SL_IntNoise}
\end{align}
Before we discuss the scaling of the signal-to-noise ratio with
$\rho$, we first note that by replacing $\rho_\psi^2 \sigma^2_s$ by
$\sigma^2_{\rm int}$ in \eref{sigma2psi}, we see that the
phase-averaging method yields the same scaling of the output noise
 and hence the signal-to-noise ratio with
$\rho$ as the Stuart-Landau model does.}

\b{\eref{SNR_SL_IntNoise} shows that increasing the coupling of the
  limit-cycle oscillator to the input raises the signal-to-noise
  ratio, as it does for the damped oscillator
  (\eref{SNR_HO_IntNoise}). However, the origin is markedly different:
  for the damped oscillator, a stronger coupling yields a larger
  amplitude (\eref{A_HO2}) while the noise $\sigma_x$
  (\eref{sigmaxsqHOIntNoise}) remains constant, whereas for the
  limit-cycle oscillator the amplitude is essentially unaffected by
  the coupling yet the noise (\eref{varLCOIntNoise}) decreases as
  $\rho$ increases, because of the larger restoring force. This
  difference manifests itself in a different scaling with $\rho$,
  which has an interesting consequence: Because the signal-to-noise
  ratio of the limit-cycle oscillator scales with $\sqrt{\rho}$ while
  that of the damped oscillator scales with $\rho$, in the
  weak-coupling regime the limit-cycle oscillator will not only be
  more robust to input noise, as discussed in the previous section,
  but will also be more resilient to internal noise.}

\b{However, this analysis also shows that the regime of weak coupling
  is not necessarily the optimal one: increasing $\rho$ enhances the
  suppression of internal noise. It should be realized, however, that
  the analysis presented here is an analysis that strictly applies
  only in the regime of weak coupling. Indeed, for large coupling
  other effects which are not captured by our analysis will inevitably
  come into play. For example, the output signal becomes
  non-sinusoidal because of the fact that the phosphorylation level
  $p(t)$ is bounded between zero and unity; these non-sinusoidal
  oscillations tend to reduce information transmission
  \cite{Monti:2018hs}. Moreover, combining the observations from the
  previous section on input-noise propagation, which decreases as the
  coupling $\rho$ is decreased, and the observations above on the
  suppression of internal noise, which increases with $\rho$, predicts
  that in the presence of both noise sources there exists an optimal
  coupling strength that maximizes the mutual information. In
  addition, it predicts that the magnitude of the optimal coupling
  strength depends on the relative amounts of input noise and internal
  noise. This is what we show in the next section.}

\subsubsection{Signal-to-noise ratio in presence of input noise and internal noise}
\label{sec:AnaIntExtNoise}
\b{{\bf Damped oscillator}
In the presence of both internal and external noise, the noise of the
output of the damped oscillator is, combining
\erefstwo{sigmaxsqHOwhite2}{sigmaxsqHOIntNoise}:
\begin{align}
\sigma^2_x = \frac{\rho^2\sigma^2_{s}}{2\gamma
  \omega_0^2} + \frac{\sigma^2_{\rm int}}{2\gamma
  \omega_0^2}.
\elabel{sigmaxsqHOTotNoise}
\end{align}
Note that for small coupling strength $\rho$ the internal noise
(second term) dominates, while for large $\rho$ the input noise
dominates.}

\b{
Combining this expression with that for the amplitude, \eref{A_HO},
yields the following signal-to-noise ratio
\begin{align}
\frac{A}{\sigma_x} &=\frac{\sqrt{2\gamma} \omega_0 \rho}{\sqrt{\rho^2
    \sigma^2_s + \sigma^2_{\rm int}}\sqrt{\gamma^2 \omega^2 + (\omega^2 -
    \omega_0^2)^2}}\sim \frac{a \rho}{\sqrt{b\rho^2 + c}},
\end{align}
where $a$, $b$, and $c$ are constants independent of $\rho$.
Hence, for small $\rho$, the signal-to-noise ratio scales linearly
with $\rho$ because in this regime the rise of the amplitude $A$ with
$\rho$ makes it possible to lift the signal above the internal
noise. Yet, for large $\rho$, the signal-to-noise ratio becomes
independent of $\rho$, because then the external noise dominates,
which scales with $\rho$ in the same way as the amplitude does. We
emphasize that these calculations pertain to the push-pull network
(PPN) and the uncoupled-hexamer model (UHM) provided that these systems
remain in the linear-response regime; as discussed in the previous
section (see also section \ref{sec:CompIntNoise}), for very large
coupling, the push-pull network and uncoupled-hexamer model will be
driven out of the linear-response regime because the output $p(t)$ is
bounded from above and below; this reduces information
transmission. We thus expect a broad plateau, precisely as the
simulation data of the PPN and UHM show
(\fref{OptCouplingIntExtNoise}A/B).}

\b{{\bf Limit-cycle oscillator: Stuart-Landau model} In the presence of
both input and internal noise, the output noise in the Stuart-Landau
model is, combining \erefstwo{varLCO}{varLCOIntNoise}:
\begin{align}
\sigma^2_v = \frac{\rho\sigma^2_sR^\ast}{2\omega} +
\frac{\sigma^2_{\rm int} R^\ast}{2\rho \omega}.
\end{align}
While the first term (the input noise) scales with $\rho$
because the coupling amplifies the input noise more than the restoring
force tames it (see discussion below \eref{varLCO}), the second term
decreases with $\rho$ because of the restoring force.  This expression
yields for the signal-to-noise ratio $A/\sigma_v = R^\ast/\sigma_v$
\begin{align}
\frac{A}{\sigma_{v}} =\sqrt\frac{2 \rho \omega
    R^\ast}{\rho^2 \sigma^2_s + \sigma^2_{\rm int}}\sim
\sqrt\frac{a \rho}{b\rho^2+c}.
\end{align}
It is seen that the signal-to-noise ratio increases with the coupling
strength for small $\rho$, scaling as $\sqrt{\rho}$, because for weak
coupling the intrinsic noise dominates over the input noise, and
increasing the coupling raises the restoring force that contains these
fluctuations. In the large coupling regime, the input noise will
dominate and then the signal-to-noise ratio will decrease with $\rho$
as $1/\sqrt{\rho}$---while the amplitude is essentially independent of
$\rho$, increasig $\rho$ amplifies the propagation of the input
fluctuations. This equation thus predicts a pronounced maximum in the
signal-to-noise ratio for the limit-cycle oscillator, as, in fact,
observed for the coupled-hexamer model, see
\fref{OptCouplingIntExtNoise}C. Since the phase-averaging method
yields the same scaling with $\rho$ for both the internal and external
noise as the Stuart-Landau model, it predicts the same behaviour.}

\b{Importantly, the optimal value of the coupling constant $\rho$
    that maximizes the mutual information depends on the relative
    amounts of internal and external noise: the optimal coupling
    constant decreases as the input noise increases with respect to
    the internal noise. The results of our coupled-hexamer model
    (\fref{OptCouplingIntExtNoiseDetuning}) indicate that at least the
    cyanobacterium {\it S. elongatus} is in the regime where the
    external noise dominates and the optimal coupling is weak. In this
    regime, the limit-cycle oscillator is superior to the damped
    oscillator, as the analysis of section \ref{sec:LCO_HO} shows.}

%
%

\end{document}